# Interactive Encoding and Decoding Based on Binary LDPC Codes with Syndrome Accumulation

Jin Meng and En-hui Yang

This work was supported in part by the Natural Sciences and Engineering Research Council of Canada under Grants RGPIN203035-02 and RGPIN203035-06, and by the Canada Research Chairs Program.

Jin Meng and En-hui Yang are with the Dept. of Electrical and Computer Engineering, University of Waterloo, Waterloo, Ontario N2L 3G1, Canada. Email: j4meng@uwaterloo.ca, ehyang@uwaterloo.ca

## Abstract

Interactive encoding and decoding based on binary low-density parity-check codes with syndrome accumulation (SA-LDPC-IED) is proposed and investigated. Assume that the source alphabet is $\mathbf{GF}(2)$, and the side information alphabet is finite. It is first demonstrated how to convert any classical universal lossless code $\mathcal{C}_n$ (with block length $n$ and side information available to both the encoder and decoder) into a universal SA-LDPC-IED scheme. It is then shown that with the word error probability approaching $0$ sub-exponentially with $n$, the compression rate (including both the forward and backward rates) of the resulting SA-LDPC-IED scheme is upper bounded by a functional of that of $\mathcal{C}_n$, which in turn approaches the compression rate of $\mathcal{C}_n$ for each and every individual sequence pair $(x^n, y^n)$ and the conditional entropy rate $\mathrm{H}(X|Y)$ for any stationary, ergodic source and side information $(X, Y)$ as the average variable node degree $\bar{l}$ of the underlying LDPC code increases without bound. When applied to the class of binary source and side information $(X, Y)$ correlated through a binary symmetrical channel with cross-over probability unknown to both the encoder and decoder, the resulting SA-LDPC-IED scheme can be further simplified, yielding even improved rate performance versus the bit error probability when $\bar{l}$ is not large. Simulation results (coupled with linear time belief propagation decoding) on binary source-side information pairs confirm the theoretic analysis, and further show that the SA-LDPC-IED scheme consistently outperforms the Slepian-Wolf coding scheme based on the same underlying LDPC code. As a by-product, probability bounds involving LDPC established in the course are also interesting on their own and expected to have implications on the performance of LDPC for channel coding as well.

## Index Terms

Belief propagation decoding, distributed source coding, entropy, interactive encoding and decoding, low-density parity-check code, rateless Slepian-Wolf coding, syndrome accumulation.

# I. Introduction

Recently, the concept of interactive encoding and decoding (IED) was formalized in [1], [2]. When applied to (near) lossless one way learning (i.e. lossless source coding) with decoder only side information, IED can be easily explained via Figure 1, where $X$ denotes a finite alphabet source to be learned at the decoder, $Y$ denotes another finite alphabet source that is correlated with $X$ and only available to the decoder as side information, and $R$ denotes the average number of bits per symbol exchanged between the encoder and the decoder measuring the rate performance of the IED scheme used. As evident from Figure 1, IED distinguishes itself from non-interactive Slepian-Wolf coding (SWC) in the fact that two-way communication is allowed in IED.

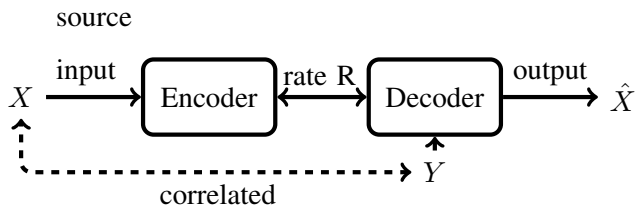

Fig. 1.  Interactive encoding and decoding for one way learning with side information at the decoder

By allowing interactions between the encoder and the decoder, IED has several advantages over SWC [1], [2]. For example, in comparison with SWC, it was shown [1], [2] that IED not only delivers better first-order (asymptotic) performance for general stationary, non-ergodic source-side information pairs, but also achieves better second-order performance for memoryless pairs with known statistics. Furthermore, in contrast to the well known fact that universal SWC does not exist, it was shown [2] that coupled with any classical universal lossless code $\mathcal{C}_n$ (with block length $n$ and with the side information available to both the encoder and decoder) such as the one in [3], one can build an IED scheme which is asymptotically optimal with respect to the class of all stationary, ergodic sources-side information pairs. Indeed, the corresponding IED scheme achieves essentially the same rate performance as that of $\mathcal{C}_n$ for each and every individual sequence pair $(x^n, y^n)$, even though the side information is not available to the encoder in the case of IED, while the word decoding error probability can be made arbitrarily small.

The above advantages make IED much more appealing than Slepian-Wolf coding to applications where the one-way learning model depicted in Figure 1 fits. However, the IED schemes constructed in [1], [2] do not have an intrinsic structure that is amenable to implement in practice. A big challenge is then how to design universal IED schemes with both low encoding and decoding complexity. To address this challenge partially, linear IED schemes, which use linear codes for encoding, were later considered in [4]. The encoder of a linear IED scheme can be conveniently described by a parity-check matrix. Based on different random matrix ensembles, two universal linear IED schemes were proposed therein. The first universal linear IED scheme proposed in [4] makes use of Gallager-type of matrix ensembles, where each matrix element is generated independently, selects randomly a matrix from such an ensemble, and then divides the selected matrix into several sub-matrices, each of which is used to generate new syndromes in each round of interaction. In the second universal linear IED scheme proposed in [4], Gallager-type ensembles are extended into vector-type ensembles, where each column of matrices is generated independently, and a matrix is generated in such way that each of its sub-matrices is randomly picked from such a vector-type ensemble; in each round of interaction, new syndromes are then generated by applying syndrome accumulation (described in [4]) once to each and every of those sub-matrices. Define the density of a linear IED scheme as the percentage of non-zero entries in its parity-check matrix. It was then shown [4] that there is no performance loss by restricting IED to linear IED and even to linear IED with density $\Omega(\frac{\ln n}{n})$, where $n$ is the block length. Thus the encoding complexity of universal IED can be kept as low as $O(n \ln n)$.

Although linear IED considered in [4] tackles its encoding complexity very well, its decoding complexity is largely untouched due to the adoption of maximum likelihood (ML) decoding, which results in exponential decoding complexity with respect to block length $n$. One of the main purposes of this paper is to address the issue of decoding complexity by building IED schemes from linear codes with low decoding complexity. This leads us to consider low-density parity-check (LDPC) codes, due to their linear complexity decoding based on belief propagation (BP) and successful application to fix-rate Slepian-Wolf coding [5] [6] [7] [8].

An LDPC code is a linear code with a sparse parity check matrix, each of whose rows and columns has only a finite number of non-zero elements with respect to its block length. Important parameters of an LDPC code include the ratio between the numbers of rows and columns (called Slepian-Wolf rate), and the portions of rows and columns with certain number of non-zero elements (called the check and variable degree distributions of the LDPC code). Given a block length $n$ and a Slepian-Wolf rate, one way to generate an LDPC code with the given Slepian-Wolf rate, is to randomly select a matrix as its parity check matrix from an ensemble in which all matrices share the same Slepian-Wolf rate, and check and variable degree distributions.

Since rows and columns of parity check matrix of an LDPC code are not generated independently, the approach of dividing the whole matrix into several sub-matrices adopted in [4] can not deliver good results from both theoretical and practical perspectives. To overcome this problem, in this paper, we shall modify syndrome accumulation (SA) used in [4] to adapt the encoding rates of the LDPC code for IED. The resulting scheme is called an interactive encoding and decoding scheme based on a binary LDPC code with syndrome accumulation (SA-LDPC-IED); its performance is then analyzed theoretically and evaluated practically based on minimum coding length decoding and BP decoding, respectively. It is shown that coupled with any classical lossless code $\mathcal{C}_n$ (with side information available to both the encoder and decoder), one can always construct an SA-LDPC-IED scheme such that

- the word decoding error probability approaches $0$ sub-exponentially with $n$; and
- the total rate (including both the forward and backward rates) of the resulting SA-LDPC-IED scheme is upper bounded by a functional of that of $\mathcal{C}_n$, which in turn approaches the compression rate of $\mathcal{C}_n$ for each and every individual sequence pair $(x^n, y^n)$ and the conditional entropy rate $\mathrm{H}(X|Y)$ for any stationary, ergodic source and side information $(X, Y)$ as the average variable node degree $\bar{l}$ of the underlying LDPC code increases without bound.

When applied to the class of binary source and side information $(X, Y)$ correlated through a binary symmetrical channel with cross-over probability unknown to both the encoder and decoder, the resulting SA-LDPC-IED scheme can be further simplified, yielding even improved rate performance versus the bit error probability when $\bar{l}$ is not large.

It should be pointed out that in the literature (see for example [9], [10], [11], and references therein), there have been several attempts towards building rateless (or rate-adaptive) SWC schemes using LDPC codes. Specifically, the technique of SA was used to construct the so-called LDPCA codes in [11]. Our SA-LDPC-IED schemes differ from the rateless SWC schemes in the following aspects:

- We are concerned with the total rate defined as the number of bits exchanged between the encoder and the decoder per symbol, while only the forward rate (from the encoder to the decoder) is considered in rateless SWC schemes.

- We assume that the joint statistics of source and side information are unknown to both the encoder and decoder, while the joint statistics are available for decoding in rateless SWC schemes.

- We provide theoretical analysis for our SA-LDPC-IED schemes, while the performance of those rateless SWC schemes has been evaluated mainly through simulation.

The rest of the paper is organized as follows. In section II, several definitions and convention are introduced to facilitate the following discussion. The concept of syndrome accumulation is revised and SA-LDPC-IED schemes are constructed in section III, while the performance analysis is performed in section IV in terms of the forward and backward rates versus the word error probability for individual sequence pairs $(x^n, y^n)$ and stationary, ergodic source-side information pairs $(X, Y)$, and in section V in terms of the forward and backward rates versus the bit error probability for binary source-side information pairs $(X, Y)$ correlated through a binary symmetrical channel. The section VI is devoted to practical implementation and simulation results, followed by the conclusion in section VII.

## II. Preliminaries and Convention

In this section, we first set out our notation for the paper and then review some concepts related to LDPC codes.

Throughout the paper, we use uppercase and lowercase letters to denote random variables and their realizations, respectively. Let $\mathcal{B}$ be the binary alphabet, and $\mathcal{B}^+$ the set of all finite strings from $\mathcal{B}$. Let $\mathcal{B}^n$ denote the set of all strings of length $n$ from $\mathcal{B}$. Similar notation applies to other alphabets (e.g. $\mathcal{Y}$) as well. A vector of dimension $n$ is represented by a letter with superscript $n$, e.g. $b^n$; a matrix of dimension $m \times n$ is represented by a bold letter with subscript $m \times n$, e.g. $\mathbf{H}_{m \times n}$. Whenever superscripts and subscripts are clear from context, they will be omitted. For example, when there is no ambiguity, we shall simply write $b^n$ as $b$ and $\mathbf{H}_{m \times n}$ as $\mathbf{H}$. The entropy function based on logarithm with bases 2 and $e$ will be denoted by $\mathrm{H}(\cdot)$ and $\mathrm{H}_e(\cdot)$, respectively. We will denote by $\mathbb{E}(\cdot)$ the expectation operator, and by $wt(\cdot)$ the Hamming weight function counting the number of non-zero elements in a vector.

For any two sequences $\{a_i\}_{i=1}^n$ and $\{b_i\}_{i=1}^n$, we write $a_n \sim b_n$ if

$$\lim_{n \to \infty} \frac{a_n}{b_n} = 1 \ .$$

Furthermore, for any positive integer $x$, define

$$\pi(x) \triangleq \begin{cases} 0 & \text{if } x \text{ is even} \\ 1 & \text{otherwise.} \end{cases} \tag{2.1}$$

Consider now a linear block code with its parity check matrix $\mathbf{H}_{m \times n}$. The tanner graph [12] of the code (or equivalently, its parity check matrix) is a bipartite graph consisting of two sets of nodes $\{v_i\}_{i=1}^n$ and $\{c_j\}_{j=1}^m$, namely, variable and check nodes, where for any $i$ and $j$ such that $1 \leq i \leq n$ and $1 \leq j \leq m$, $v_i$ and $c_j$, representing the $i$-th column and $j$-th row of $\mathbf{H}_{m \times n}$ respectively, are connected if and only if the element $h_{ji}$ of $\mathbf{H}_{m \times n}$ located at $i$-th column and $j$-th row is equal to 1. Note that the degree of a node in a graph is the number of edges connected to it. Let $\{l_i : 1 \leq i \leq L\}$ ($\{r_j : 1 \leq j \leq R\}$, respectively) be the set of degrees of all variable nodes (check nodes, respectively) in the tanner graph of $\mathbf{H}_{m \times n}$. Furthermore, let $\Lambda_i$ ($P_j$, respectively) denote the number of variable nodes (check nodes, respectively)

 

with degree $l_i$ ( $r_j$, respectively) in the tanner graph of $\mathbf{H}_{m \times n}$. Then we call $(\{\Lambda_i\}, \{l_i\})$ $((\{P_j\}, \{r_j\})$, respectively) the variable (check, respectively) degree distribution from a node perspective of $\mathbf{H}_{m \times n}$ (and its tanner graph) [13]. Define polynomials $\Lambda(z)$ and $P(z)$ as

$$\Lambda(z) = \sum_{i=1}^{L} \Lambda_i z^{l_i}$$

and

$$P(z) = \sum_{j=1}^{R} P_j z^{r_j}.$$

The tanner graph is said to be sparse and accordingly its corresponding code is said to be a low-density parity-check code if $\Lambda'(1)$ is in the order of $O(n)$, where $\Lambda'(1) = \sum_{i=1}^{L} \Lambda_i l_i$ is the total number of edges in the tanner graph. Normalizing $\{\Lambda_i\}$ and $\{P_j\}$ by the total numbers of variable nodes and check nodes respectively, we get normalized variable and check degree distributions $L(z)$ and $R(z)$:

$$L(z) = \sum_i L_i z^{l_i} = \frac{\Lambda(z)}{\Lambda(1)}$$

and

$$R(z) = \sum_j R_j z^{r_j} = \frac{P(z)}{P(1)}$$

where $L_i$ and $R_i$ represent the percentages of variable and check nodes with degrees $l_i$ and $r_i$ respectively.

Given $m$, $n$, and (normalized) variable and check degree distributions $L(z)$ and $R(z)$ satisfying $nL'(1) = mR'(1)$, let $\mathcal{H}_{m,n,L(z),R(z)}$ (simply $\mathcal{H}_{n,L(z),R(z)}$ if $m = n$) denote the collection of all $m \times n$ parity check matrices with normalized variable and check degree distributions $L(z)$ and $R(z)$. Without loss of generality, we only consider those matrices such that the degrees of rows and columns do not decrease with their indices. (In other words, $i > j$ implies the degree of the $i$-th row (or column) is not less than that of the $j$-th row (or column).) Then an LDPC code of designed rate $1 - m/n$ is said to be randomly generated from the ensemble with degree distributions $L(z)$ and $R(z)$ if its parity check matrix $\mathbf{H}_{m \times n}$ is uniformly picked from $\mathcal{H}_{m,n,L(z),R(z)}$. In this paper, we consider only such generated LDPC codes.

The performance of an LDPC code (under ML and BP decoding) depends largely on degree distributions of the ensemble it is picked from. According to the analysis in [13], a class of degree distributions, called check-concentrated degree distributions, are of special interest due to their superior performance, where given a variable node degree distribution, the check node degree distribution is made as concentrated as possible. In this case of $\mathcal{H}_{n,L(z),R(z)}$, given $L(z)$, $R(z)$ is determined as follows:

$$R(z) = R_1 z^{r_1} + R_2 z^{r_2}$$

 

where

$$r_1 = \lfloor \bar{l} \rfloor$$

$$r_2 = \lceil \bar{l} \rceil$$

$$R_1 = 1 + \lfloor \bar{l} \rfloor - \bar{l}$$

$$R_2 = \bar{l} - \lfloor \bar{l} \rfloor$$

and

$$\bar{l} = L'(1) = \sum_{i=1}^{L} L_i l_i \ .$$

Under this circumstance, $\mathcal{H}_{n,L(z),R(z)}$ is simply referred to as $\mathcal{H}_{n,L(z)}$.

## III. Interactive Encoding and Decoding based on Syndrome Accumulation

### A. Syndrome Accumulation

The concept of syndrome accumulation has been introduced in [4]. To clarify our following discussion, we revise this concept here.

Suppose a syndrome vector $s^n = \mathbf{H}_{n \times n} x^n$ is given, where $s^n$ consists of $n$ syndromes $s_1 s_2 \ldots s_n$, and $\mathbf{H}_{n \times n}$ is an $n \times n$ matrix. To facilitate the discussion below, we assume that $n$ is a power of 2, i.e. $2^T$ for some positive integer $T$. Let $\mathcal{N} = \{1, 2, \ldots, n\}$ and $\mathcal{P} = \{\Lambda_1, \Lambda_2, \ldots, \Lambda_{|\mathcal{P}|}\}$ where $\mathcal{P}$ forms a partition on $\mathcal{N}$ with each $\Lambda_i$ as a subset of $\mathcal{N}$ and $|\mathcal{P}|$ as the number of elements in $\mathcal{P}$. $\Lambda_i$ is also called a cell in $\mathcal{P}$, and we use $|\Lambda_i|$ to represent the cardinality of $\Lambda_i$, i.e. the number of indices in $\Lambda_i$. Now given $s^n$ and $\mathcal{P}$, we can form a new syndrome vector $\tilde{s}^{|\mathcal{P}|}$, which is called an accumulated syndrome vector, in the following way:

$$\tilde{s}^{|\mathcal{P}|} = \begin{pmatrix} \tilde{s}_1 \\ \tilde{s}_2 \\ \vdots \\ \tilde{s}_{|\mathcal{P}|} \end{pmatrix}$$

$$\tilde{s}_i = \sum_{j \in \Lambda_i} s_j \text{ for } 1 \leq i \leq |\mathcal{P}|$$

The derivation below shows that $\tilde{s}^{|\mathcal{P}|}$ is indeed a syndrome vector:

$$
\tilde{s}^{|\mathcal{P}|} = \begin{pmatrix} \tilde{s}_1 \\ \tilde{s}_2 \\ \vdots \\ \tilde{s}_{|\mathcal{P}|} \end{pmatrix}
$$

$$
= \begin{pmatrix} \sum_{j \in \Lambda_1} s_j \\ \sum_{j \in \Lambda_2} s_j \\ \vdots \\ \sum_{j \in \Lambda_{|\mathcal{P}|}} s_j \end{pmatrix}
$$

$$
= \begin{pmatrix} \sum_{j \in \Lambda_1} \sum_{k=1}^{n} h_{jk} x_k \\ \sum_{j \in \Lambda_2} \sum_{k=1}^{n} h_{jk} x_k \\ \vdots \\ \sum_{j \in \Lambda_{|\mathcal{P}|}} \sum_{k=1}^{n} h_{jk} x_k \end{pmatrix}
$$

$$
= \begin{pmatrix} \sum_{k=1}^{n} \sum_{j \in \Lambda_1} h_{jk} x_k \\ \sum_{k=1}^{n} \sum_{j \in \Lambda_2} h_{jk} x_k \\ \vdots \\ \sum_{k=1}^{n} \sum_{j \in \Lambda_{|\mathcal{P}|}} h_{jk} x_k \end{pmatrix}
$$

$$
= \left( \sum_{j \in \Lambda_i} h_{jk} \right)_{1 \le i \le |\mathcal{P}|, 1 \le k \le n|} \begin{pmatrix} x_1 \\ x_2 \\ \vdots \\ x_n \end{pmatrix}
$$

$$
\stackrel{\triangle}{=} \mathbf{H}_{\mathcal{P}} x^n
$$

where $h_{jk}$ is the element in the $j$-th row and $k$-th column of $\mathbf{H}_{n \times n}$, and $x_k$ is the $k$-th element in $x^n$. Also, $\mathbf{H}_{\mathcal{P}}$ defined above is the parity check matrix corresponding to the partition $\mathcal{P}$.

To proceed, we introduce a sequence of partitions $\mathcal{P}_1 \mathcal{P}_2 \cdots \mathcal{P}_n$. (Later on, it can be seen that this sequence effectively represents the procedure of encoding of SA-LDPC-IED schemes.) The sequence $\mathcal{P}_1 \mathcal{P}_2 \cdots \mathcal{P}_n$ is generated in a recursive manner, depicted below:

- $\mathcal{P}_1 = \{\mathcal{N}\}$.
- Suppose $\mathcal{P}_i = \{\Lambda_{i,1}, \Lambda_{i,2}, \ldots, \Lambda_{i,i}\}$ has been generated. Let $j_i = 2(i - 2^{\lfloor \log_2 i \rfloor}) + 1$. Split $\Lambda_{i,j_i}$ equally into two parts, $\Lambda_{i,j_i+}$ and $\Lambda_{i,j_i-}$, where $\Lambda_{i,j_i+}$ ($\Lambda_{i,j_i-}$) consists of the first (second) half of elements in $\Lambda_{i,j_i}$, ordered by their values.

- $\mathcal{P}_{i+1} = \{\Lambda_{i+1,1}, \Lambda_{i+1,2}, \ldots, \Lambda_{i+1,i+1}\}$ is generated as below:

    - $\Lambda_{i+1,k} = \Lambda_{i,k}$ for $1 \leq k < j_i$.

    - $\Lambda_{i+1,j_i} = \Lambda_{i,j_i+}$.

    - $\Lambda_{i+1,j_i+1} = \Lambda_{i,j_i-}$.

    - $\Lambda_{i+1,k} = \Lambda_{i,k-1}$ for $j_i + 1 < k \leq i+1$.

Note that since we assume $n = 2^T$ for some integer $T$, $|\Lambda_{i,k}|$ is also a power of 2 for $1 \leq i \leq n, 1 \leq k \leq i$. Moreover, for $1 < i < n$, $|\Lambda_{i,k_1}| = 2|\Lambda_{i,k_2}| = 2^{T-\lfloor \log_2 i \rfloor}$ always holds for $j_i \leq k_1 \leq i$ and $1 \leq k_2 \leq j_i - 1$. Therefore, the splitting of $\Lambda_{i,j_i}$ can always be applied. In fact,

$$\Lambda_{i,k} = \left\{ (k-1)2^{T-\lceil \log_2 i \rceil} + 1, \ldots, k2^{T-\lceil \log_2 i \rceil} \right\}$$

for $1 \leq k < j_i$, and

$$\Lambda_{i,k} = \left\{ (j_i - 1)2^{T-\lceil \log_2 i \rceil} + (k - j_i)2^{T-\lfloor \log_2 i \rfloor} + 1, \right.$$
$$\left. \ldots, (j_i - 1)2^{T-\lceil \log_2 i \rceil} + (k - j_i + 1)2^{T-\lfloor \log_2 i \rfloor} \right\}$$

for $j_i \leq k \leq i$.

Now given $s^n = \mathbf{H}_{n \times n} x^n$ and $\mathcal{P}_1 \mathcal{P}_2 \cdots \mathcal{P}_n$, we can generate a sequence of accumulated syndrome vectors $\tilde{s}_1^1 \tilde{s}_2^2 \ldots \tilde{s}_n^n$, where the upper scripts represent the dimension and lower scripts indicate which partitions the syndromes are associated with. The upper scripts, which always equal to the lower scripts, are dropped for simplicity. Now for any $\tilde{s}_i$, we use $\tilde{s}_{i,j}$ to represent its $j$-th element. In fact, this procedure can be done recursively as above, where

$$\tilde{s}_1 = \tilde{s}_{1,1} = \sum_{j \in \mathcal{N}} s_j$$

and $\tilde{s}_{i+1}$ is generated by replacing $\tilde{s}_{i,j_i}$ with $\tilde{s}_{i+1,j_i}$ and $\tilde{s}_{i+1,j_i+1}$. Moreover, since $\{\Lambda_{i+1,j_i}, \Lambda_{i+1,j_i+1}\}$ is a partition on $\Lambda_{i,j_i}$, we have

$$\tilde{s}_{i,j_i} = \tilde{s}_{i+1,j_i} + \tilde{s}_{i+1,j_i+1}$$

and therefore, if $\tilde{s}_i$ is known, only one of $\tilde{s}_{i+1,j_i}$ and $\tilde{s}_{i+1,j_i+1}$ is needed to calculate $\tilde{s}_{i+1}$. We call $\tilde{s}_{i+1,j_i}$ as the augmenting syndrome from $\tilde{s}_i$ to $\tilde{s}_{i+1}$, denoted by $a_{i+1}$. We also adopt the convention that $a_1 = \tilde{s}_{1,1}$ for convenience. In addition, according to the discussion above, $\tilde{s}_i = \mathbf{H}_{\mathcal{P}_i} x^n$, where $\mathbf{H}_{\mathcal{P}_i}$ can be determined by $\mathbf{H}_{n \times n}$ and $\mathcal{P}_i$. For clarification, we refer to $\mathbf{H}_{\mathcal{P}_i}$ as $\mathbf{H}_{i \times n}^{(i)}$, where the lower script indicates its dimension.

 

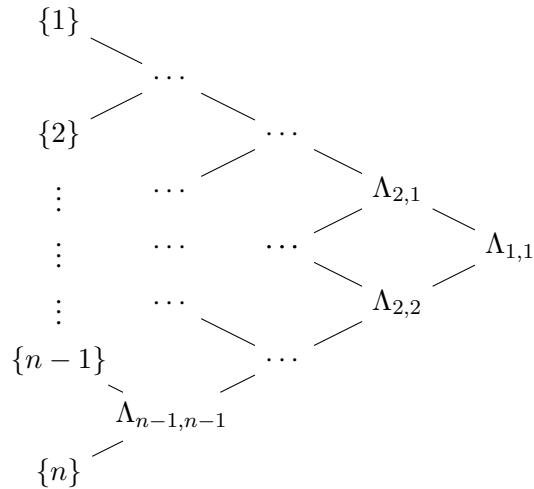

Fig. 2.   Binary Tree Structure of Syndrome Accumulation

By remark 7 in [4], a binary tree can be associated with $\mathcal{P}_1 \mathcal{P}_2 \cdots \mathcal{P}_n$ or $\tilde{s}_1 \tilde{s}_2 \cdots \tilde{s}_n$, shown in figure 2, where each node represents a subset of $\mathcal{N}$. Let $v$ and $\Lambda(v)$ be a node and its associated set. $\{\Lambda(v_l), \Lambda(v_r)\}$ forms a partition of $\Lambda(v)$ when $v_l$ and $v_r$ are the left and right child nodes of $v$. Moreover, let $v(\Lambda)$ be the node associated with the set $\Lambda$, and $d_v$ be the depth of a node $v$. Then $|\Lambda| = 2^{T - d_{v(\Lambda)}}$.

*B. Interactive Encoding and Decoding Schemes*

In light of LDPC codes, we consider only binary sources. That is, the source alphabet $\mathcal{X}$ is binary. However, the side information alphabet $\mathcal{Y}$ could be arbitrary. For any $x^n \in \mathcal{X}^n$, let $\bar{x}^n$ be the complement sequence of $x^n$, i.e., the sequence having hamming distance $n$ from $x^n$. Let $\mathbf{H}_{n \times n}$ be the parity check matrix of a LDPC code randomly generated from the ensemble $\mathcal{H}_{n,L(z)}$ for some $L(z)$. Let $\mathbf{H}'_{\eta_n n \times n}$ and $\mathbf{H}''_{(n\mathrm{H}(\epsilon) + \Delta) \times n}$ be matrices from Gallager parity check ensemble (the set of matrices with each element generated independently and uniformly from $\mathcal{B}$), where $0 < \eta_n < 1$, $0 < \epsilon < 0.5$, and $n\mathrm{H}(\epsilon)$ is assumed to be an integer. Furthermore, let $\mathcal{P}_1 \mathcal{P}_2 \cdots \mathcal{P}_n$ be the partition sequence described in the previous subsection. Based on the concepts introduced above, we are now ready to describe our SA-LDPC-IED scheme $\mathcal{I}_n$, which is presented in details in Algorithm 1 below, where $x^n$ is the source sequence to be encoded, $y^n \in \mathcal{Y}^n$ is the side information sequence available only to the decoder, and $\Delta$ is an integer to be specified later such that $\frac{n}{\Delta}$ is also an integer. Moreover, the specification of $\Gamma_b$, $\eta_n$ and the function $\gamma_n : \mathcal{X}^n \times \mathcal{Y}^n \to (0, +\infty)$ depends on $L(z)$, and will be discussed in the next section.

As in [1] [2] [4], given any $(x^n, y^n) \in \mathcal{X}^n \times \mathcal{Y}^n$, the performance of $\mathcal{I}_n$ is measured by the number of

---

**Algorithm 1** SA-LDPC-IED scheme $\mathcal{I}_n$

---

1: Based on $\mathcal{P}_1\mathcal{P}_2\cdots\mathcal{P}_n$ and $s^n = \mathbf{H}_{n\times n}x^n$, the encoder generates accumulated syndromes $\tilde{s}_1\tilde{s}_2\cdots\tilde{s}_n$ and augmenting syndromes $a_1a_2\cdots a_n$.

2: Based on $\mathcal{P}_1\mathcal{P}_2\cdots\mathcal{P}_n$ and $\mathbf{H}_{n\times n}$, the decoder calculates matrices $\mathbf{H}^{(\Delta)}_{\Delta\times n}\mathbf{H}^{(2\Delta)}_{2\Delta\times n}\cdots\mathbf{H}^{(n)}_{n\times n}$.

3: $b \leftarrow 0$.

4: **while** The encoder does not receive bit 1 from the decoder **do**

5:    $b \leftarrow b + 1$.

6:    **if** $b \leq \frac{n}{\Delta}$ **then**

7:       The encoder sends augmenting syndromes $a_{(b-1)\Delta+1}\cdots a_{b\Delta}$ to the decoder by $\Delta$ bits.

8:    **else**

9:       The encoder sends syndromes $s'_{\eta_n n} = \mathbf{H}'_{\eta_n n\times n}x^n$ to the decoder by $\eta_n n$ bits.

10:    **end if**

11:    Upon receiving syndromes sent from the encoder, the decoder calculates $\hat{x}^n$ by solving the optimization problem

$$\hat{x}^n = \begin{cases} \arg\min_{z^n:\mathbf{H}^{(b\Delta)}_{b\Delta\times n}z^n=\tilde{s}_{b\Delta}} \gamma_n(z^n, y^n) & \text{if } b \leq \frac{n}{\Delta} \\ \arg\min_{z^n:\mathbf{H}_{n\times n}z^n=s^n,\mathbf{H}'_{\eta_n n\times n}z^n=s'_{\eta_n n}} \gamma_n(z^n, y^n) & \text{otherwise} \end{cases}$$

12:    **if** $\gamma_n(\hat{x}^n|y^n) \leq \Gamma_b$ or $b > \frac{n}{\Delta}$ **then**

13:       The decoder sends bit 1 to the encoder.

14:    **else**

15:       The decoder sends bit 0 to the encoder.

16:    **end if**

17: **end while**

18: Upon receiving bit 1 from the decoder, the encoder sends $s''_{n\mathrm{H}(\epsilon)+\Delta} = \mathbf{H}''_{(n\mathrm{H}(\epsilon)+\Delta)\times n}x^n$ to the decoder.

19: Upon receiving $s''_{n\mathrm{H}(\epsilon)+\Delta}$, the decoder calculates the set

$$D = \left\{ z^n : \mathbf{H}''_{(n\mathrm{H}(\epsilon)+\Delta)\times n}z^n = s''_{n\mathrm{H}(\epsilon)+\Delta}, wt(z^n - \hat{x}^n) \leq \epsilon \text{ or } wt(z^n - \hat{x}^n) \geq 1 - \epsilon \right\}.$$

If $D$ contains a unique element $\tilde{x}^n$, the decoder outputs $\tilde{x}^n$ as the estimate of $x^n$. Otherwise, decoding failure is declared.

---

                                                                 

bits per symbol from the encoder to the decoder $r_f(x^n, y^n | \mathcal{I}_n)$, the number of bits per symbol from the decoder to the encoder $r_b(x^n, y^n | \mathcal{I}_n)$, and the conditional error probability $P(\mathcal{I}_n | x^n, y^n)$ of $\mathcal{I}_n$ given $x^n$ and $y^n$. Let $j(x^n, y^n)$ be the number of interactions at the time the decoder sends bit 1 to the encoder. It follows from the description of Algorithm 1 that

$$r_f(x^n, y^n | \mathcal{I}_n) = \begin{cases} \frac{j(x^n, y^n)\Delta}{n} + \mathrm{H}(\epsilon) + \frac{\Delta}{n} & \text{if } j(x^n, y^n) \leq \Delta/n \\ 1 + \eta_n + \mathrm{H}(\epsilon) + \frac{\Delta}{n} & \text{otherwise} \end{cases} \tag{3.1}$$

and

$$r_b(x^n, y^n | \mathcal{I}_n) = \frac{j(x^n, y^n)}{n}. \tag{3.2}$$

Moreover, let $(X, Y) = \{(X_i, Y_i)\}_{i=1}^{\infty}$ be a stationary source pair. We further define

$$r_f(\mathcal{I}_n) \overset{\Delta}{=} \mathbb{E}\left[r_f(X^n, Y^n | \mathcal{I}_n)\right]$$

$$r_b(\mathcal{I}_n) \overset{\Delta}{=} \mathbb{E}\left[r_b(X^n, Y^n | \mathcal{I}_n)\right]$$

and

$$P_e(\mathcal{I}_n) \overset{\Delta}{=} \Pr\{\tilde{X}^n \neq X^n\}$$

## IV. Performance of SA-LDPC-IED: General Case

This section is devoted to the theoretical performance analysis of our proposed SA-LDPC-IED scheme $\mathcal{I}_n$ for both individual sequences $x^n$ and $y^n$ and stationary, ergodic sources. Throughout this section, we assume that $\Delta \sim \sqrt{n}$.

### A. Specification of $\gamma_n(\cdot, \cdot)$, $\eta_n$, and $\{\Gamma_b\}$, and Probability Bounds

In order for our proposed SA-LDPC-IED scheme $\mathcal{I}_n$ to be truly universal, i.e., to achieve good performance for each and every individual source and side information pair $(x^n, y^n)$, we associate $\gamma_n(\cdot, \cdot)$ with a classical universal lossless code $\mathcal{C}_n$ (with block length $n$ and the side information available to both the encoder and decoder), where $\mathcal{C}_n$ is a mapping from $\mathcal{X}^n \times \mathcal{Y}^n$ to $\{0, 1\}^*$ satisfying that for any $y^n \in \mathcal{Y}^n$, the set $\{\mathcal{C}_n(x^n, y^n) : x^n \in \mathcal{X}^n\}$ is a prefix set. Specifically, we define

$$\gamma_n(x^n, y^n) = h_n(x^n | y^n)$$

where $nh_n(x^n | y^n)$ is the number of bits resulting from applying $\mathcal{C}_n$ to encode $x^n$ from $\mathcal{X}$ given the side information sequence $y^n$ from $\mathcal{Y}$ available to both the encoder and decoder.

Following the approach adopted in [2] [4], it is essential to calculate the following probabilities $\Pr\left\{\mathbf{H}'_{\eta_n n \times n} x^n = 0^{\eta_n n}\right\}$, $\Pr\left\{\mathbf{H}''_{(n\mathrm{H}(\epsilon)+\Delta) \times n} x^n = 0^{n\mathrm{H}(\epsilon)+\Delta}\right\}$ and $\Pr\left\{\mathbf{H}^{(b\Delta)}_{b\Delta \times n} x^n = 0^{b\Delta}\right\}$ for $1 \leq b \leq \frac{n}{\Delta}$,

 

given $x^n \neq 0^n$. In addition, in our case, the specification of $\eta_n$, and $\{\Gamma_b\}$ is also related to the probability $\Pr\left\{\mathbf{H}_{b\Delta\times n}^{(b\Delta)}x^n = 0^{b\Delta}\right\}$. Since $\mathbf{H}'_{\eta_n n \times n}$ and $\mathbf{H}''_{(n\mathrm{H}(\epsilon)+\Delta)\times n}$ are obtained from Gallager parity check ensemble, it can be easily shown that

$$\Pr\left\{\mathbf{H}'_{\eta_n n \times n}x^n = 0^{\eta_n n}\right\} = 2^{-\eta_n n}$$

$$\Pr\left\{\mathbf{H}''_{(n\mathrm{H}(\epsilon)+\Delta)\times n}x^n = 0^{n\mathrm{H}(\epsilon)+\Delta}\right\} = 2^{-n\mathrm{H}(\epsilon)-\Delta}$$

for any $x^n \neq 0^n$. However, calculating $\Pr\left\{\mathbf{H}_{b\Delta\times n}^{(b\Delta)}x^n = 0^{b\Delta}\right\}$ is much harder.

It can be seen that

$$\Pr\left\{\mathbf{H}_{b\Delta\times n}^{(b\Delta)}x^n = 0^{b\Delta}\right\}$$

depends on the support set of $x^n$, i.e., the positions of non-zero elements in $x^n$. Let $\varkappa(x^n)$ represent the support set of $x^n$, and we write $\varkappa(x^n)$ simply as $\varkappa$ whenever $x^n$ is generic or can be determined from context. Let $\mathbf{H}_{n\times|\varkappa|}^{\varkappa}$ be the matrix consisting of those columns of $\mathbf{H}_{n\times n}$ with indices in $\varkappa$. The degree polynomial of $\varkappa$, denoted by $L^{\varkappa}(z)$, is defined by

$$L^{\varkappa}(z) \triangleq \sum_{i}^{L} L_i^{\varkappa} z^{l_i}$$

where $L_i^{\varkappa} n$ is the number of columns with degree $l_i$ within $\mathbf{H}_{n\times|\varkappa|}^{\varkappa}$. And define

$$\bar{l}^{\varkappa} \triangleq \sum_{i=1}^{L} L_i^{\varkappa} l_i.$$

Now let

$$
\begin{aligned}
t_{b\Delta}^{(1)} &= \min\left\{2b\Delta - 2^{\lceil\log_2 b\Delta\rceil}, R_1 2^{\lceil\log_2 b\Delta\rceil}\right\}, \\
t_{b\Delta}^{(2)} &= \max\left\{R_1 2^{\lceil\log_2 b\Delta\rceil-1} - \left(b\Delta - 2^{\lceil\log_2 b\Delta\rceil-1}\right), 0\right\}, \\
t_{b\Delta}^{(3)} &= \max\left\{R_2 2^{\lceil\log_2 b\Delta\rceil} - 2\left(2^{\lceil\log_2 b\Delta\rceil} - b\Delta\right), 0\right\}, \\
t_{b\Delta}^{(4)} &= \min\left\{2^{\lceil\log_2 b\Delta\rceil} - b\Delta, R_2 2^{\lceil\log_2 b\Delta\rceil-1}\right\}.
\end{aligned}
$$

To understand the meaning of $\left\{t_{b\Delta}^{(i)}\right\}_{i=1}^{4}$, let us focus on $\mathcal{P}_{b\Delta} = \{\Lambda_{b\Delta,i}\}_{i=1}^{b\Delta}$. By the binary tree representation in the previous section,

$$
\begin{aligned}
t_{b\Delta}^{(1)} &= \text{ \# of } \Lambda_{b\Delta,i} \text{ s.t. } \Lambda_{b\Delta,i} \subseteq \{1\cdots R_1 n\} \text{ and } d_{v(\Lambda_{b\Delta,i})} = 2^{\lceil\log_2 b\Delta\rceil} \\
t_{b\Delta}^{(2)} &= \text{ \# of } \Lambda_{b\Delta,i} \text{ s.t. } \Lambda_{b\Delta,i} \subseteq \{1\cdots R_1 n\} \text{ and } d_{v(\Lambda_{b\Delta,i})} = 2^{\lceil\log_2 b\Delta\rceil-1} \\
t_{b\Delta}^{(3)} &= \text{ \# of } \Lambda_{b\Delta,i} \text{ s.t. } \Lambda_{b\Delta,i} \subseteq \{R_1 n+1\cdots n\} \text{ and } d_{v(\Lambda_{b\Delta,i})} = 2^{\lceil\log_2 b\Delta\rceil} \\
t_{b\Delta}^{(4)} &= \text{ \# of } \Lambda_{b\Delta,i} \text{ s.t. } \Lambda_{b\Delta,i} \subseteq \{R_1 n+1\cdots n\} \text{ and } d_{v(\Lambda_{b\Delta,i})} = 2^{\lceil\log_2 b\Delta\rceil-1}
\end{aligned}
$$

Since the block length $n$ is assumed to be a power of 2, it follows that

$$\frac{t_{b\Delta}^{(1)}}{n} = \min\left\{\frac{2b\Delta}{n} - 2^{\left\lceil \log_2 \frac{b\Delta}{n} \right\rceil}, R_1 2^{\left\lceil \log_2 \frac{b\Delta}{n} \right\rceil}\right\}$$

$$\frac{t_{b\Delta}^{(2)}}{n} = \max\left\{R_1 2^{\left\lceil \log_2 \frac{b\Delta}{n} \right\rceil - 1} - \left(\frac{b\Delta}{n} - 2^{\left\lceil \log_2 \frac{b\Delta}{n} \right\rceil - 1}\right), 0\right\}$$

$$\frac{t_{b\Delta}^{(3)}}{n} = \max\left\{R_2 2^{\left\lceil \log_2 \frac{b\Delta}{n} \right\rceil} - 2\left(2^{\left\lceil \log_2 \frac{b\Delta}{n} \right\rceil} - \frac{b\Delta}{n}\right), 0\right\}$$

$$\frac{t_{b\Delta}^{(4)}}{n} = \min\left\{2^{\left\lceil \log_2 \frac{b\Delta}{n} \right\rceil} - \frac{b\Delta}{n}, R_2 2^{\left\lceil \log_2 \frac{b\Delta}{n} \right\rceil - 1}\right\}$$

and hence $\frac{t_{b\Delta}^{(i)}}{n}$, $i = 1, 2, 3, 4$, all depend only on $b\Delta/n$.

We have the following result, which is proved in Appendix A.

**Lemma 1.** *Let $L(z)$ be a normalized variable node degree distribution from a node perspective with minimum degree $l_1 \geq 2$. Let $c_{b\Delta} = 2^{-\left\lceil \log_2 \frac{b\Delta}{n} \right\rceil}$ and $g(\tau, k) \triangleq (1+\tau)^k + (1-\tau)^k$ for any $\tau$ and $k$. Suppose $\mathbf{H}_{n \times n}$ is uniformly picked from ensemble $\mathcal{H}_{n, L(z)}$. Then for any $x^n \neq 0$ with its support set $\varkappa$,*

$$\Pr\left\{\mathbf{H}_{b\Delta \times n}^{(b\Delta)} x^n = 0^{b\Delta}\right\} \leq \exp\left\{nP\left(\frac{b\Delta}{n}, \bar{l}, \bar{l}^{\varkappa}\right) + \frac{3n\lceil \bar{l} \rceil}{b\Delta}\ln(n\hat{l}^{\varkappa}) + \frac{1}{2}\ln n\bar{l}^{\varkappa}(1 - \frac{\bar{l}^{\varkappa}}{\bar{l}}) + O(1)\right\}$$

*where*

$$\hat{l}^{\varkappa} = \max\left\{\frac{1}{n}, \min\{\bar{l}^{\varkappa}, \bar{l} - \bar{l}^{\varkappa}\}\right\}$$

*and for any $\frac{b\Delta}{n}, \bar{l}$ and $\xi \in (0, \bar{l}]$, $P\left(\frac{b\Delta}{n}, \bar{l}, \xi\right)$ is defined as*

$$P\left(\frac{b\Delta}{n}, \bar{l}, \xi\right)$$

$$\triangleq -\bar{l}\mathrm{H}_e\left(\xi/\bar{l}\right) - \xi\ln\tau$$

$$+ \frac{t_{b\Delta}^{(1)}}{n}\ln\frac{g(\tau, r_1 c_{b\Delta})}{2}$$

$$+ \frac{t_{b\Delta}^{(2)}}{n}\ln\frac{g(\tau, 2r_1 c_{b\Delta})}{2}$$

$$+ \frac{t_{b\Delta}^{(3)}}{n}\ln\frac{g(\tau, r_2 c_{b\Delta})}{2}$$

$$+ \frac{t_{b\Delta}^{(4)}}{n}\ln\frac{g(\tau, 2r_2 c_{b\Delta})}{2} \tag{4.1}$$

*in which $\tau$ is the solution to*

$$r_1 c_{b\Delta} \frac{t_{b\Delta}^{(1)}}{n} \frac{g(\tau, r_1 c_{b\Delta} - 1)}{g(\tau, r_1 c_{b\Delta})}$$

$$+2r_1 c_{b\Delta} \frac{t_{b\Delta}^{(2)}}{n} \frac{g(\tau, 2r_1 c_{b\Delta} - 1)}{g(\tau, 2r_1 c_{b\Delta})}$$

$$+r_2 c_{b\Delta} \frac{t_{b\Delta}^{(3)}}{n} \frac{g(\tau, r_2 c_{b\Delta} - 1)}{g(\tau, r_2 c_{b\Delta})}$$

$$+2r_2 c_{b\Delta} \frac{t_{b\Delta}^{(4)}}{n} \frac{g(\tau, 2r_2 c_{b\Delta} - 1)}{g(\tau, 2r_1 c_{b\Delta})}$$

$$= \bar{l} - \xi. \tag{4.2}$$

*for $\xi \in \left[0, \bar{l} - \frac{t_{b\Delta}^{(1)}}{n} \pi \left(c_{b\Delta} r_1\right) - \frac{t_{b\Delta}^{(3)}}{n} \pi \left(c_{b\Delta} r_2\right)\right]$, and*

$$P\left(\frac{b\Delta}{n}, \bar{l}, \xi\right) \triangleq -\infty \tag{4.3}$$

*for $\xi \in \left(\bar{l} - \frac{t_{b\Delta}^{(1)}}{n} \pi \left(c_{b\Delta} r_1\right) - \frac{t_{b\Delta}^{(3)}}{n} \pi \left(c_{b\Delta} r_2\right), \bar{l}\right]$ with the convention that $e^{-\infty} = 0$.*

**Remark 1.** *When $\xi = \bar{l} - \frac{t_{b\Delta}^{(1)}}{n} \pi \left(c_{b\Delta} r_1\right) - \frac{t_{b\Delta}^{(3)}}{n} \pi \left(c_{b\Delta} r_2\right)$, the solution $\tau$ to (4.2) is $\tau = +\infty$. In this case, the expression in (4.1) should be understood as its limit as $\tau \to +\infty$, i.e.,*

$$P\left(\frac{b\Delta}{n}, \bar{l}, \xi\right)$$

$$\triangleq -\bar{l}\mathrm{H}_e\left(\xi/\bar{l}\right) + \lim_{\tau \to +\infty} \left[-\xi \ln \tau + \frac{t_{b\Delta}^{(1)}}{n} \ln \frac{g(\tau, r_1 c_{b\Delta})}{2} + \frac{t_{b\Delta}^{(2)}}{n} \ln \frac{g(\tau, 2r_1 c_{b\Delta})}{2}\right.$$

$$\left. + \frac{t_{b\Delta}^{(3)}}{n} \ln \frac{g(\tau, r_2 c_{b\Delta})}{2} + \frac{t_{b\Delta}^{(4)}}{n} \ln \frac{g(\tau, 2r_2 c_{b\Delta})}{2}\right]$$

$$= -\bar{l}\mathrm{H}_e\left(\xi/\bar{l}\right) + \frac{t_{b\Delta}^{(1)}}{n} \pi(c_{b\Delta} r_1) \ln[c_{b\Delta} r_1] + \frac{t_{b\Delta}^{(3)}}{n} \pi(c_{b\Delta} r_2) \ln[c_{b\Delta} r_2] \tag{4.4}$$

*when $\xi = \bar{l} - \frac{t_{b\Delta}^{(1)}}{n} \pi \left(c_{b\Delta} r_1\right) - \frac{t_{b\Delta}^{(3)}}{n} \pi \left(c_{b\Delta} r_2\right)$.*

**Remark 2.** *Replace $\frac{b\Delta}{n}$ by any real number $R \in (0,1]$ in $\frac{t_{b\Delta}^{(i)}}{n}$, $i = 1, 2, 3$, and $4$, $c_{b\Delta}$, and $P\left(\frac{b\Delta}{n}, \bar{l}, \xi\right)$. It is not hard to verify that $\frac{t_{b\Delta}^{(i)}}{n}$, $i = 1, 2, 3$, and $4$, $c_{b\Delta}$, and $P\left(R, \bar{l}, \xi\right)$ as a respective function of $R \in (0,1]$ are all well defined. One can further verify that as a function of $R \in (0,1]$, the following identities hold:*

$$\sum_{i=1}^{4} \frac{t_{b\Delta}^{(i)}}{n} = R \tag{4.5}$$

*and*

$$r_1 c_{b\Delta} \frac{t_{b\Delta}^{(1)}}{n} + 2r_1 c_{b\Delta} \frac{t_{b\Delta}^{(2)}}{n} + r_2 c_{b\Delta} \frac{t_{b\Delta}^{(3)}}{n} + 2r_2 c_{b\Delta} \frac{t_{b\Delta}^{(4)}}{n} = \bar{l}. \tag{4.6}$$

As illustrated in Figure 3, the function $P\left(R,\bar{l},\xi\right)$ has several interesting properties including

PR1  given $(R,\bar{l})$, $P\left(R,\bar{l},\xi\right)$ is a strictly decreasing function of $\xi$ over $\xi\in(0,\bar{l}/2]$;

PR2  given $0<\xi\leq\bar{l}/2$, $P\left(R,\bar{l},\xi\right)$ as a function of $R$ is continuous and strictly decreasing over $R\in(0,1]$, and furthermore

$$P\left(0,\bar{l},\xi\right)\overset{\Delta}{=}\lim_{R\to0}P\left(R,\bar{l},\xi\right)=0$$

PR3  and $P\left(R,\bar{l},\xi\right)$ is close to $-R\ln 2$ when $\xi\leq\bar{l}/2$ is not too far away from $\bar{l}/2$.

These and other properties of $P\left(R,\bar{l},\xi\right)$ are needed in the performance analysis of our proposed SA-LDPC-IED Scheme $\mathcal{I}_n$. Their exact statements and respective proofs will be relegated to Appendix B.

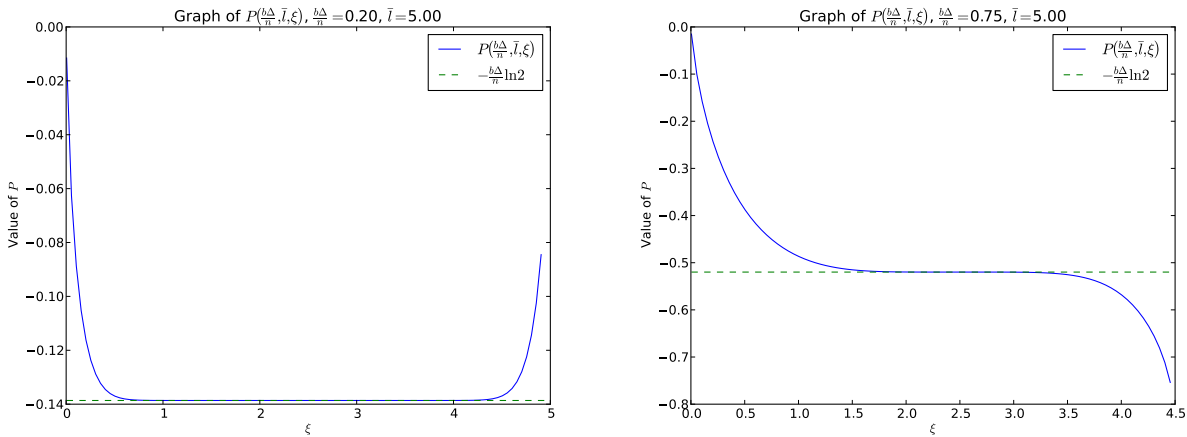

Fig. 3.  Graphical Illustration of $P\left(\frac{b\Delta}{n},\bar{l},\xi\right)$

Based on the function $P\left(\frac{b\Delta}{n},\bar{l},\xi\right)$, we are now ready to specify $\eta_n$ and $\{\Gamma_b\}$ for any $1\leq b\leq\frac{n}{\Delta}$ in our proposed SA-LDPC-IED Scheme $\mathcal{I}_n$, which are defined respectively as

$$\eta_n=1+\frac{1}{\ln 2}\left[P\left(1,\bar{l},l_1\epsilon\right)+\frac{3\lceil\bar{l}\rceil}{n}\ln\frac{n\bar{l}}{2}+\frac{1}{2n}\ln\frac{n\bar{l}}{4}\right]+\frac{\Delta}{n}$$

and

$$\Gamma_b=\frac{1}{\ln 2}\left[-P\left(\frac{b\Delta}{n},\bar{l},l_1\epsilon\right)-\frac{3\lceil\bar{l}\rceil}{\Delta}\ln\frac{n\bar{l}}{2}-\frac{1}{2n}\ln\frac{n\bar{l}}{4}\right]-\frac{\Delta}{n}$$

where $\epsilon>0$ is the same as in the description of the SA-LDPC-IED Scheme $\mathcal{I}_n$.

## B. Performance for Individual Sequences

We now analyze the performance of the SA-LDPC-IED scheme $\mathcal{I}_n$ in terms of the performance of the classical universal code $\mathcal{C}_n$ for any individual sequences $x^n$ and $y^n$. We have the following theorem, which is proved in Appendix C.

**Theorem 1.** *Let $L(z)$ represent a normalized variable node degree distribution from a node perspective with minimum degree $l_1 \geq 2$. Then for any $(x^n, y^n) \in \mathcal{X}^n \times \mathcal{Y}^n$,*

$$r_f(x^n, y^n | \mathcal{I}_n) \leq R^{(\Delta)}_{L(z)} \left( \epsilon, h_n(x^n|y^n) \right) + \mathrm{H}(\epsilon) + \frac{2\Delta}{n} \tag{4.7}$$

$$r_b(x^n, y^n | \mathcal{I}_n) = O\left( \frac{1}{\sqrt{n}} \right) \tag{4.8}$$

*and*

$$P_e(\mathcal{I}_n | x^n, y^n) \leq 2^{-\Delta + \log_2\left(\frac{n}{\Delta} + 1\right) + O(1)} \tag{4.9}$$

*where $P_e(\mathcal{I}_n | x^n, y^n)$ denotes the conditional error probability of $\mathcal{I}_n$ given $x^n$ and $y^n$, and $R^{(\Delta)}_{L(z)} \left( \epsilon, h_n(x^n|y^n) \right)$ is the positive solution $R$ to*

$$- P\left( R, \bar{l}, l_1 \epsilon \right) = \left[ h_n(x^n|y^n) + \frac{\Delta}{n} \right] \ln 2 + \frac{3\lceil \bar{l} \rceil}{\Delta} \ln \frac{n\bar{l}}{2} + \frac{1}{2n} \ln \frac{n\bar{l}}{4} \tag{4.10}$$

*if $h_n(x^n|y^n) \leq \Gamma_{\frac{n}{\Delta}}$, and*

$$R^{(\Delta)}_{L(z)} \left( \epsilon, h_n(x^n|y^n) \right) = 2 + \frac{1}{\ln 2} \left[ P\left( 1, \bar{l}, l_1 \epsilon \right) + \frac{3\lceil \bar{l} \rceil}{n} \ln \frac{n\bar{l}}{2} + \frac{1}{2n} \ln \frac{n\bar{l}}{4} \right] \tag{4.11}$$

*otherwise.*

In order to analyze the asymptotical performance of the SA-LDPC-IED scheme $\mathcal{I}_n$ first as $n \to \infty$ and then as the average degree $\bar{l}$ of $L(z)$ goes to $\infty$, we define for any $h \in [0, 1]$

$$R_{L(z)} \left( \epsilon, h \right) \triangleq \lim_{n \to \infty} R^{(\Delta)}_{L(z)} \left( \epsilon, h \right)$$

and

$$r_{L(z)} \left( \epsilon, h \right) \triangleq R_{L(z)} \left( \epsilon, h \right) + \mathrm{H}(\epsilon) - h.$$

Clearly, $r_{L(z)} \left( \epsilon, h \right)$ represents the redundancy of $\mathcal{I}_n$, i.e., the gap between the asymptotical total rate of $\mathcal{I}_n$ and the desired rate $h$. We have the following two results, which will be proved in Appendix D.

**Proposition 1.** *Let $L(z)$ be a normalized degree distribution with $l_1 \geq 2$ and $\epsilon$ be a real number where $\frac{\bar{l}}{l_1 \lfloor \bar{l} \rfloor} \leq \epsilon < 0.5$. Then for any $h \geq 0$,*

$$r_{L(z)} \left( \epsilon, h \right) \leq \mathrm{H}(\epsilon) + \left( 1 + \mathrm{I} \left( h \ln 2 \geq -P(1, \bar{l}, l_1 \epsilon) \right) \right) \left\{ \frac{2l_1 \epsilon}{\ln 2} \exp\left[ -\frac{2l_1 \epsilon}{\bar{l}} \left( \lfloor \bar{l} \rfloor - 1 \right) \right] + \frac{1}{\ln 2} \exp\left( -\frac{2l_1 \epsilon}{\bar{l}} \lfloor \bar{l} \rfloor \right) \right\}$$

*where* $\mathrm{I}(\cdot)$ *is the indicator function such that*

$$\mathrm{I}\left(h\ln 2 \geq -P(1,\bar{l},l_1\epsilon)\right) = \begin{cases} 1 & \text{if } h\ln 2 \geq -P(1,\bar{l},l_1\epsilon) \\ 0 & \text{otherwise} \end{cases}$$

**Proposition 2.** *Let* $L(z)$ *be a normalized degree distribution with* $l_1 \geq 2$. *Then*

$$r_{L(z^k)}\left(\frac{\ln k}{2k}, h\right) = O\left(\frac{\ln^2 k}{k}\right)$$

*for any* $k \geq e^{\frac{2}{l_1}}$ *and* $h \geq 0$.

### C. Performance for Stationary, Ergodic Sources

In this subsection, we analyze the performance of the SA-LDPC-IED scheme $\mathcal{I}_n$ for any stationary, ergodic source-side information pair $(X,Y) = \{(X_i,Y_i)\}_{i=1}^{\infty}$ with alphabet $\mathcal{X} \times \mathcal{Y}$. To this end, we select $\{\mathcal{C}_n\}_{n=1}^{\infty}$ to be a sequence of universal (classical) prefix codes with side information available to both the encoder and decoder such that

$$\lim_{n\to\infty} h_n(X^n|Y^n) = \mathrm{H}(X|Y) \quad \text{with probability one} \tag{4.12}$$

for any stationary, ergodic source-side information pair $(X,Y)$. (Note that from the literature of classical universal lossless source coding (see, for example, [3], [14], [15], [16], [17], and the references therein), such a sequence exists.) To bring out the dependence of $\mathcal{I}_n$ on $L(z)$ and $\epsilon$, we shall write $\mathcal{I}_n$ as $\mathcal{I}_n(L(z),\epsilon)$. Then we have the following result, which is proved in Appendix D.

**Theorem 2.** *Let* $L(z)$ *be a normalized variable node degree distribution. Then for any stationary, ergodic source side information pair* $(X,Y)$,

$$\lim_{k\to\infty}\lim_{n\to\infty} r_f\left(X^n,Y^n\left|\mathcal{I}_n\left(L(z^k),\frac{\ln k}{2k}\right)\right.\right) = \mathrm{H}(X|Y) \quad \text{with probability one} \tag{4.13}$$

$$r_b\left(X^n,Y^n\left|\mathcal{I}_n\left(L(z^k),\frac{\ln k}{2k}\right)\right.\right) = O\left(\frac{1}{\sqrt{n}}\right) \tag{4.14}$$

*and*

$$P_e\left(\mathcal{I}_n\left(L(z^k),\frac{\ln k}{2k}\right)\right) \leq 2^{-\Delta+\log_2\left(\frac{n}{\Delta}+1\right)+O(1)} \tag{4.15}$$

*whenever* $k \geq 9$.

## V. Performance of SA-LDPC-IED: Binary Case and Bit Error Probability

Theorems 1 and 2 show the performance of our proposed SA-LDPC-IED scheme $\mathcal{I}_n$ in terms of the forward and backward rates versus the word error probability for both individual sequences $x^n$ and $y^n$ and stationary, ergodic sources. In this section, we consider instead the forward and backward rates versus the bit error probability by focusing on independent and identically distributed (i.i.d) source-side information pairs $(X, Y) = \{(X_i, Y_i)\}_{i=1}^{\infty}$, where the source $X$ and side-information $Y$ are correlated through a binary symmetric channel with a cross-over probability $p_0 \in (0, 0.5)$, which is unknown to both the decoder and the encoder. Limiting ourselves to this smaller class of source-side information pairs allows us to illustrate the SA-LDPC-IED scheme $\mathcal{I}_n$ by using a specific and simple function $\gamma(\cdot, \cdot)$, which in turn leads to further simplification of the SA-LDPC-IED scheme $\mathcal{I}_n$ itself and paves the way for the belief propagation (BP) decoding to be used as a decoding method in IED in the next section.

Note that in this binary case

$$\mathrm{H}(X|Y) = \mathrm{H}(p_0) \ .$$

Define $\mathrm{H}^{-1}(\cdot) : [0, 1] \to [0, 0.5]$ as the inverse function of $\mathrm{H}(\cdot)$ such that $x = \mathrm{H}^{-1}(h)$ if and only if $h = \mathrm{H}(x)$ for $x \in [0, 0.5]$ and $h \in [0, 1]$. Now specify $\gamma(\cdot, \cdot)$ as

$$\gamma(x^n, y^n) = \begin{cases} \frac{\ln n + 1}{n} + \mathrm{H}\left(\frac{1}{n} wt(x^n - y^n)\right) & \text{if } \frac{1}{n} wt(x^n - y^n) \leq 0.5 \\ \frac{1}{n} + 1 & \text{otherwise.} \end{cases} \tag{5.1}$$

It is easy to see that $\gamma(x^n, y^n)$ is actually the normalized code length function of the classical prefix code $\mathcal{C}_n$ with side information available to both the encoder and decoder as described in Algorithm 2. With the assumption on the correlation between the source $X$ and side information $Y$ and with this specific function $\gamma(\cdot, \cdot)$, we can further get rid of the last round of transmission from the encoder to the decoder in $\mathcal{I}_n$, yielding a simplified version $\tilde{\mathcal{I}}_n$ as described in Algorithm 3.

Now let us analyze the performance of the SA-LDPC-IED scheme $\tilde{\mathcal{I}}_n$ in terms of the forward and backward rates versus the bit error probability $P_b$, where

$$P_b \overset{\Delta}{=} \frac{1}{n} \mathbb{E}\left[wt(\hat{X}^n - X^n)\right] \ .$$

Then we have the following theorem, which is proved in Appendix E.

**Theorem 3.** *Let $L(z)$ be a normalized variable node degree distribution from a node perspective with minimum degree $l_1 \geq 2$ and average degree $\bar{l}$ being an odd integer. Select $\epsilon > 0$ such that $\epsilon \leq 0.5 - \mathrm{H}^{-1}(0.75)$. Then for any i.i.d source-side information pair $(X, Y)$ correlated through a binary symmetric*

 

---

**Algorithm 2** A classical prefix code $\mathcal{C}_n$ with side information available to both the encoder and decoder

1: The encoder calculates $w = wt(x^n - y^n)$.

2: **if** $w \leq 0.5n$ **then**

3:     The encoder sends bit 0 followed by a codeword of fixed-length $\ln n$ specifying $w$ and then by a codeword of length $n\mathrm{H}\left(\frac{w}{n}\right)$ specifying the index of $x^n - y^n$ in the set $\{z^n : wt(z^n) = w\}$ sorted by the lexicographical order.

4: **else**

5:     The encoder sends bit 1 followed by $x^n$ itself.

6: **end if**

---

*channel with cross-over probability $p_0 \in (0, 0.5)$ and for sufficiently large $n$,*

$$r_f(\tilde{\mathcal{I}}_n) \quad \leq \quad R_{L(z)}^{(\Delta)}\left(\epsilon, \mathrm{H}(p_0) + \frac{\ln n + 1}{n} + \log_2\left(\frac{1 - p_0}{p_0}\right)\sqrt{\frac{\ln n}{n}}\right) + n^{-2}R_{L(z)}^{(\Delta)}(\epsilon, 1) + \frac{\Delta}{n}$$

$$(5.2)$$

$$r_b(\tilde{\mathcal{I}}_n) = O\left(\frac{1}{\sqrt{n}}\right) \tag{5.3}$$

*and*

$$P_b(\tilde{\mathcal{I}}_n) \leq \epsilon + e^{-2n(0.5 - p_0)^2} + 2^{-\Delta + \log_2\left(\frac{n}{\Delta} + 1\right) + O(1)}. \tag{5.4}$$

By defining

$$\tilde{r}_{L(z)}(\epsilon, p_0) \stackrel{\triangle}{=} R_{L(z)}\left(\epsilon, \mathrm{H}(p_0)\right) - \mathrm{H}(p_0)$$

we have the following proposition, the proof of which is omitted due to its similarity to that of Proposition 2.

**Proposition 3.** *Let $L(z)$ be a normalized degree distribution with $l_1 \geq 2$ and $k \geq 2$. For $p_0 \in (0, 0.5)$,*

$$\tilde{r}_{L(z^k)}\left(\frac{1}{2\sqrt{k}}, p_0\right) = O\left(e^{-\sqrt{k} + \frac{1}{2}\ln k}\right).$$

We conclude this section by providing the following theorem (proved in Appendix F), which analyzes the performance of the modified SA-LDPC-IED scheme $\tilde{\mathcal{I}}_n$ when $L(z^k)$ is used. Once again, to bring out the dependence of $\tilde{\mathcal{I}}_n$ on $(L(z), \epsilon)$, we write $\tilde{\mathcal{I}}_n$ as $\tilde{\mathcal{I}}_n(L(z), \epsilon)$.

**Theorem 4.** *Let $L(z)$ be a normalized variable node degree distribution with minimum degree $l_1 \geq 2$. For any i.i.d source-side information pair $(X, Y)$ correlated through a binary symmetric channel with*

 

---

**Algorithm 3** SA-LDPC-IED scheme $\tilde{\mathcal{I}}_n$ for i.i.d source-side information pairs

---

1: Based on $\mathcal{P}_1\mathcal{P}_2\cdots\mathcal{P}_n$ and $s^n = \mathbf{H}_{n\times n}x^n$, the encoder generates accumulated syndromes $\tilde{s}_1\tilde{s}_2\cdots\tilde{s}_n$ and augmenting syndromes $a_1a_2\cdots a_n$.

2: Based on $\mathcal{P}_1\mathcal{P}_2\cdots\mathcal{P}_n$ and $\mathbf{H}_{n\times n}$, the decoder calculates matrices $\mathbf{H}_{\Delta\times n}^{(\Delta)}\mathbf{H}_{2\Delta\times n}^{(2\Delta)}\cdots\mathbf{H}_{n\times n}^{(n)}$.

3: $b \leftarrow 0$.

4: **while** The encoder does not receive bit 1 from the decoder **do**

5:    $b \leftarrow b + 1$.

6:    **if** $b \leq \frac{n}{\Delta}$ **then**

7:       The encoder sends augmenting syndromes $a_{(b-1)\Delta+1}\cdots a_{b\Delta}$ to the decoder by $\Delta$ bits.

8:    **else**

9:       The encoder sends syndromes $s'_{\eta_n n} = \mathbf{H}'_{\eta_n n\times n}x^n$ to the decoder by $\eta_n n$ bits.

10:    **end if**

11:    Upon receiving syndromes sent from the encoder, the decoder calculates $\hat{x}^n$ by solving the optimization problem

$$\hat{x}^n = \begin{cases} \arg\min_{z^n : \mathbf{H}_{b\Delta\times n}^{(b\Delta)}z^n = \tilde{s}_{b\Delta}} \gamma_n(z^n, y^n) & \text{if } b \leq \frac{n}{\Delta} \\ \arg\min_{z^n : \mathbf{H}_{n\times n}z^n = s^n, \mathbf{H}'_{\eta_n n\times n}z^n = s'_{\eta_n n}} \gamma_n(z^n, y^n) & \text{otherwise.} \end{cases}$$

12:    **if** $\gamma_n(\hat{x}^n|y^n) \leq \Gamma_b$ or $b > \frac{n}{\Delta}$ **then**

13:       The decoder sends bit 1 to the encoder, and outputs $\hat{x}^n$ as the estimate of $x^n$.

14:    **else**

15:       The decoder sends bit 0 to the encoder and leaves the estimate of $x^n$ undecided.

16:    **end if**

17: **end while**

---

*cross-over probability* $p_0 \in (0, 0.5)$,

$$\lim_{k\to\infty}\lim_{n\to\infty} r_f\left(\tilde{\mathcal{I}}_n\left(L(z^k), \frac{1}{2\sqrt{k}}\right)\right) = \mathrm{H}(p_0) \tag{5.5}$$

$$r_b\left(\tilde{\mathcal{I}}_n\left(L(z^k), \frac{1}{2\sqrt{k}}\right)\right) = O\left(\frac{1}{\sqrt{n}}\right) \tag{5.6}$$

*and*

$$P_b\left(\tilde{\mathcal{I}}_n\left(L(z^k), \frac{1}{2\sqrt{k}}\right)\right) \leq \frac{1}{2\sqrt{k}} + e^{-2n\left(0.5 - \frac{1}{4\sqrt{k}} - p_0\right)^2} + 2^{-\Delta + \log_2\left(\frac{n}{\Delta} + 1\right) + O(1)} \tag{5.7}$$

*whenever $k > \left(\frac{1}{2(1-2p_0)}\right)^2$.*

## VI. Implementation and Simulation Results

To verify our theoretical analysis in the last two sections, we have implemented our proposed SA-LDPC-IED schemes with some modification, namely by adopting the BP decoding in the place of the minimum coding length or Hamming distance decoding. In this section, we report their performance for binary source-side information pairs $(X, Y)$, where $X$ and $Y$ are correlated through a binary channel with probability transition matrix (from $Y$ to $X$) given by

$$\left( \begin{array}{cc} 1-p_1 & p_2 \\ p_1 & 1-p_2 \end{array} \right)$$

and where $p_1, p_2 \in (0, 0.5]$ are assumed unknown to both the encoder and decoder. Since the standard BP decoding algorithm applies only to fix-rate LDPC codes with known statistics of source and side information pairs, we first have to modify the BP decoding algorithm so that it fits into our variable-rate and unknown statistics situation as well while maintaining its low complexity.

### A. Modified BP Decoding Algorithm

The BP decoding algorithm can be considered as a sum-product algorithm [18] on a Tanner graph, which represents the parity check matrix of the LDPC code, with variable nodes corresponding to bits of the source, and check nodes corresponding to syndromes. Generally speaking, it tries to marginalize the distribution of each bit of the source based on local calculations. Specifically, it iteratively calculates messages from variable nodes to their connected check nodes, and vice versa, i.e.

$$m_{v_i \to c_j} = \log \frac{\Pr\{X_i = 0|Y_i\}}{\Pr\{X_i = 1|Y_i\}} + \sum_{c_k \neq c_j : c_k \text{ is connected to } v_i} m_{c_k \to v_i} \quad (6.1)$$

$$m_{c_j \to v_i} = 2\tanh^{-1}(1 - 2s_j) \prod_{v_k \neq v_i : v_k \text{ is connected to } c_j} \tanh\left(\frac{m_{v_k \to c_j}}{2}\right) \quad (6.2)$$

where $m_{v_i \to c_j}$ and $m_{c_j \to v_i}$ are messages passed from the variable node $v_i$ to the check node $c_j$ and vice versa, respectively, and $s_j$ is the syndrome corresponding to $c_j$. After certain iterations, assuming the calculation converges to a stationary point, the marginal distribution of each variable node is calculated based on the messages sent from its connected check nodes, and the decision on each bit is made according to the distribution in the following way

$$\hat{x}_i = \left\{ \begin{array}{ll} 0 & \text{if } \frac{\Pr\{X_i=0|Y_i\}}{\Pr\{X_i=1|Y_i\}} + \sum_{c_k : c_k \text{ is connected to } v_i} m_{c_k \to v_i} \geq 0 \\ 1 & \text{otherwise.} \end{array} \right. \quad (6.3)$$

To initialize the iterative procedure, for each variable node $X_i$, the marginal distribution is assumed to be $(\Pr\{X_i = 0|Y_i\}, \Pr\{X_i = 1|Y_i\})$. Therefore, the standard BP decoding algorithm needs the statistics of source and side information as inputs.

However, in our case, the statistics of source-side information are unavailable, i.e., $p_1$ and $p_2$ are unknown. To deal with this problem, let us first consider the case $p_1 = p_2 = p_0$, i.e. $X$ and $Y$ are correlated through a binary symmetrical channel. Now let

$$p_b = \mathrm{H}^{-1}\left(\max\left\{0, \Gamma_b - \frac{\ln n + 1}{n}\right\}\right)$$

where $p_b$ can be interpreted as the maximum cross-over probability of the binary symmetrical channel correlating $X$ and $Y$, such that the error probability of the SA-LDPC-IED scheme $\tilde{\mathcal{I}}_n$ can be maintained asymptotically zero at the $b$-th interaction. Therefore, we will use $p_b$ as the input to the BP decoding at the $b$-th interaction. Moreover, at each interaction, decoding failure is detected and the decoder will send bit 0 to the encoder for more syndromes if one of the following two situations occurs:

- the number of bits with significant log-likelihood (larger than certain value) is less than a threshold within first several iterations of BP decoding;
- or the number of syndrome constaints satisfied by the codeword calculated using (6.3) at the end of each iteration does not increase for several iterations.

On the other hand, successful decoding is identified when the modified BP decoding algorithm converges to a codeword satisfying all syndrome constraints without encountering those two situations listed above. Simulation shows that under this decoding rule, the bit error probability is still very small. Moreover, since this decoding rule is more aggressive than threshold decoding used in section V, for some $(X, Y)$ the rate achieved by the SA-LDPC-IED scheme implemented in this way can be smaller than that given in Theorem 3.

To further consider a general memoryless source-side information pair, i.e. $p_1 \neq p_2$, at the $b$-th interaction, we can quantize $p_1$ into a quantized value, say $q_1$, then calculate the quantized value $q_2$ of $p_2$ according to

$$\Pr\{Y = 0\}\mathrm{H}(q_1) + \Pr\{Y = 1\}\mathrm{H}(q_2) = \mathrm{H}(p_b)$$

and finally apply the modified BP decoding algorithm for each such quantized pair $(q_1, q_2)$. Successful decoding is claimed whenever there is one such quantized $(q_1, q_2)$ that makes the BP decoding algorithm converge to a source sequence satisfying syndrome constraints. When there is a tie, i.e. more than one pair $(q_1, q_2)$ that make the BP decoding algorithm succeed with different outputs, we will choose the one with the smaller value of $q_1$. Here we assume that the distribution of side information $Y$ is known to the decoder. Otherwise, the empirical distribution can be calculated, since the decoder has the full access to side information.

## B. Simulation Results

We first consider the case where the source and side information are correlated through a binary symmetrical channel with unknown cross-over probability, and the side information is uniformly distributed. Figure 4 shows the performance of our implemented scheme (referred to as the simulation rate) along with the conditional entropy rate and the performance upper bound established in Theorem 3, where the blue solid line represents the simulation rate with bit error probabilities below or around $2 \times 10^{-5}$, and the green dashed line represents the upper bound established in Theorem 3 with $\epsilon = 0.1$. The block length is 8000, and the variable degree distribution (from an edge prospective) used is shown below:

$$
\begin{aligned}
\lambda(x) \quad = \quad & 0.178704x + 0.176202x^2 + 0.102845x^5 \\
& + 0.114789x^6 + 0.0122023x^{12} + 0.0479225x^{13} \\
& + 0.115911x^{14} + 0.251424x^{39}
\end{aligned}
$$

which is designed for rate 0.5, and obtained from [19]. It can be seen that our implemented SA-LDPC-IED scheme can indeed adapt to the entropy rate $H(X|Y)$ well in a large rate region. To interpret the upper bound $R_{L(z)}(\epsilon, H(p_0))$ also shown in Figure 4 better, an explanation on $\epsilon$ is needed here. The reason that $\epsilon >>$ bit error probability in the simulation is due to the minimum hamming distance $d_{\min}^{(b)}$ of the code generated by $\mathbf{H}_{b\Delta \times n}^{(b\Delta)}$. From the proof of Theorem 3, it follows that with high probability, $\frac{1}{n}wt(\hat{X}^n - X^n) \leq \epsilon$. On the other hand, $\frac{1}{n}wt(\hat{X}^n - X^n) \leq \epsilon$ implies that $\hat{X}^n = X^n$ if $d_{\min}^{(b)} > \epsilon n$ when the coding procedure terminates at the $b$-th interaction. Moreover, since the implemented decoding algorithm only checks syndrome constraints to determine the decoding success, instead of using thresholds given in Theorem 3, the bound on rate can be improved if the choice of $\epsilon$ for the $b$-th interaction depends on $d_{\min}^{(b)}$, especially for the high rate case as $d_{\min}^{(b)}$ increases with $b$. However, since $d_{\min}^{(b)}$ can not be expressed in a neat way and does not affect redundancy with respect to $k$ when $L(z^k)$ is used, we do not include the corresponding result in this paper. In the meantime, by using the same degree distribution $L(z)$ in Figure 4, Figure 5 shows how fast $R_{L(z^k)}\left(\frac{1}{2\sqrt{k}}, H(p_0)\right)$ converges to $H(p_0)$, where the gap is always less than $0.02$ when $k = 5$.

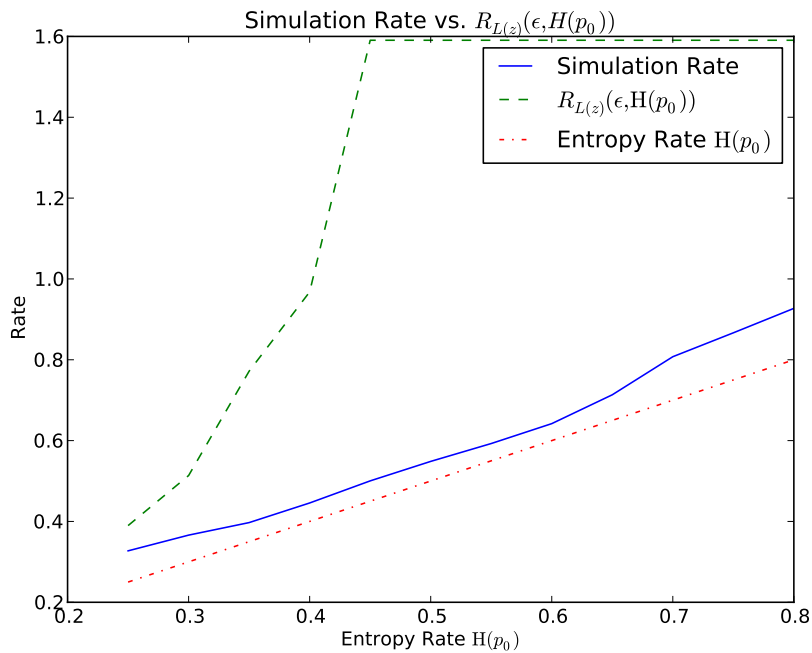

Fig. 4.   Performance of SA-LDPC-IED: Symmetrical Channel

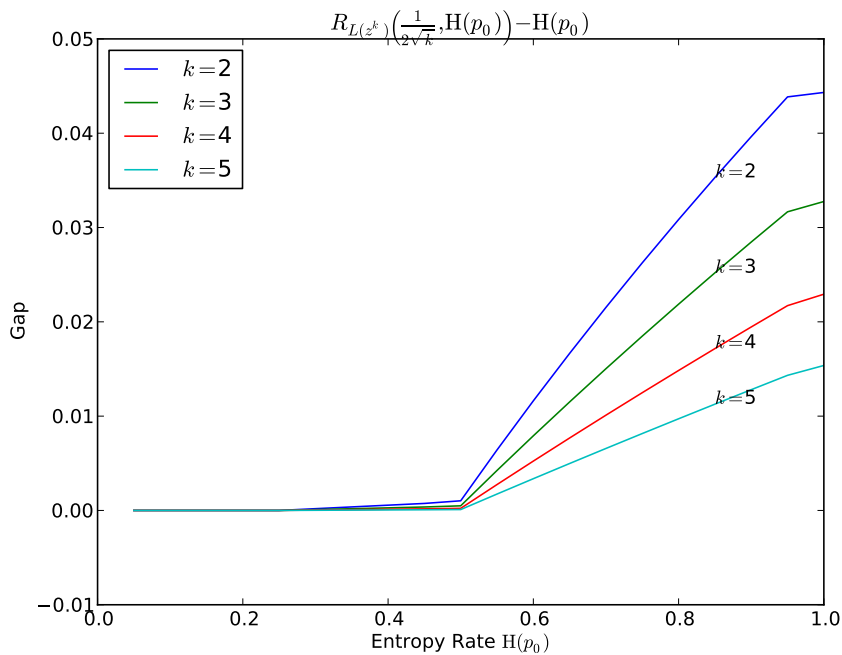

Fig. 5. Redundancy bound with different $k$

We next consider source and side-information pairs correlated through binary asymmetrical channels.

Table I lists our simulation results, where the side information $Y$ is still assumed to be uniformly

| $\Pr\{X = 1|Y = 0\}$ | $\Pr\{X = 0|Y = 1\}$ | Rate |
|:---:|:---:|:---:|
| 0.05 | 0.1959 | 0.541 |
| 0.1 | 0.1206 | 0.544 |
| 0.15 | 0.0766 | 0.543 |
| 0.2 | 0.0481 | 0.540 |

TABLE I

PERFORMANCE OF SA-LDPC-IED: ASYMMETRICAL CHANNEL

distributed, and the transition probabilities are selected such that $\mathrm{H}(X|Y) = 0.5$ for all cases. In our simulation, we did not see any error in 1000 blocks, each block being 8000 bits. As can be seen, our implemented SA-LDPC-IED scheme also works very well in this situation too.

To make a comparison with SWC, a SWC scheme using the same LDPC code (LDPC-SWC) was also implemented for the source and side information correlated through a binary symmetrical channel. The respective results are shown in Table II, where bit error probabilities are maintained below $10^{-5}$ for

| $H(X|Y)$ | $R_{\mathrm{SA\text{-}IED}}$ | $R_{\mathrm{SW}}$ |
|:---:|:---:|:---:|
| 0.426 | 0.473 | 0.5 |

TABLE II

SA-LDPC-IED vs. LDPC-SWC

both SA-LDPC-IED and LDPC-SWC schemes. Note that $R_{\mathrm{SW}}$ is deliberately chosen to be 0.5, since the degree distribution of the LDPC code used here is designed for rate 0.5. Moreover, in the simulation of the LDPC-SWC scheme, we assumed that the cross-over probability $p_0$ is known to the decoder, while in our implemented SA-LDPC-IED scheme, $p_0$ is unknown. Clearly, simulation results show that SA-LDPC-IED outperforms LDPC-SWC.

## VII. CONCLUSION

In this paper, interactive encoding and decoding based on binary low-density parity-check codes with syndrome accumulation (SA-LDPC-IED) has been proposed and investigated. Given any classical universal lossless code $\mathcal{C}_n$ (with block length $n$ and side information available to both the encoder and decoder) and an LDPC code, we have demonstrated, with the help of syndrome accumulation, how to convert $\mathcal{C}_n$ into a universal SA-LDPC-IED scheme. With its word error probability approaching 0 sub-exponentially with $n$, the resulting SA-LDPC-IED scheme has been shown to achieve roughly the same rate performance as does $\mathcal{C}_n$ for each and every individual sequence pair $(x^n, y^n)$ and the conditional entropy rate $\mathrm{H}(X|Y)$ for any stationary, ergodic source and side information $(X, Y)$ as the average variable node degree $\bar{l}$ of the underlying LDPC code increases without bound. When applied to the class of binary source and side information $(X, Y)$ correlated through a binary symmetrical channel with cross-over probability unknown to both the encoder and decoder, the SA-LDPC-IED scheme has been further simplified, resulting in even improved rate performance versus the bit error probability when $\bar{l}$ is not large. Coupled with linear time belief propagation decoding, the SA-LDPC-IED scheme has been implemented for binary source-side information pairs, which confirms the theoretic analysis, and further shows that the SA-LDPC-IED scheme consistently outperforms the Slepian-Wolf coding scheme based on the same underlying LDPC code. In the course of analyzing the performance of the SA-LDPC-IED scheme, probability bounds involving LDPC have been established, and it has been shown that their exponent as a function of the SWC coding rate, the average node degree $\bar{l}$, and a weighted Hamming weight of a codeword has several interesting properties. It is believed that these properties can be applied to analyze the capacity-achieving performance of LDPC for channel coding as well, which will be investigated in the future.

## Acknowledgment

Discussions with Da-ke He on the subject of this paper are hereby acknowledged.

## Appendix A

### Proof of **Lemma 1**

We consider only the case in which $\bar{l}$ is not an integer. The case where $\bar{l}$ is an integer is a bit easier and can be dealt with in a similar manner.

Although there is thorough analysis of the probability $\Pr\{\mathbf{H}_{m \times n} x^n = 0^m\}$ for $\mathbf{H}_{m \times n}$ from $\mathcal{H}_{m,n,L(z),R(z)}$ in [20], [21], [22], and [23], the result therein in general is not applicable to $\mathbf{H}_{b\Delta \times n}^{(b\Delta)}$, the matrix obtained from syndrome accumulation on $\mathbf{H}_{n \times n}$. Towards analyzing $\Pr\left\{\mathbf{H}_{b\Delta \times n}^{(b\Delta)} x^n = 0^{b\Delta}\right\}$, we focus on $\{\mathcal{P}_{b\Delta}\}_{b=1}^{\frac{n}{\Delta}}$ defined in section III-A. Given $\mathcal{P}_{b\Delta} = \{\Lambda_{b\Delta,i}\}_{i=1}^{b\Delta}$, one can classify $\Lambda_{b\Delta,i}$ into three categories:

- $\Lambda_{b\Delta,i} \subseteq \{1, 2, \dots, R_1 n\}$,
- $\Lambda_{b\Delta,i} \subseteq \{R_1 n + 1, R_1 n + 2, \dots, n\}$, or

- $\Lambda_{b\Delta,i} \nsubseteq \{1, 2, \ldots, R_1 n\}$, and $\Lambda_{b\Delta,i} \nsubseteq \{R_1 n + 1, R_1 n + 2, \ldots, n\}$.

To avoid complicating the analysis unnecessarily, we assume that there does not exist $\Lambda_{b\Delta,i}$ falling into the third category. Further effort reveals that this assumption holds if and only if $2^{T - \lfloor \log_2 \Delta \rfloor} | R_1 n$, or in other words,

$$R_1 = \frac{C}{2^{\lfloor \log_2 \Delta \rfloor}}$$

for some positive integer $C$, where the parameter $\Delta$ is a function of block length $n$. In fact, in this paper we only consider the case where $\Delta \sim \sqrt{n}$, which implies $2^{\lfloor \log_2 \Delta \rfloor} \sim \sqrt{n}$, and therefore the assumption above always holds for sufficiently large $n$ if $\bar{l}$ is a fractional number with a power of 2 as its denominator.

Consequently, each $\Lambda_{b\Delta,i}$ can be further categorized into one of four cases:

- $\Lambda_{b\Delta,i} \subseteq \{1, 2, \ldots, R_1 n\}$, and $|\Lambda_{b\Delta,i}| = 2^{T - \lceil \log_2 b\Delta \rceil}$;
- $\Lambda_{b\Delta,i} \subseteq \{1, 2, \ldots, R_1 n\}$, and $|\Lambda_{b\Delta,i}| = 2^{T - \lceil \log_2 b\Delta \rceil + 1}$;
- $\Lambda_{b\Delta,i} \subseteq \{R_1 n + 1, R_1 n + 2, \ldots, n\}$, and $|\Lambda_{b\Delta,i}| = 2^{T - \lceil \log_2 b\Delta \rceil}$; or
- $\Lambda_{b\Delta,i} \subseteq \{R_1 n + 1, R_1 n + 2, \ldots, n\}$, and $|\Lambda_{b\Delta,i}| = 2^{T - \lceil \log_2 b\Delta \rceil + 1}$.

Now we use $\left\{ t_{b\Delta}^{(i)} \right\}_{i=1}^{4}$ to represent the number of $\Lambda_{b\Delta,i}$'s falling into each category, which are given by the following formulas:

$$
\begin{aligned}
t_{b\Delta}^{(1)} &= \min\left\{ 2b\Delta - 2^{\lceil \log_2 b\Delta \rceil}, R_1 2^{\lceil \log_2 b\Delta \rceil} \right\}, \\
t_{b\Delta}^{(2)} &= \max\left\{ R_1 2^{\lceil \log_2 b\Delta \rceil - 1} - \left( b\Delta - 2^{\lceil \log_2 b\Delta \rceil - 1} \right), 0 \right\}, \\
t_{b\Delta}^{(3)} &= \max\left\{ R_2 2^{\lceil \log_2 b\Delta \rceil} - 2 \left( 2^{\lceil \log_2 b\Delta \rceil} - b\Delta \right), 0 \right\}, \\
t_{b\Delta}^{(4)} &= \min\left\{ 2^{\lceil \log_2 b\Delta \rceil} - b\Delta, R_2 2^{\lceil \log_2 b\Delta \rceil - 1} \right\}.
\end{aligned}
$$

Note that we assume that block length $n = 2^T$ for some integer $T$. It then follows that

$$
\begin{aligned}
\frac{t_{b\Delta}^{(1)}}{n} &= \min\left\{ \frac{2b\Delta}{n} - 2^{\lceil \log_2 \frac{b\Delta}{n} \rceil}, R_1 2^{\lceil \log_2 \frac{b\Delta}{n} \rceil} \right\} \\
\frac{t_{b\Delta}^{(2)}}{n} &= \max\left\{ R_1 2^{\lceil \log_2 \frac{b\Delta}{n} \rceil - 1} - \left( \frac{b\Delta}{n} - 2^{\lceil \log_2 \frac{b\Delta}{n} \rceil - 1} \right), 0 \right\} \\
\frac{t_{b\Delta}^{(3)}}{n} &= \max\left\{ R_2 2^{\lceil \log_2 \frac{b\Delta}{n} \rceil} - 2 \left( 2^{\lceil \log_2 \frac{b\Delta}{n} \rceil} - \frac{b\Delta}{n} \right), 0 \right\} \\
\frac{t_{b\Delta}^{(4)}}{n} &= \min\left\{ 2^{\lceil \log_2 \frac{b\Delta}{n} \rceil} - \frac{b\Delta}{n}, R_2 2^{\lceil \log_2 \frac{b\Delta}{n} \rceil - 1} \right\}
\end{aligned}
$$

Recall that

$$c_{b\Delta} = 2^{T - \lceil \log_2 b\Delta \rceil} = 2^{-\lceil \log_2 \frac{b\Delta}{n} \rceil}.$$

Therefore $c_{b\Delta}$ also depends only on $\frac{b\Delta}{n}$.

Now define $\mathcal{H}_{n,L(z),\varkappa,b\Delta}$ as a subset of $\mathcal{H}_{n,L(z)}$ such that

$$\mathbf{H}_{n\times n} \in \mathcal{H}_{n,L(z),\varkappa,b\Delta}$$

if and only if

$$\mathbf{H}_{n\times n} \in \mathcal{H}_{n,L(z)} \text{ and } \mathbf{H}_{b\Delta \times n}^{(b\Delta)} x^n = 0^{b\Delta}$$

where $\varkappa$ is the support set of $x^n$. It is easy to see that given $x^n$ (and therefore $\varkappa$), these subsets $\mathcal{H}_{n,L(z),\varkappa,b\Delta}$ are nested with each other

$$\mathcal{H}_{n,L(z),\varkappa,s\Delta} \subseteq \mathcal{H}_{n,L(z),\varkappa,b\Delta}$$

if $s \geq b$. Furthermore, let $\Lambda_{n,L(z),\varkappa,b\Delta} = |\mathcal{H}_{n,L(z),\varkappa,b\Delta}|$. Then we have

$$\Pr\left\{\mathbf{H}_{b\Delta \times n}^{(b\Delta)} x^n = 0^{b\Delta}\right\} = \frac{\Lambda_{n,L(z),\varkappa,b\Delta}}{|\mathcal{H}_{n,L(z)}|} \tag{A.1}$$

where $\mathbf{H}_{n\times n}$ is uniformly picked from $\mathcal{H}_{n,L(z)}$. Therefore the main issue is to derive asymptotic formulas for $|\mathcal{H}_{n,L(z)}|$ and $\Lambda_{n,L(z),\varkappa,b\Delta}$. At this point, we invoke the following result from Mineev and Pavlov [24] (see also [25] for a stronger version).

**Theorem 5** (Mineev-Pavlov). *Suppose $\mathcal{H}_{\vec{r},\vec{l}}$ is the ensemble of $m \times n$ 0-1 matrices with $i$-th row sum $r_i$ and $j$-th column sum $l_j$ satisfying $\max\{r_i, l_j : 1 \leq i \leq m \text{ and } 1 \leq j \leq n\} \leq \log^{1/4-\epsilon} m$, where $\epsilon$ is an arbitrarily small positive constant. Then*

$$|\mathcal{H}_{\vec{r},\vec{l}}| = \frac{(\sum_{i=1}^m r_i)!}{(\prod_{i=1}^m r_i!)(\prod_{i=1}^n l_i!)} \left[\exp\left\{-\frac{2}{(2\sum_{i=1}^m r_i)^2}\left(\sum_{i=1}^m r_i(r_i-1)\sum_{j=1}^n l_j(l_j-1)\right)\right\} + o(m^{-0.5+\delta})\right] \tag{A.2}$$

*where $0 < \delta < 0.5$ is an arbitrarily small constant.*

First of all, applying Theorem 5 to $|\mathcal{H}_{n,L(z)}|$, we have

$$|\mathcal{H}_{n,L(z)}| = \frac{(\bar{l}n)!}{(r_1!)^{R_1 n}(r_2!)^{R_2 n}\prod_{i=1}^L (l_i!)^{L_i n}}(C_{L(z)} + o(n^{-0.5+\delta}))$$

where

$$C_{L(z)} = \exp\left\{-\frac{(R_1 r_1(r_1-1) + R_2 r_2(r_2-1))\sum_{i=1}^L L_i l_i(l_i-1)}{2\bar{l}^2}\right\}$$

Towards calculating $\Lambda_{n,L(z),\varkappa,b\Delta}$, note that each $\mathbf{H}_{n\times n}$ consists of two sub-matrices $\mathbf{H}_{n\times|\varkappa|}^{\varkappa}$ and $\mathbf{H}_{n\times(n-|\varkappa|)}^{\varkappa^c}$, where $\varkappa^c$ is the complement of $\varkappa$. Suppose $\{r_i^{\varkappa}\}_{i=1}^n$ is the row-sum profile of $\mathbf{H}_{n\times|\varkappa|}^{\varkappa}$. Then the row-sum

 

profile $\{r_i^{\varkappa^c}\}_{i=1}^n$ of $\mathbf{H}_{n\times(n-|\varkappa|)}^{\varkappa^c}$ is given by

$$r_i^{\varkappa^c} \;=\; r_1 - r_i^{\varkappa} \text{ for } 1 \le i \le R_1 n$$

$$r_i^{\varkappa^c} \;=\; r_2 - r_i^{\varkappa} \text{ for } R_1 n + 1 \le i \le n$$

For each $\mathbf{H}_{n\times n} \in \mathcal{H}_{n,L(z)}$, its $\mathbf{H}_{n\times|\varkappa|}^{\varkappa}$ and $\mathbf{H}_{n\times(n-|\varkappa|)}^{\varkappa^c}$ should have $L^{\varkappa}(z)$ and $L^{\varkappa^c}(z)$ as their column-sum profiles. Therefore

$$0 \le r_i^{\varkappa} \le r_1 \text{ for } 1 \le i \le R_1 n \tag{A.3}$$

$$0 \le r_i^{\varkappa} \le r_2 \text{ for } R_1 n + 1 \le i \le n \tag{A.4}$$

$$\sum_{i=1}^{n} r_i^{\varkappa} = \bar{l}^{\varkappa} n \tag{A.5}$$

Note that

$$
\mathbf{H}_{b\Delta\times n}^{(b\Delta)} x^n \;=\; \begin{pmatrix} \sum_{i\in\Lambda_1} h_{i,1} & \sum_{i\in\Lambda_1} h_{i,2} & \cdots & \sum_{i\in\Lambda_1} h_{i,n} \\ \vdots & \vdots & \ddots & \vdots \\ \sum_{i\in\Lambda_{b\Delta}} h_{i,1} & \sum_{i\in\Lambda_{b\Delta}} h_{i,2} & \cdots & \sum_{i\in\Lambda_{b\Delta}} h_{i,n} \end{pmatrix} x^n
$$

$$
\;=\; \underbrace{\begin{pmatrix} \sum_{i\in\Lambda_1} h_{i,j_1} & \sum_{i\in\Lambda_1} h_{i,j_2} & \cdots & \sum_{i\in\Lambda_1} h_{i,j_{|\varkappa|}} \\ \vdots & \vdots & \ddots & \vdots \\ \sum_{i\in\Lambda_{b\Delta}} h_{i,j_1} & \sum_{i\in\Lambda_{b\Delta}} h_{i,j_2} & \cdots & \sum_{i\in\Lambda_{b\Delta}} h_{i,j_{|\varkappa|}} \end{pmatrix}}_{\varkappa} \begin{pmatrix} 1 \\ 1 \\ \vdots \\ 1 \end{pmatrix}
$$

$$
\;=\; \begin{pmatrix} \sum_{j\in\varkappa} \sum_{i\in\Lambda_1} h_{ij} \\ \vdots \\ \sum_{j\in\varkappa} \sum_{i\in\Lambda_{b\Delta}} h_{ij} \end{pmatrix}
$$

$$
\;=\; \begin{pmatrix} \sum_{i\in\Lambda_1} \sum_{j\in\varkappa} h_{ij} \\ \vdots \\ \sum_{i\in\Lambda_{b\Delta}} \sum_{j\in\varkappa} h_{ij} \end{pmatrix}
$$

$$
\;=\; \begin{pmatrix} \sum_{i\in\Lambda_1} r_i^{\varkappa} \\ \vdots \\ \sum_{i\in\Lambda_{b\Delta}} r_i^{\varkappa} \end{pmatrix}
$$

Then $\mathbf{H}_{n \times n} \in \mathcal{H}_{n, L(z), \varkappa, b\Delta}$ if and only if

$$2 \left| \sum_{u=1}^{c_{b\Delta}} r^{\varkappa}_{c_{b\Delta} j + u} \right| \quad \text{for } 0 \le j \le t^{(1)}_{b\Delta} - 1 \tag{A.6}$$

$$2 \left| \sum_{u=1}^{2c_{b\Delta}} r^{\varkappa}_{t^{(1)}_{b\Delta} c_{b\Delta} + 2 c_{b\Delta} j + u} \right| \quad \text{for } 0 \le j \le t^{(2)}_{b\Delta} - 1 \tag{A.7}$$

$$2 \left| \sum_{u=1}^{c_{b\Delta}} r^{\varkappa}_{t^{(1)}_{b\Delta} c_{b\Delta} + 2 t^{(2)}_{b\Delta} c_{b\Delta} + c_{b\Delta} j + u} \right| \quad \text{for } 0 \le j \le t^{(3)}_{b\Delta} - 1 \tag{A.8}$$

$$2 \left| \sum_{u=1}^{2c_{b\Delta}} r^{\varkappa}_{t^{(1)}_{b\Delta} c_{b\Delta} + 2 t^{(2)}_{b\Delta} c_{b\Delta} + t^{(3)}_{b\Delta} c_{b\Delta} + 2 c_{b\Delta} j + u} \right| \quad \text{for } 0 \le j \le t^{(4)}_{b\Delta} - 1 \tag{A.9}$$

Let $\mathcal{R}_{b\Delta, \bar{l}\varkappa}$ denote the set of all row-sum profiles $\{r^{\varkappa}_i\}_{i=1}^{n}$ which satisfy the constraints (A.3) to (A.9). Furthermore, let $\Lambda^{\varkappa}_{\{r^{\varkappa}_i\}_{i=1}^{n}}$ and $\Lambda^{\varkappa}_{\{r^{\varkappa^c}_i\}_{i=1}^{n}}$ denote the number of $\mathbf{H}^{\varkappa}_{n \times |\varkappa|}$'s and $\mathbf{H}^{\varkappa^c}_{n \times (n - |\varkappa|)}$'s with the given row profile $\{r^{\varkappa}_i\}^n$ and $\{r^{\varkappa^c}_i\}^n$, respectively. Then it is easy to see that

$$\Lambda_{n, L(z), \varkappa, b\Delta} = \sum_{\{r_i\}_{i=1}^{n} \in \mathcal{R}_{b\Delta, \bar{l}\varkappa}} \Lambda^{\varkappa}_{\{r^{\varkappa}_i\}_{i=1}^{n}} \Lambda^{\varkappa^c}_{\{r^{\varkappa^c}_i\}_{i=1}^{n}} \tag{A.10}$$

Applying Theorem 5 to $\Lambda^{\varkappa}_{\{r^{\varkappa}_i\}_{i=1}^{n}}$ and $\Lambda^{\varkappa}_{\{r^{\varkappa}_i\}_{i=1}^{n}}$, we have

$$\begin{aligned}
\Lambda^{\varkappa^c}_{\{r^{\varkappa}_i\}_{i=1}^{n}} &= \frac{(\sum_{i=1}^{n} r^{\varkappa}_i)!}{(\prod_{i=1}^{n} r^{\varkappa}_i!) \prod_{i=1}^{L} (l_i!)^{L^{\varkappa}_i n}} (C_{r^{\varkappa}} + o(n^{-0.5+\delta})) \\
&= \frac{(\bar{l}^{\varkappa} n)!}{(\prod_{i=1}^{n} r^{\varkappa}_i!) \prod_{i=1}^{L} (l_i!)^{L^{\varkappa}_i n}} (C_{r^{\varkappa}} + o(n^{-0.5+\delta}))
\end{aligned} \tag{A.11}$$

where

$$\exp\left\{ -\frac{r_2(l_L - 1)}{2} \right\} \le \exp\left\{ -\frac{r_2 \sum_{i=1}^{L} L^{\varkappa}_i l_i (l_i - 1)}{2 \bar{l}^{\varkappa}} \right\} \le C_{r^{\varkappa}} \le 1 \ .$$

Similarly,

$$\Lambda^{\varkappa^c}_{\{r^{\varkappa}_i\}_{i=1}^{n}} = \frac{((\bar{l} - \bar{l}^{\varkappa}) n)!}{\left( \prod_{i=1}^{R_1 n} (r_1 - r^{\varkappa}_i) \prod_{i=R_1 n + 1}^{n} (r_2 - r^{\varkappa}_i) \right) \prod_{i=1}^{L} (l_i!)^{(L_i - L^{\varkappa}_i) n}} (C_{r^{\varkappa^c}} + o(n^{-0.5+\delta})) \tag{A.12}$$

where

$$\exp\left\{ -\frac{r_2(l_L - 1)}{2} \right\} \le \exp\left\{ -\frac{r_2 \sum_{i=1}^{L} (L_i - L^{\varkappa}_i) l_i (l_i - 1)}{2(\bar{l} - \bar{l}^{\varkappa})} \right\} \le C_{r^{\varkappa^c}} \le 1 \ .$$

Combining (A.1) with (A.10) to (A.12) yields

$$\Pr\left\{\mathbf{H}_{b\Delta\times n}^{(b\Delta)}x^n = 0^{b\Delta}\right\} = \frac{\Lambda_{n,L(z),\varkappa,b\Delta}}{|\mathcal{H}_{n,L(z)}|}$$

$$\leq \frac{2}{C_{L(z)}} \frac{\sum_{\{r_i^\varkappa\}_{i=1}^n \in \mathcal{R}_{b\Delta,\varkappa}} \frac{C_{r^\varkappa}C_{r,\varkappa^c}(n\bar{l}^\varkappa)!\left(n(\bar{l}-\bar{l}^\varkappa)\right)!}{\prod_{i=1}^L (l_i!)^{L_i n}\prod_{i=1}^{R_1 n} r_i^\varkappa!(r_1-r_i^\varkappa)!\prod_{i=R_1 n+1}^n r_i^\varkappa!(r_2-r_i^\varkappa)!}}{\frac{(n\bar{l})!}{(r_1!)^{R_1 n}(r_2)^{R_2 n}\prod_{i=1}^L (l_i!)^{L_i n}}}$$

$$\leq \frac{2}{C_{L(z)}} \begin{pmatrix} n\bar{l} \\ n\bar{l}^\varkappa \end{pmatrix}^{-1} \sum_{\{r_i^\varkappa\}_{i=1}^n \in \mathcal{R}_{b\Delta,\varkappa}} \prod_{i=1}^{R_1 n} \begin{pmatrix} r_1 \\ r_i^\varkappa \end{pmatrix} \prod_{i=R_1 n+1}^n \begin{pmatrix} r_2 \\ r_i^\varkappa \end{pmatrix} \quad (A.13)$$

for sufficiently large $n$. To further evaluate $\Pr\{\mathbf{H}_{b\Delta\times n}^{(b\Delta)}x^n = 0^{b\Delta}\}$, we define the type $\left(m^{(1)}, m^{(2)}, m^{(3)}, m^{(4)}\right)$ of $\{r_i^\varkappa\}_{i=1}^n$ as follows:

$$m_s^{(1)} \triangleq \sum_{j=0}^{t_{b\Delta}^{(1)}-1} \delta\left(\sum_{u=1}^{c_{b\Delta}} r_{c_{b\Delta}j+u}^\varkappa - s\right) \text{ for } 0 \leq s \leq c_{b\Delta}r_1$$

$$m_s^{(2)} \triangleq \sum_{j=0}^{t_{b\Delta}^{(2)}-1} \delta\left(\sum_{u=1}^{2c_{b\Delta}} r_{c_{b\Delta}t_{b\Delta}^{(1)}+2c_{b\Delta}j+u}^\varkappa - s\right) \text{ for } 0 \leq s \leq 2c_{b\Delta}r_1$$

$$m_s^{(3)} \triangleq \sum_{j=0}^{t_{b\Delta}^{(3)}-1} \delta\left(\sum_{u=1}^{c_{b\Delta}} r_{t_{b\Delta}^{(1)}+2t_{b\Delta}^{(2)}c_{b\Delta}+c_{b\Delta}j+u}^\varkappa - s\right) \text{ for } 0 \leq s \leq c_{b\Delta}r_2$$

$$m_s^{(4)} \triangleq \sum_{j=0}^{t_{b\Delta}^{(4)}-1} \delta\left(\sum_{u=1}^{2c_{b\Delta}} r_{t_{b\Delta}^{(1)}c_{b\Delta}+2t_{b\Delta}^{(2)}c_{b\Delta}+t_{b\Delta}^{(3)}c_{b\Delta}+2c_{b\Delta}j+u}^\varkappa - s\right) \text{ for } 0 \leq s \leq 2c_{b\Delta}r_2$$

where

$$\delta(x) \triangleq \begin{cases} 1 & \text{if } x = 0 \\ 0 & \text{otherwise.} \end{cases}$$

Now we can see that $\{r_i^\varkappa\}_{i=1}^n$ belongs to $\mathcal{R}_{b\Delta,\varkappa}$ if and only if its type $\left(m^{(1)}, m^{(2)}, m^{(3)}, m^{(4)}\right)$ satisfies

$$\sum_{j=0}^{\lfloor\frac{c_{b\Delta}r_1}{2}\rfloor} m_{2j}^{(1)} = t_{b\Delta}^{(1)} \quad (A.14)$$

$$\sum_{j=0}^{c_{b\Delta}r_1} m_{2j}^{(2)} = t_{b\Delta}^{(2)} \quad (A.15)$$

$$\sum_{j=0}^{\lfloor\frac{c_{b\Delta}r_2}{2}\rfloor} m_{2j}^{(3)} = t_{b\Delta}^{(3)} \quad (A.16)$$

$$\sum_{j=0}^{c_{b\Delta}r_2} m_{2j}^{(4)} = t_{b\Delta}^{(4)} \quad (A.17)$$

and

$$\sum_{j=0}^{\lfloor \frac{c_{b\Delta}r_1}{2}\rfloor} 2j \cdot m_{2j}^{(1)} + \sum_{j=0}^{c_{b\Delta}r_1} 2j \cdot m_{2j}^{(2)} + \sum_{j=0}^{\lfloor \frac{c_{b\Delta}r_2}{2}\rfloor} 2j \cdot m_{2j}^{(3)} + \sum_{j=0}^{c_{b\Delta}r_2} 2j \cdot m_{2j}^{(4)} = \bar{l}^{\varkappa} n \quad \text{(A.18)}$$

Denote the set of types $\left\{m^{(1)}, m^{(2)}, m^{(3)}, m^{(4)}\right\}$ satisfying the above constraints (A.14) to (A.18) by $\mathcal{M}_{b\Delta, \varkappa}$. If $\mathcal{M}_{b\Delta, \varkappa} \neq \emptyset$, then the constraints (A.14) to (A.18) implies

$$0 \leq$$
$$\sum_{j=0}^{\lfloor \frac{c_{b\Delta}r_1}{2}\rfloor} (c_{b\Delta}r_1 - \pi(c_{b\Delta}r_1) - 2j)m_{2j}^{(1)} + \sum_{j=0}^{c_{b\Delta}r_1} (2c_{b\Delta}r_1 - 2j)m_{2j}^{(2)}$$
$$+ \sum_{j=0}^{\lfloor \frac{c_{b\Delta}r_2}{2}\rfloor} (c_{b\Delta}r_2 - \pi(c_{b\Delta}r_2) - 2j)m_{2j}^{(3)} + \sum_{j=0}^{c_{b\Delta}r_2} (2c_{b\Delta}r_2 - 2j)m_{2j}^{(4)}$$
$$= t_{b\Delta}^{(1)}(c_{b\Delta}r_1 - \pi(c_{b\Delta}r_1)) + 2t_{b\Delta}^{(2)}c_{b\Delta}r_1 + t_{b\Delta}^{(3)}(c_{b\Delta}r_2 - \pi(c_{b\Delta}r_2)) + 2t_{b\Delta}^{(4)}c_{b\Delta}r_2 - \bar{l}^{\varkappa} n$$
$$= n\bar{l} - t_{b\Delta}^{(1)}\pi(c_{b\Delta}r_1) - t_{b\Delta}^{(3)}\pi(c_{b\Delta}r_2) - \bar{l}^{\varkappa} n \quad \text{(A.19)}$$

and therefore

$$\bar{l}^{\varkappa} \leq \bar{l} - \frac{t_{b\Delta}^{(1)}}{n}\pi(c_{b\Delta}r_1) - \frac{t_{b\Delta}^{(3)}}{n}\pi(c_{b\Delta}r_2).$$

On the other hand, $\mathcal{M}_{t,\theta} = \emptyset$ implies $\Pr\left\{\mathbf{H}_{b\Delta \times n}^{(b\Delta)} x^n = 0^{b\Delta}\right\} = 0$, and hence the lemma is proved when

$$\bar{l}^{\varkappa} > \bar{l} - \frac{t_{b\Delta}^{(1)}}{n}\pi(c_{b\Delta}r_1) - \frac{t_{b\Delta}^{(3)}}{n}\pi(c_{b\Delta}r_2).$$

Now suppose

$$\bar{l}^{\varkappa} < \bar{l} - \frac{t_{b\Delta}^{(1)}}{n}\pi(c_{b\Delta}r_1) - \frac{t_{b\Delta}^{(3)}}{n}\pi(c_{b\Delta}r_2).$$

For convenience, define

$$k^{(1)} = \frac{c_{b\Delta}r_1 - \pi(c_{b\Delta}r_1)}{2}$$
$$k^{(2)} = c_{b\Delta}r_1$$
$$k^{(3)} = \frac{c_{b\Delta}r_2 - \pi(c_{b\Delta}r_2)}{2}$$
$$k^{(4)} = c_{b\Delta}r_2$$

To proceed, we can group $\{r_i^\varkappa\}_{i=1}^n$ with the same type together, and therefore have

$$\sum_{\{r_i^\varkappa\}_{i=1}^n \in \mathcal{R}_{b\Delta,\varkappa}} \prod_{i=1}^{R_1 n} \begin{pmatrix} r_1 \\ r_i^\varkappa \end{pmatrix} \prod_{i=R_1 n+1}^{n} \begin{pmatrix} r_2 \\ r_i^\varkappa \end{pmatrix} =$$

$$\sum_{\{m^{(1)},m^{(2)},m^{(3)},m^{(4)}\} \in \mathcal{M}_{b\Delta,\varkappa}} \prod_{i=1}^{4} \begin{pmatrix} t_{b\Delta}^{(i)} \\ m_0^{(i)}, m_2^{(i)}, \ldots, m_{2k^{(i)}}^{(i)} \end{pmatrix}$$

$$\times \prod_{j=0}^{k^{(1)}} \left( \sum_{\{r_u^\varkappa\}_{u=1}^{c_{b\Delta}} : \sum_{u=1}^{c_{b\Delta}} r_u = 2j} \prod_{u=1}^{c_{b\Delta}} \begin{pmatrix} r_1 \\ r_u^\varkappa \end{pmatrix} \right)^{m_{2j}^{(1)}} \prod_{j=0}^{k^{(2)}} \left( \sum_{\{r_u^\varkappa\}_{u=1}^{2c_{b\Delta}} : \sum_{u=1}^{2c_{b\Delta}} r_u^\varkappa = 2j} \prod_{u=1}^{2c_{b\Delta}} \begin{pmatrix} r_1 \\ r_u^\varkappa \end{pmatrix} \right)^{m_{2j}^{(2)}}$$

$$\prod_{j=0}^{k^{(3)}} \left( \sum_{\{r_u^\varkappa\}_{u=1}^{c_{b\Delta}} : \sum_{u=1}^{c_{b\Delta}} r_u^\varkappa = 2j} \prod_{u=1}^{c_{b\Delta}} \begin{pmatrix} r_2 \\ r_u^\varkappa \end{pmatrix} \right)^{m_{2j}^{(3)}} \prod_{j=0}^{k^{(4)}} \left( \sum_{\{r_u^\varkappa\}_{u=1}^{2c_{b\Delta}} : \sum_{u=1}^{2c_{b\Delta}} r_u^\varkappa = 2j} \prod_{u=1}^{2c_{b\Delta}} \begin{pmatrix} r_2 \\ r_u^\varkappa \end{pmatrix} \right)^{m_{2j}^{(4)}}$$

Now define for any $j \geq 0$

$$\xi_j^{(1)} \triangleq \sum_{\{r_u^\varkappa\}_{u=1}^{c_{b\Delta}} : \sum_{u=1}^{c_{b\Delta}} r_u^\varkappa = j} \prod_{u=1}^{c_{b\Delta}} \begin{pmatrix} r_1 \\ r_u^\varkappa \end{pmatrix}$$

$$\xi_j^{(2)} \triangleq \sum_{\{r_u^\varkappa\}_{u=1}^{2c_{b\Delta}} : \sum_{u=1}^{2c_{b\Delta}} r_u^\varkappa = j} \prod_{u=1}^{2c_{b\Delta}} \begin{pmatrix} r_1 \\ r_u^\varkappa \end{pmatrix}$$

$$\xi_j^{(3)} \triangleq \sum_{\{r_u^\varkappa\}_{u=1}^{c_{b\Delta}} : \sum_{u=1}^{c_{b\Delta}} r_u^\varkappa = j} \prod_{u=1}^{c_{b\Delta}} \begin{pmatrix} r_2 \\ r_u^\varkappa \end{pmatrix}$$

$$\xi_j^{(4)} \triangleq \sum_{\{r_u^\varkappa\}_{u=1}^{2c_{b\Delta}} : \sum_{u=1}^{2c_{b\Delta}} r_u^\varkappa = j} \prod_{u=1}^{2c_{b\Delta}} \begin{pmatrix} r_2 \\ r_u^\varkappa \end{pmatrix}$$

Furthermore, we define

$$M_{\{m^{(i)}\}_{i=1}^4} \triangleq \prod_{i=1}^{4} \left[ \begin{pmatrix} t_{b\Delta}^{(i)} \\ m_0^{(i)}, m_2^{(i)}, \ldots, m_{2k^{(i)}}^{(i)} \end{pmatrix} \prod_{j=0}^{k^{(i)}} \left( \xi_{2j}^{(i)} \right)^{m_{2j}^{(i)}} \right]$$

Therefore

$$\sum_{\{r_i^\varkappa\}_{i=1}^n \in \mathcal{R}_{b\Delta,\varkappa}} \prod_{i=1}^{R_1 n} \begin{pmatrix} r_1 \\ r_i^\varkappa \end{pmatrix} \prod_{i=R_1 n+1}^{n} \begin{pmatrix} r_2 \\ r_i^\varkappa \end{pmatrix} = \sum_{\{m^{(i)}\}_{i=1}^4 \in \mathcal{M}_{b\Delta,\varkappa}} M_{\{m^{(i)}\}_{i=1}^4}$$

In view of (A.14) to (A.18), we can get a trivial bound on $|\mathcal{M}_{b\Delta,\kappa}|$ as follows:

$$|\mathcal{M}_{b\Delta,\varkappa}| \leq \left( \frac{n\bar{l}^\varkappa}{2} + 1 \right)^{k^{(1)}} \left( \frac{n\bar{l}^\varkappa}{2} + 1 \right)^{k^{(2)}} \left( \frac{n\bar{l}^\varkappa}{2} + 1 \right)^{k^{(3)}} \left( \frac{n\bar{l}^\varkappa}{2} + 1 \right)^{k^{(4)}}$$
$$\leq \left( n\bar{l}^\varkappa \right)^{k^{(1)}+k^{(2)}+k^{(3)}+k^{(4)}} \leq \left( n\bar{l}^\varkappa \right)^{3\lceil \bar{l} \rceil c_{b\Delta}}$$

In a similar manner, in view of (A.14) to (A.17) and (A.19, we have

$$\left| \mathcal{M}_{b\Delta, \varkappa} \right| \leq \left( n(\bar{l} - \bar{l}^{\varkappa}) \right)^{3\lceil \bar{l} \rceil c_{b\Delta}}$$

Define

$$\hat{l}^{\varkappa} = \max \left\{ \frac{1}{n}, \min\{\bar{l}^{\varkappa}, \bar{l} - \bar{l}^{\varkappa}\} \right\} .$$

Then we have

$$\sum_{\{m^{(i)}\}_{i=1}^4 \in \mathcal{M}_{b\Delta, \varkappa}} M_{\{m^{(i)}\}_{i=1}^4}$$

$$\leq \quad (n\hat{l}^{\varkappa})^{3\lceil \bar{l} \rceil c_{b\Delta}} \max_{\{m^{(i)}\}_{i=1}^4 \in \mathcal{M}_{b\Delta, \varkappa}} M_{\{m^{(i)}\}_{i=1}^4}$$

$$\leq \quad \exp\left\{ \frac{3n\lceil \bar{l} \rceil}{b\Delta} \ln(n\hat{l}^{\varkappa}) \right\} \max_{\{m^{(i)}\}_{i=1}^4 \in \mathcal{M}_{b\Delta, \varkappa}} M_{\{m^{(i)}\}_{i=1}^4}$$

where the last inequality is due to the fact that $c_{b\Delta} \leq \frac{n}{b\Delta}$. This, coupled with (A.13), implies

$$\Pr\left\{ \mathbf{H}_{b\Delta \times n}^{(b\Delta)} x^n = 0^{b\Delta} \right\}$$

$$\leq \quad \exp\left\{ \frac{3n\lceil \bar{l} \rceil}{b\Delta} \ln(n\hat{l}^{\varkappa}) + O(1) \right\} \begin{pmatrix} n\bar{l} \\ n\bar{l}^{\varkappa} \end{pmatrix}^{-1} \max_{\{m^{(i)}\}_{i=1}^4 \in \mathcal{M}_{b\Delta, \varkappa}} M_{\{m^{(i)}\}_{i=1}^4} . \quad (A.20)$$

To continue, we now upper bound

$$\max_{\{m^{(i)}\}_{i=1}^4 \in \mathcal{M}_{b\Delta, \varkappa}} M_{\{m^{(i)}\}_{i=1}^4}$$

under the conditions (A.14) to (A.18). By the type bound [26, Lemma 2.3],

$$\max \ln M_{\{m^{(i)}\}_{i=1}^4} = \max \ln \prod_{i=1}^4 \left( \frac{t_{b\Delta}^{(i)}!}{\prod_{j=0}^{k^{(i)}} m_{2j}^{(i)}!} \prod_{j=0}^{k^{(i)}} \left( \xi_{2j}^{(i)} \right)^{m_{2j}^{(i)}} \right)$$

$$\leq \quad \max\left\{ \sum_{i=1}^4 t_{b\Delta}^{(i)} \ln t_{b\Delta}^{(i)} - \sum_{i=1}^4 \sum_{j=0}^{k^{(i)}} \left( m_{2j}^{(i)} \ln m_{2j}^{(i)} \right) + \sum_{i=1}^4 \sum_{j=0}^{k^{(i)}} m_{2j}^{(i)} \ln \xi_{2j}^{(i)} \right\}$$

$$\leq \quad \max G\left( \left\{ m^{(i)} \right\}_{i=1}^4 \right) \quad (A.21)$$

where

$$G\left( \left\{ m^{(i)} \right\}_{i=1}^4 \right) \triangleq \sum_{i=1}^4 t_{b\Delta}^{(i)} \ln t_{b\Delta}^{(i)} - \sum_{i=1}^4 \sum_{j=0}^{k^{(i)}} \left( m_{2j}^{(i)} \ln m_{2j}^{(i)} \right) + \sum_{i=1}^4 \sum_{j=0}^{k^{(i)}} m_{2j}^{(i)} \ln \xi_{2j}^{(i)} \quad (A.22)$$

in which $m_{2j}^{(i)}$ can take any non-negative real number with constraints (A.14) to (A.18). Since the function

$$f(x) = -x \ln x + cx$$

is concave in the region $x > 0$, it follows that $G\left(\left\{m^{(i)}\right\}_{i=1}^4\right)$ is a concave function, and hence the maximum can be calculated by using K.K.T condition, which is shown as follows.

Define the function $F\left(\left\{m^{(i)}\right\}_{i=1}^4, \left\{\alpha_i\right\}_{i=1}^4, \beta\right)$ as

$$F\left(\left\{m^{(i)}\right\}_{i=1}^4, \left\{\alpha_i\right\}_{i=1}^4, \beta\right) = G\left(\left\{m^{(i)}\right\}_{i=1}^4\right) + \sum_{i=1}^4 \alpha_i \sum_{j=0}^{k^{(i)}} m_{2j}^{(i)} + \beta \sum_{i=1}^4 \sum_{j=0}^{k^{(i)}} 2j m_{2j}^{(i)}$$

Now by taking the derivative of $F\left(\left\{m^{(i)}\right\}_{i=1}^4, \left\{\alpha_i\right\}_{i=1}^4, \beta\right)$ with respect to $m^{(i)}$, we have

$$\frac{\partial F}{\partial m_{2j}^{(i)}} = -\ln m_{2j}^{(i)} - 1 + \ln \xi_{2j}^{(i)} + \alpha_i + 2j\beta.$$

According to K.K.T condition, let this derivative be zero, and we have

$$m_{2j}^{(i)} = e^{\alpha_i - 1 + 2j\beta} \xi_{2j}^{(i)}$$

Since

$$\sum_{j=0}^{k^{(i)}} m_{2j}^{(i)} = t_{b\Delta}^{(i)}$$

it follows that

$$e^{\alpha_i - 1} \sum_{j=0}^{k^{(i)}} \xi_{2j}^{(i)} \left(e^\beta\right)^{2j} = t_{b\Delta}^{(i)}$$

For convenience, define

$$g^{(i)}(\tau) \triangleq \sum_{j=0}^{k^{(i)}} \xi_{2j}^{(i)} \tau^{2j}$$

Then

$$e^{\alpha_i - 1} = \frac{t_{b\Delta}^{(i)}}{g^{(i)}(e^\beta)}$$

which implies

$$m_{2j}^{(i)} = \frac{t_{b\Delta}^{(i)}}{g^{(i)}(e^\beta)} e^{2j\beta} \xi_{2j}^{(i)} \tag{A.23}$$

Now by taking into account the condition

$$\sum_{i=1}^4 \sum_{j=0}^{k^{(i)}} 2j m_{2j}^{(i)} = \bar{l}^\varkappa n$$

we have

$$\sum_{i=1}^4 \frac{t_{b\Delta}^{(i)}}{g^{(i)}(e^\beta)} \sum_{j=0}^{k^{(i)}} 2j e^{2j\beta} \xi_{2j}^{(i)} = \bar{l}^\varkappa n$$

It is easy to see that

$$\sum_{j=0}^{k^{(i)}} 2j \tau^{2j} \xi_{2j}^{(i)} = \tau g'^{(i)}(\tau) \tag{A.24}$$

where

$$g'^{(i)}(\tau) = \frac{dg^{(i)}(\tau)}{d\tau}$$

Therefore $e^\beta$ is the solution to

$$\sum_{i=1}^{4} t_{b\Delta}^{(i)} \frac{e^\beta g'^{(i)}(e^\beta)}{g^{(i)}(e^\beta)} = \bar{l}^{\varkappa} n \tag{A.25}$$

Putting (A.22) to (A.25) together yields

$$
\begin{aligned}
\max G\left(\left\{m^{(i)}\right\}_{i=1}^{4}\right) &= \sum_{i=1}^{4}\left[t_{b\Delta}^{(i)} \ln t_{b\Delta}^{(i)} - \sum_{j=0}^{k^{(i)}} \frac{t_{b\Delta}^{(i)}}{g^{(i)}(e^\beta)} e^{2j\beta} \xi_{2j}^{(i)} \ln \frac{t_{b\Delta}^{(i)}}{g^{(i)}(e^\beta)} e^{2j\beta}\right] \\
&= \sum_{i=1}^{4}\left[t_{b\Delta}^{(i)} \ln g^{(i)}(e^\beta) - \beta t_{b\Delta}^{(i)} \frac{e^\beta g'^{(i)}(e^\beta)}{g^{(i)}(e^\beta)}\right] \\
&= \sum_{i=1}^{4} t_{b\Delta}^{(i)} \ln g^{(i)}(e^\beta) - \bar{l}^{\varkappa} n \beta
\end{aligned}
$$

Substituting $e^\beta$ by $\tau$, we have

$$\max G\left(\left\{m^{(i)}\right\}_{i=1}^{4}\right) = \sum_{i=1}^{4} t_{b\Delta}^{(i)} \ln g^{(i)}(\tau) - \bar{l}^{\varkappa} n \ln \tau \tag{A.26}$$

where $\tau$ is the solution to

$$\sum_{i=1}^{4} t_{b\Delta}^{(i)} \frac{\tau g'^{(i)}(\tau)}{g^{(i)}(\tau)} = \bar{l}^{\varkappa} n \tag{A.27}$$

Notice that

$$
\begin{aligned}
(1+\tau)^{c_{b\Delta} r_1} &= \left((1+\tau)^{r_1}\right)^{c_{b\Delta}} \\
&= \prod_{u=1}^{c_{b\Delta}}\left(\sum_{r_u^{\varkappa}=0}^{r_1}\binom{r_1}{r_u^{\varkappa}} \tau^{r_u^{\varkappa}}\right) \\
&= \sum_{j=0}^{c_{b\Delta} r_1} \xi_j^{(1)} \tau^j
\end{aligned}
$$

Meanwhile

$$(1-\tau)^{c_{b\Delta} r_1} = \sum_{j=0}^{c_{b\Delta} r_1} \xi_j^{(1)} (-1)^j \tau^j$$

Therefore

$$
\begin{aligned}
g^{(1)}(\tau) &= \sum_{j=0}^{k^{(1)}} \xi_{2j}^{(1)} (\tau)^{2j} \\
&= \frac{(1+\tau)^{c_{b\Delta} r_1} + (1-\tau)^{c_{b\Delta} r_1}}{2} \\
&= \frac{g(\tau, c_{b\Delta} r_1)}{2}
\end{aligned}
$$

where $g(\tau, k)$ is defined in the lemma. Similarly, we can show that

$$
\begin{aligned}
g^{(2)}(\tau) &= \frac{g(\tau, 2c_{b\Delta}r_1)}{2} \\
g^{(3)}(\tau) &= \frac{g(\tau, c_{b\Delta}r_2)}{2} \\
g^{(4)}(\tau) &= \frac{g(\tau, 2c_{b\Delta}r_2)}{2}
\end{aligned}
$$

It is not hard to verify that

$$
t_{b\Delta}^{(1)} \frac{\tau g'(\tau, c_{b\Delta}r_1)}{g(\tau, c_{b\Delta}r_1)} + t_{b\Delta}^{(2)} \frac{\tau g'(\tau, 2c_{b\Delta}r_1)}{g(\tau, 2c_{b\Delta}r_1)} + t_{b\Delta}^{(3)} \frac{\tau g'(\tau, c_{b\Delta}r_2)}{g(\tau, c_{b\Delta}r_2)} + t_{b\Delta}^{(4)} \frac{\tau g'(\tau, 2c_{b\Delta}r_2)}{g(\tau, 2c_{b\Delta}r_2)}
$$

$$
= n\bar{l} - t_{b\Delta}^{(1)}c_{b\Delta}r_1 \frac{g(\tau, c_{b\Delta}r_1 - 1)}{g(\tau, c_{b\Delta}r_1)} - 2t_{b\Delta}^{(2)}c_{b\Delta}r_1 \frac{g(\tau, 2c_{b\Delta}r_1 - 1)}{g(\tau, 2c_{b\Delta}r_1)}
$$

$$
- t_{b\Delta}^{(3)}c_{b\Delta}r_2 \frac{g(\tau, c_{b\Delta}r_2 - 1)}{g(\tau, c_{b\Delta}r_2)} - 2t_{b\Delta}^{(4)}c_{b\Delta}r_2 \frac{g(\tau, 2c_{b\Delta}r_2 - 1)}{g(\tau, 2c_{b\Delta}r_2)}
$$

which, together with (A.26) and (A.27), implies

$$
\begin{aligned}
\max G\left(\left\{m^{(i)}\right\}_{i=1}^4\right) =\ & -n\bar{l}^\varkappa \ln \tau \\
& + t_{b\Delta}^{(1)} \ln \frac{g(\tau, r_1 c_{b\Delta})}{2} \\
& + t_{b\Delta}^{(2)} \ln \frac{g(\tau, 2r_1 c_{b\Delta})}{2} \\
& + t_{b\Delta}^{(3)} \ln \frac{g(\tau, r_2 c_{b\Delta})}{2} \\
& + t_{b\Delta}^{(4)} \ln \frac{g(\tau, 2r_2 c_{b\Delta})}{2}
\end{aligned} \tag{A.28}
$$

where $\tau$ is the solution to

$$
\begin{aligned}
& r_1 c_{b\Delta} \frac{t_{b\Delta}^{(1)}}{n} \frac{g(\tau, r_1 c_{b\Delta} - 1)}{g(\tau, r_1 c_{b\Delta})} \\
& + 2r_1 c_{b\Delta} \frac{t_{b\Delta}^{(2)}}{n} \frac{g(\tau, 2r_1 c_{b\Delta} - 1)}{g(\tau, 2r_1 c_{b\Delta})} \\
& + r_2 c_{b\Delta} \frac{t_{b\Delta}^{(3)}}{n} \frac{g(\tau, r_2 c_{b\Delta} - 1)}{g(\tau, r_2 c_{b\Delta})} \\
& + 2r_2 c_{b\Delta} \frac{t_{b\Delta}^{(4)}}{n} \frac{g(\tau, 2r_2 c_{b\Delta} - 1)}{g(\tau, 2r_1 c_{b\Delta})} \\
& = \bar{l} - \bar{l}^\varkappa \ .
\end{aligned} \tag{A.29}
$$

Putting (A.20), (A.21), (A.28), and (A.29) together, we then have

$$\Pr\left\{\mathbf{H}_{b\Delta\times n}^{(b\Delta)}x^n = 0^{b\Delta}\right\}$$

$$\leq \exp\left\{\frac{3n\lceil\bar{l}\rceil}{b\Delta}\ln(n\hat{l}^\varkappa) + O(1)\right\}\left(\begin{array}{c}n\bar{l}\\n\bar{l}^\varkappa\end{array}\right)^{-1}\max_{\{m^{(i)}\}_{i=1}^4\in\mathcal{M}_{b\Delta,\varkappa}} M_{\{m^{(i)}\}_{i=1}^4}$$

$$\leq \exp\left\{\max G\left(\left\{m^{(i)}\right\}_{i=1}^4\right) + \frac{3n\lceil\bar{l}\rceil}{b\Delta}\ln(n\hat{l}^\varkappa) + O(1)\right\}\left(\begin{array}{c}n\bar{l}\\n\bar{l}^\varkappa\end{array}\right)^{-1}$$

$$\leq \exp\left\{-n\bar{l}\mathrm{H}_\mathrm{e}\left(\frac{\bar{l}^\varkappa}{\bar{l}}\right) + \frac{1}{2}\ln n\bar{l}^\varkappa\left(1-\frac{\bar{l}^\varkappa}{\bar{l}}\right) + \max G\left(\left\{m^{(i)}\right\}_{i=1}^4\right) + \frac{3n\lceil\bar{l}\rceil}{b\Delta}\ln(n\hat{l}^\varkappa) + O(1)\right\}$$

$$= \exp\left\{nP\left(\frac{b\Delta}{n},\bar{l},\bar{l}^\varkappa\right) + \frac{3n\lceil\bar{l}\rceil}{b\Delta}\ln(n\hat{l}^\varkappa) + \frac{1}{2}\ln n\bar{l}^\varkappa\left(1-\frac{\bar{l}^\varkappa}{\bar{l}}\right) + O(1)\right\}$$

where the last inequality above is due to the fact that

$$\ln\left(\begin{array}{c}n\bar{l}\\n\bar{l}^\varkappa\end{array}\right)^{-1} \leq -n\bar{l}\mathrm{H}_\mathrm{e}(\bar{l}^\varkappa/\bar{l}) + \frac{1}{2}\ln n\bar{l}^\varkappa\left(1-\frac{\bar{l}^\varkappa}{\bar{l}}\right) + O(1)$$

which can be derived from Sterling formula. This competes the proof of Lemma 1 when $\bar{l}^\varkappa < \bar{l} - \frac{t_{b\Delta}^{(1)}}{n}\pi(c_{b\Delta}r_1) - \frac{t_{b\Delta}^{(3)}}{n}\pi(c_{b\Delta}r_2)$.

Finally, let us look at the case when $\bar{l}^\varkappa = \bar{l} - \frac{t_{b\Delta}^{(1)}}{n}\pi(c_{b\Delta}r_1) - \frac{t_{b\Delta}^{(3)}}{n}\pi(c_{b\Delta}r_2)$. In this case, it follows from (A.19) that $\mathcal{M}_{t,\theta}$ contains only one type, i.e., the type given by

$$m_j^{(i)} = \left\{\begin{array}{ll}t_{b\Delta}^{(i)} & \text{if } j = 2k^{(i)}\\0 & \text{otherwise}\end{array}\right. \tag{A.30}$$

for $i = 1, 2, 3$, and $4$. Combining this with (A.21), one can verify that in this case

$$\max\ln M_{\{m^{(i)}\}_{i=1}^4} = t_{b\Delta}^{(1)}\pi(c_{b\Delta}r_1)\ln[c_{b\Delta}r_1] + t_{b\Delta}^{(3)}\pi(c_{b\Delta}r_3)\ln[c_{b\Delta}r_3]\ . \tag{A.31}$$

Plugging (A.31) into (A.20) then leads to the desired result. This competes the proof of Lemma 1.

# Appendix B

## Properties of $P\left(R, \bar{l}, \xi\right)$

This Appendix is devoted to several lemmas related to the function $P\left(R, \bar{l}, \xi\right)$, which are needed in our performance analysis. To keep our notation consistent as in Lemma 1, only $R = \frac{b\Delta}{n}$ appears explicitly in the statements of these lemmas. However, in view of Remark 2, (4.5), and (4.6), by replacing $\frac{b\Delta}{n}$ by any real number $R \in (0, 1]$, all lemmas in this appendix (Lemmas 2 to 6) remain valid. Their respective proofs are the same whether or not $R \in (0, 1]$ is in the form of $R = \frac{b\Delta}{n}$.

In view of (4.2), we define

$$\tilde{l}\left(\frac{b\Delta}{n},\bar{l},\tau\right) \triangleq \bar{l} - \frac{t_{b\Delta}^{(1)}c_{b\Delta}r_1}{n}\frac{g(\tau,c_{b\Delta}r_1-1)}{g(\tau,c_{b\Delta}r_1)} - \frac{2t_{b\Delta}^{(2)}c_{b\Delta}r_1}{n}\frac{g(\tau,2c_{b\Delta}r_1-1)}{g(\tau,2c_{b\Delta}r_1)}$$
$$- \frac{t_{b\Delta}^{(3)}c_{b\Delta}r_2}{n}\frac{g(\tau,c_{b\Delta}r_2-1)}{g(\tau,c_{b\Delta}r_2)} - \frac{2t_{b\Delta}^{(4)}c_{b\Delta}r_2}{n}\frac{g(\tau,2c_{b\Delta}r_2-1)}{g(\tau,2c_{b\Delta}r_2)}$$

**Lemma 2.** *Given $\frac{b\Delta}{n}$ and $\bar{l}$, the following properties hold:*

P1    *As a function of $\tau$, $\tilde{l}\left(\frac{b\Delta}{n},\bar{l},\tau\right)$ is strictly increasing over the interval $[0,+\infty)$.*

P2    *For any $\bar{l}^{\star} \in [0,\bar{l}-\frac{t_{b\Delta}^{(1)}}{n}\pi\left(c_{b\Delta}r_1\right)-\frac{t_{b\Delta}^{(3)}}{n}\pi\left(c_{b\Delta}r_2\right))$, there is a unique solution of $\tau$ to $\tilde{l}\left(\frac{b\Delta}{n},\bar{l},\tau\right) = \bar{l}^{\star}$.*

*Proof of Lemma 2*: In view of the definition of $\tilde{l}\left(\frac{b\Delta}{n},\bar{l},\tau\right)$, for Property P1, it is sufficient to prove that $\frac{g(\tau,k-1)}{g(\tau,k)}$ as function of $\tau$ is strictly decreasing over $\tau \in [0,\infty)$ for any positive value $k > 1$. To this end, take the first derivative of $\frac{g(\tau,k-1)}{g(\tau,k)}$ with respect to $\tau$, yielding

$$\frac{-(1+\tau)^{2k-2}+(k-1)(1+\tau)^{k-2}(1-\tau)^k - (k-1)(1-\tau)^{k-2}(1+\tau)^k + (1-\tau)^{2k-2}}{g^2(\tau,k)} \quad (B.1)$$

Denote the enumerator of (B.1) by $f(\tau)$. It is easy to see that $f(0) = 0$. Since the denominator of (B.1) is always positive, it suffices to show that $f(\tau) < 0$ for any $\tau > 0$.

To continue, one can verify that

$$
\begin{aligned}
f(\tau) &= -(1+\tau)^{2k-2}+(1-\tau)^{2k-2}+(k-1)(1-\tau^2)^{k-2}[(1-\tau)^2-(1+\tau)^2] \\
&= -(1+\tau)^{2k-2}+(1-\tau)^{2k-2}-4\tau(k-1)(1-\tau^2)^{k-2} \\
&= -2\sum_{i=0}^{k-2}\binom{2k-2}{2i+1}\tau^{2i+1}-4\tau(k-1)(1-\tau^2)^{k-2} \\
&= -2\tau\left[\sum_{i=0}^{k-2}\binom{2k-2}{2i+1}\tau^{2i}+2(k-1)\sum_{i=0}^{k-2}\binom{k-2}{i}(-1)^i\tau^{2i}\right] \\
&= -2\tau\left[\sum_{0\le i\le k-2:\text{ even}}\left(\binom{2k-2}{2i+1}+2(k-1)\binom{k-2}{i}\right)\tau^{2i}\right.\\
&\qquad\qquad \left.+\sum_{0\le i\le k-2:\text{ odd}}\left(\binom{2k-2}{2i+1}-2(k-1)\binom{k-2}{i}\right)\tau^{2i}\right] \\
&\le -2\tau\sum_{0\le i\le k-2:\text{ even}}\left(\binom{2k-2}{2i+1}+2(k-1)\binom{k-2}{i}\right)\tau^{2i} \\
&< 0 \qquad\qquad\qquad\qquad\qquad\qquad\qquad\qquad\qquad\qquad\qquad\qquad (B.2)
\end{aligned}
$$

for any $\tau > 0$. In (B.2), the first inequality is due to the fact that for any odd $i < k - 2$

$$
\begin{aligned}
\binom{2k-2}{2i+1} &= \binom{2k-3}{2i+1} + \binom{2k-3}{2i} \\
&\geq \binom{k-2}{i}\binom{k-1}{i+1} + \binom{k-2}{i}\binom{k-1}{i} \\
&\geq 2(k-1)\binom{k-2}{i}
\end{aligned}
$$

and for $i = k - 2$ when $k$ is odd,

$$
\binom{2k-2}{2i+1} - 2(k-1)\binom{k-2}{i} = 0.
$$

From (B.2), Property P1 follows.

Since $c_{b\Delta}r_2 \geq c_{b\Delta}r_1 > 1$, it is easy to see that

$$
\begin{aligned}
\tilde{l}\left(\frac{b\Delta}{n}, \bar{l}, 0\right) &= \bar{l} - \frac{t_{b\Delta}^{(1)}c_{b\Delta}r_1}{n} - \frac{2t_{b\Delta}^{(2)}c_{b\Delta}r_1}{n} - \frac{t_{b\Delta}^{(3)}c_{b\Delta}r_2}{n} - \frac{2t_{b\Delta}^{(4)}c_{b\Delta}r_2}{n} \\
&= \bar{l} - R_1 r_1 - r_2 R_2 \\
&= 0.
\end{aligned}
\tag{B.3}
$$

On the other hand, one can verify that for any $k \geq 1$,

$$
\lim_{\tau \to +\infty} \frac{g(\tau, k-1)}{g(\tau, k)} = \frac{\pi(k)}{k}
$$

which implies that

$$
\lim_{\tau \to +\infty} \tilde{l}\left(\frac{b\Delta}{n}, \bar{l}, \tau\right) = \bar{l} - \frac{t_{b\Delta}^{(1)}}{n}\pi\left(c_{b\Delta}r_1\right) - \frac{t_{b\Delta}^{(3)}}{n}\pi\left(c_{b\Delta}r_2\right).
\tag{B.4}
$$

Property P2 now follows from (B.3), (B.4), and Property P1. This completes the proof of Lemma 2.

**Lemma 3.** *For fixed $\frac{b\Delta}{n}$ and $\bar{l}$, $P\left(\frac{b\Delta}{n}, \bar{l}, \xi\right)$ as a function of $\xi$ is strictly decreasing over $\xi \in (0, \bar{l}/2)$.*

*Proof of **Lemma 3***: To show that $P\left(\frac{b\Delta}{n}, \bar{l}, \xi\right)$ is strictly decreasing over $\xi \in (0, \bar{l}/2)$, take its first

derivative, yielding

$$
\begin{aligned}
\frac{\partial P}{\partial \xi} \;=\;& -\ln \frac{1 - \xi/\bar{l}}{\xi/\bar{l}} - \ln \tau - \frac{\xi}{\tau}\frac{\partial \tau}{\partial \xi} \\[6pt]
& + r_1 c_{b\Delta} \frac{t_{b\Delta}^{(1)}}{n}\frac{(1+\tau)^{r_1 c_{b\Delta}-1} - (1-\tau)^{r_1 c_{b\Delta}-1}}{g(\tau, r_1 c_{b\Delta})}\frac{\partial \tau}{\partial \xi} \\[6pt]
& + 2 r_1 c_{b\Delta}\frac{t_{b\Delta}^{(2)}}{n}\frac{(1+\tau)^{2 r_1 c_{b\Delta}-1} - (1-\tau)^{2 r_1 c_{b\Delta}-1}}{g(\tau, 2 r_1 c_{b\Delta})}\frac{\partial \tau}{\partial \xi} \\[6pt]
& + r_2 c_{b\Delta}\frac{t_{b\Delta}^{(3)}}{n}\frac{(1+\tau)^{r_2 c_{b\Delta}-1} - (1-\tau)^{r_2 c_{b\Delta}-1}}{g(\tau, r_2 c_{b\Delta})}\frac{\partial \tau}{\partial \xi} \\[6pt]
& + 2 r_2 c_{b\Delta}\frac{t_{b\Delta}^{(4)}}{n}\frac{(1+\tau)^{2 r_2 c_{b\Delta}-1} - (1-\tau)^{2 r_2 c_{b\Delta}-1}}{g(\tau, 2 r_2 c_{b\Delta})}\frac{\partial \tau}{\partial \xi}.
\end{aligned}
\tag{B.5}
$$

Note that

$$
\begin{aligned}
g(\tau, k) \;=\;& (1+\tau)^k + (1-\tau)^k \\[4pt]
=\;& (1+\tau)^{k-1}(1+\tau) + (1-\tau)^{k-1}(1-\tau) \\[4pt]
=\;& \tau\left[(1+\tau)^{k-1} - (1-\tau)^{k-1}\right] + g(\tau, k-1)
\end{aligned}
$$

and hence

$$
(1+\tau)^{k-1} - (1-\tau)^{k-1} = \frac{g(\tau, k) - g(\tau, k-1)}{\tau}
$$

Plugging the above equality into (B.5) yields

$$
\begin{aligned}
\frac{\partial P}{\partial \xi} \;=\;& -\ln \frac{1 - \xi/\bar{l}}{\xi/\bar{l}} - \ln \tau - \frac{\xi}{\tau}\frac{\partial \tau}{\partial \xi} \\[6pt]
& + \frac{r_1 c_{b\Delta}}{\tau}\frac{t_{b\Delta}^{(1)}}{n}\left[1 - \frac{g(\tau, r_1 c_{b\Delta}-1)}{g(\tau, r_1 c_{b\Delta})}\right]\frac{\partial \tau}{\partial \xi} \\[6pt]
& + \frac{2 r_1 c_{b\Delta}}{\tau}\frac{t_{b\Delta}^{(2)}}{n}\left[1 - \frac{g(\tau, 2 r_1 c_{b\Delta}-1)}{g(\tau, 2 r_1 c_{b\Delta})}\right]\frac{\partial \tau}{\partial \xi} \\[6pt]
& + \frac{r_2 c_{b\Delta}}{\tau}\frac{t_{b\Delta}^{(3)}}{n}\left[1 - \frac{g(\tau, r_2 c_{b\Delta}-1)}{g(\tau, r_2 c_{b\Delta})}\right]\frac{\partial \tau}{\partial \xi} \\[6pt]
& + \frac{2 r_2 c_{b\Delta}}{\tau}\frac{t_{b\Delta}^{(4)}}{n}\left[1 - \frac{g(\tau, 2 r_2 c_{b\Delta}-1)}{g(\tau, 2 r_2 c_{b\Delta})}\right]\frac{\partial \tau}{\partial \xi} \\[6pt]
=\;& -\ln \frac{1 - \xi/\bar{l}}{\xi/\bar{l}} - \ln \tau
\end{aligned}
\tag{B.6}
$$

where the second step comes from the fact that $\tau$ is the solution to (4.2) and from the identity (4.6). Note that $\tau = 1$ is the solution to (4.2) when $\xi = \frac{\bar{l}}{2}$, and therefore by Lemma 2, $0 < \tau < 1$ whenever

$\xi \in (0, \bar{l}/2)$. Furthermore, it can be verified that for any $\tau \in (0, 1)$

$$\frac{g(\tau, k-1)}{g(\tau, k)} = \frac{(1+\tau)^{k-1} + (1-\tau)^{k-1}}{(1+\tau)^k + (1-\tau)^k} > \frac{1}{1+\tau}$$

which, coupled with (4.2), implies

$$\frac{\bar{l}}{1+\tau} < \bar{l} - \xi$$

or

$$\tau > \frac{\xi/\bar{l}}{1 - \xi/\bar{l}}$$

for $\xi \in (0, \bar{l}/2)$. Plugging the above inequality into (B.6), we have

$$\frac{\partial P}{\partial \xi} < 0$$

for $\xi \in (0, \bar{l}/2)$. This completes the proof of Lemma 3.

**Lemma 4.** *For fixed $\frac{b\Delta}{n}$ and $\bar{l}$, $P\left(\frac{b\Delta}{n}, \bar{l}, \xi\right) \geq P\left(\frac{b\Delta}{n}, \bar{l}, \bar{l} - \xi\right)$ for $0 < \xi \leq \bar{l}/2$.*

*Proof of **Lemma 4***: First, we consider the case where

$$\frac{t^{(1)}}{n} \pi(c_{b\Delta} r_1) + \frac{t^{(3)}}{n} \pi(c_{b\Delta} r_2) < \xi \leq \frac{\bar{l}}{2}$$

Define

$$
\begin{aligned}
P\left(\frac{b\Delta}{n}, \bar{l}, \xi, \tau\right) &= -\bar{l} \mathrm{H_e}(\xi/\bar{l}) - \xi \ln \tau \\
&\quad + \frac{t_{b\Delta}^{(1)}}{n} \ln \frac{g(\tau, r_1 c_{b\Delta})}{2} + \frac{t_{b\Delta}^{(2)}}{n} \ln \frac{g(\tau, 2r_1 c_{b\Delta})}{2} \\
&\quad + \frac{t_{b\Delta}^{(3)}}{n} \ln \frac{g(\tau, r_2 c_{b\Delta})}{2} + \frac{t_{b\Delta}^{(4)}}{n} \ln \frac{g(\tau, 2r_2 c_{b\Delta})}{2}
\end{aligned}
$$

and $\tau_\xi$ as the solution to

$$\xi = \tilde{l}\left(\frac{b\Delta}{n}, \bar{l}, \tau\right)$$

Then it is easy to observe that

$$P\left(\frac{b\Delta}{n}, \bar{l}, \xi\right) = P\left(\frac{b\Delta}{n}, \bar{l}, \xi, \tau_\xi\right)$$

Note that when $\xi \leq \bar{l}/2$, $\tau_\xi \leq 1$. For $\tau \leq 1$,

$$g(\tau^{-1}, k) = \frac{(1+\tau)^k + (\tau-1)^k}{\tau^k} \leq \frac{g(\tau, k)}{\tau^k}$$

and

$$
\begin{aligned}
P\left(\frac{b\Delta}{n}, \bar{l}, \bar{l}-\xi, \tau^{-1}\right) &= -\bar{l}\mathrm{H_e}((\bar{l}-\xi)/\bar{l}) - (\bar{l}-\xi)\ln\tau^{-1} \\
&\quad + \frac{t_{b\Delta}^{(1)}}{n}\ln\frac{g(\tau^{-1}, r_1 c_{b\Delta})}{2} + \frac{t_{b\Delta}^{(2)}}{n}\ln\frac{g(\tau^{-1}, 2r_1 c_{b\Delta})}{2} \\
&\quad + \frac{t_{b\Delta}^{(3)}}{n}\ln\frac{g(\tau^{-1}, r_2 c_{b\Delta})}{2} + \frac{t_{b\Delta}^{(4)}}{n}\ln\frac{g(\tau^{-1}, 2r_2 c_{b\Delta})}{2} \\
&\leq -\bar{l}\mathrm{H_e}(\xi/\bar{l}) + (\bar{l}-\xi)\ln\tau \\
&\quad + \frac{t_{b\Delta}^{(1)}}{n}\ln\frac{g(\tau, r_1 c_{b\Delta})}{2\tau^{r_1 c_{b\Delta}}} + \frac{t_{b\Delta}^{(2)}}{n}\ln\frac{g(\tau, 2r_1 c_{b\Delta})}{2\tau^{2r_1 c_{b\Delta}}} \\
&\quad + \frac{t_{b\Delta}^{(3)}}{n}\ln\frac{g(\tau, r_2 c_{b\Delta})}{2\tau^{r_2 c_{b\Delta}}} + \frac{t_{b\Delta}^{(4)}}{n}\ln\frac{g(\tau, 2r_2 c_{b\Delta})}{2\tau^{2r_2 c_{b\Delta}}} \\
&= -\bar{l}\mathrm{H_e}(\xi/\bar{l}) - \xi\ln\tau \\
&\quad + \frac{t_{b\Delta}^{(1)}}{n}\ln\frac{g(\tau, r_1 c_{b\Delta})}{2} + \frac{t_{b\Delta}^{(2)}}{n}\ln\frac{g(\tau, 2r_1 c_{b\Delta})}{2} \\
&\quad + \frac{t_{b\Delta}^{(3)}}{n}\ln\frac{g(\tau, r_2 c_{b\Delta})}{2} + \frac{t_{b\Delta}^{(4)}}{n}\ln\frac{g(\tau, 2r_2 c_{b\Delta})}{2} \\
&= P\left(\frac{b\Delta}{n}, \bar{l}, \xi, \tau\right)
\end{aligned}
$$

where the third step is due to (4.6). Therefore,

$$
P\left(\frac{b\Delta}{n}, \bar{l}, \xi, \tau_\xi\right) \geq P\left(\frac{b\Delta}{n}, \bar{l}, \bar{l}-\xi, \tau_\xi^{-1}\right)
$$

Now it can be verified that

$$
\frac{\partial P\left(\frac{b\Delta}{n}, \bar{l}, \xi, \tau\right)}{\partial\tau} = \frac{-\xi + \tilde{l}\left(\frac{b\Delta}{n}, \bar{l}, \tau\right)}{\tau}
$$

and since $\tilde{l}\left(\frac{b\Delta}{n}, \bar{l}, \tau\right)$ is an increasing function of $\tau$, it is easy to see that $\frac{\partial P\left(\frac{b\Delta}{n}, \bar{l}, \xi, \tau\right)}{\partial\tau} < 0$ for $\tau < \tau_\xi$ and $\frac{\partial P\left(\frac{b\Delta}{n}, \bar{l}, \xi, \tau\right)}{\partial\tau} > 0$ for $\tau > \tau_\xi$. Therefore, $\tau_\xi$ is the value that minimizes the function $P\left(\frac{b\Delta}{n}, \bar{l}, \xi, \tau\right)$ given $\xi$. In the other words,

$$
P\left(\frac{b\Delta}{n}, \bar{l}, \xi, \tau_\xi\right) \leq P\left(\frac{b\Delta}{n}, \bar{l}, \xi, \tau\right)
$$

for any $\tau > 0$. In total, we have

$$
\begin{aligned}
P\left(\frac{b\Delta}{n}, \bar{l}, \xi\right) &= P\left(\frac{b\Delta}{n}, \bar{l}, \xi, \tau_\xi\right) \\
&\geq P\left(\frac{b\Delta}{n}, \bar{l}, \bar{l}-\xi, \tau_\xi^{-1}\right) \\
&\geq P\left(\frac{b\Delta}{n}, \bar{l}, \bar{l}-\xi, \tau_{\bar{l}-\xi}\right) \\
&= P\left(\frac{b\Delta}{n}, \bar{l}, \bar{l}-\xi\right)
\end{aligned}
$$

Now if
$$\xi < \frac{t^{(1)}}{n}\pi(c_b\Delta r_1) + \frac{t^{(3)}}{n}\pi(c_b\Delta r_2),$$
then $P\left(\frac{b\Delta}{n}, \bar{l}, \bar{l} - \xi\right) = -\infty$, and $P\left(\frac{b\Delta}{n}, \bar{l}, \xi\right) \geq P\left(\frac{b\Delta}{n}, \bar{l}, \bar{l} - \xi\right)$ is obvious. For
$$\xi = \frac{t^{(1)}}{n}\pi(c_b\Delta r_1) + \frac{t^{(3)}}{n}\pi(c_b\Delta r_2),$$
it can be shown that
$$\frac{\partial P\left(\frac{b\Delta}{n}, \bar{l}, \bar{l} - \xi, \tau\right)}{\partial \tau} < 0$$
for $\tau > 0$. Then

$$
\begin{aligned}
P\left(\frac{b\Delta}{n}, \bar{l}, \xi\right) &= P\left(\frac{b\Delta}{n}, \bar{l}, \xi, \tau_\xi\right) \\
&\geq P\left(\frac{b\Delta}{n}, \bar{l}, \bar{l} - \xi, \tau_\xi^{-1}\right) \\
&\geq \lim_{\tau \to \infty} P\left(\frac{b\Delta}{n}, \bar{l}, \bar{l} - \xi, \tau\right) \\
&= P\left(\frac{b\Delta}{n}, \bar{l}, \bar{l} - \xi\right)
\end{aligned}
$$

where the last equality is due to (4.4). This completes the proof of Lemma 4.

**Lemma 5.** *For* $\frac{\bar{l}}{\lfloor \bar{l} \rfloor} \leq \xi \leq \frac{\bar{l}}{2}$,
$$P\left(\frac{b\Delta}{n}, \bar{l}, \xi\right) \leq -\frac{b\Delta}{n}\ln 2 + 2\xi\exp\left[-\frac{2\xi}{\bar{l}}\left(c_b\Delta r_1 - 1\right)\right] + \frac{b\Delta}{n}\exp\left(-\frac{2\xi}{\bar{l}}r_1c_b\Delta\right)$$

*Proof of Lemma 5*: Let $\tau_\xi$ be the solution to the equation
$$\tilde{l}\left(\frac{b\Delta}{n}, \bar{l}, \tau\right) = \xi.$$
From the proof of Lemma 4, we know that
$$\frac{\xi/\bar{l}}{1 - \xi/\bar{l}} \leq \tau_\xi \leq 1 \text{ or } \frac{\xi}{\bar{l}} \leq \frac{\tau_\xi}{1 + \tau_\xi} \leq \frac{1}{2}$$
whenever $\xi \leq \frac{\bar{l}}{2}$. Furthermore, it can be verified that
$$f(x) = \frac{1 + (1-x)^{k-1}}{1 + (1-x)^k}$$
is strictly decreasing for $\frac{2}{k} \leq x \leq 1$, where $k$ is an integer no less than 2. To see this is the case, we have

$$
\begin{aligned}
f'(x) &= \frac{-\left[1 + (1-x)^k\right](k-1)(1-x)^{k-2} + \left[1 + (1-x)^{k-1}\right]k(1-x)^{k-1}}{\left[1 + (1-x)^k\right]^2} \\
&= \frac{(1-x)^{k-2}}{\left[1 + (1-x)^k\right]^2}\left[1 - kx + (1-x)^k\right] < 0
\end{aligned}
$$

 

for $\frac{2}{k} \leq x \leq 1$. Now assume that

$$\frac{r_1}{\bar{l}}\xi = \frac{\lfloor \bar{l} \rfloor}{\bar{l}}\xi \geq 1.$$

Then

$$\frac{\tau_\xi}{1+\tau_\xi}r_1 c_{b\Delta} \geq \frac{\xi}{\bar{l}}r_1 c_{b\Delta} \geq 1.$$

Therefore,

$$
\begin{aligned}
\frac{g\left(\tau_\xi, c_{b\Delta}r_1-1\right)}{g\left(\tau_\xi, c_{b\Delta}r_1\right)} &= \frac{1}{1+\tau_\xi}\frac{1+\left(1-\frac{2\tau_\xi}{1+\tau_\xi}\right)^{c_{b\Delta}r_1-1}}{1+\left(1-\frac{2\tau_\xi}{1+\tau_\xi}\right)^{c_{b\Delta}r_1}} \\
&\leq \frac{1}{1+\tau_\xi}\frac{1+\left(1-\frac{2\xi}{\bar{l}}\right)^{c_{b\Delta}r_1-1}}{1+\left(1-\frac{2\xi}{\bar{l}}\right)^{c_{b\Delta}r_1}} \\
&= \frac{1}{1+\tau_\xi}\left[1+\frac{\left(1-\frac{2\xi}{\bar{l}}\right)^{c_{b\Delta}r_1-1}-\left(1-\frac{2\xi}{\bar{l}}\right)^{c_{b\Delta}r_1}}{1+\left(1-\frac{2\xi}{\bar{l}}\right)^{c_{b\Delta}r_1}}\right] \\
&= \frac{1}{1+\tau_\xi}\left[1+\frac{\frac{2\xi}{\bar{l}}\left(1-\frac{2\xi}{\bar{l}}\right)^{c_{b\Delta}r_1-1}}{1+\left(1-\frac{2\xi}{\bar{l}}\right)^{c_{b\Delta}r_1}}\right] \\
&\leq \frac{1}{1+\tau_\xi}\left[1+\frac{2\xi}{\bar{l}}\left(1-\frac{2\xi}{\bar{l}}\right)^{c_{b\Delta}r_1-1}\right].
\end{aligned}
$$

Similarly,

$$
\begin{aligned}
\frac{g\left(\tau_\xi, 2c_{b\Delta}r_1-1\right)}{g\left(\tau_\xi, 2c_{b\Delta}r_1\right)} &\leq \frac{1}{1+\tau_\xi}\left[1+\frac{2\xi}{\bar{l}}\left(1-\frac{2\xi}{\bar{l}}\right)^{2c_{b\Delta}r_1-1}\right] \\
&\leq \frac{1}{1+\tau_\xi}\left[1+\frac{2\xi}{\bar{l}}\left(1-\frac{2\xi}{\bar{l}}\right)^{c_{b\Delta}r_1-1}\right] \\
\frac{g\left(\tau_\xi, c_{b\Delta}r_2-1\right)}{g\left(\tau_\xi, c_{b\Delta}r_2\right)} &\leq \frac{1}{1+\tau_\xi}\left[1+\frac{2\xi}{\bar{l}}\left(1-\frac{2\xi}{\bar{l}}\right)^{c_{b\Delta}r_2-1}\right] \\
&\leq \frac{1}{1+\tau_\xi}\left[1+\frac{2\xi}{\bar{l}}\left(1-\frac{2\xi}{\bar{l}}\right)^{c_{b\Delta}r_1-1}\right] \\
\frac{g\left(\tau_\xi, 2c_{b\Delta}r_2-1\right)}{g\left(\tau_\xi, 2c_{b\Delta}r_2\right)} &\leq \frac{1}{1+\tau_\xi}\left[1+\frac{2\xi}{\bar{l}}\left(1-\frac{2\xi}{\bar{l}}\right)^{2c_{b\Delta}r_2-1}\right] \\
&\leq \frac{1}{1+\tau_\xi}\left[1+\frac{2\xi}{\bar{l}}\left(1-\frac{2\xi}{\bar{l}}\right)^{c_{b\Delta}r_1-1}\right].
\end{aligned}
$$

Thus,

$$
\begin{aligned}
\xi &= \tilde{l}\left(\frac{b\Delta}{n}, \bar{l}, \tau_\xi\right) \\
&= \bar{l} - \frac{t_{b\Delta}^{(1)} c_{b\Delta} r_1}{n} \frac{g(\tau, c_{b\Delta} r_1 - 1)}{g(\tau, c_{b\Delta} r_1)} - \frac{2t_{b\Delta}^{(2)} c_{b\Delta} r_1}{n} \frac{g(\tau, 2c_{b\Delta} r_1 - 1)}{g(\tau, 2c_{b\Delta} r_1)} \\
&\quad - \frac{t_{b\Delta}^{(3)} c_{b\Delta} r_2}{n} \frac{g(\tau, c_{b\Delta} r_2 - 1)}{g(\tau, c_{b\Delta} r_2)} - \frac{2t_{b\Delta}^{(4)} c_{b\Delta} r_2}{n} \frac{g(\tau, 2c_{b\Delta} r_2 - 1)}{g(\tau, 2c_{b\Delta} r_2)} \\
&\geq \bar{l} - \frac{\bar{l}}{1 + \tau_\xi}\left[1 + \frac{2\xi}{\bar{l}}\left(1 - \frac{2\xi}{\bar{l}}\right)^{c_{b\Delta} r_1 - 1}\right]
\end{aligned}
$$

where in the last step, the identity (4.6) was applied. This implies that

$$
1 + \tau_\xi \leq \frac{\bar{l}}{\bar{l} - \xi}\left[1 + \frac{2\xi}{\bar{l}}\left(1 - \frac{2\xi}{\bar{l}}\right)^{c_{b\Delta} r_1 - 1}\right].
$$

On the other hand,

$$
\begin{aligned}
g\left(\tau_\xi, r_1 c_{b\Delta}\right) &= (1 + \tau_\xi)^{r_1 c_{b\Delta}} + (1 - \tau_\xi)^{r_1 c_{b\Delta}} \\
&= (1 + \tau_\xi)^{r_1 c_{b\Delta}}\left[1 + \left(1 - \frac{2\tau_\xi}{1 + \tau_\xi}\right)^{r_1 c_{b\Delta}}\right] \\
&\leq (1 + \tau_\xi)^{r_1 c_{b\Delta}}\left[1 + \left(1 - \frac{2\xi}{\bar{l}}\right)^{r_1 c_{b\Delta}}\right].
\end{aligned}
$$

Again,

$$
\begin{aligned}
g\left(\tau_\xi, 2r_1 c_{b\Delta}\right) &\leq (1 + \tau_\xi)^{2r_1 c_{b\Delta}}\left[1 + \left(1 - \frac{2\xi}{\bar{l}}\right)^{2r_1 c_{b\Delta}}\right] \\
&\leq (1 + \tau_\xi)^{2r_1 c_{b\Delta}}\left[1 + \left(1 - \frac{2\xi}{\bar{l}}\right)^{r_1 c_{b\Delta}}\right] \\
g\left(\tau_\xi, r_2 c_{b\Delta}\right) &\leq (1 + \tau_\xi)^{r_2 c_{b\Delta}}\left[1 + \left(1 - \frac{2\xi}{\bar{l}}\right)^{r_2 c_{b\Delta}}\right] \\
&\leq (1 + \tau_\xi)^{r_2 c_{b\Delta}}\left[1 + \left(1 - \frac{2\xi}{\bar{l}}\right)^{r_1 c_{b\Delta}}\right] \\
g\left(\tau_\xi, 2r_2 c_{b\Delta}\right) &\leq (1 + \tau_\xi)^{2r_2 c_{b\Delta}}\left[1 + \left(1 - \frac{2\xi}{\bar{l}}\right)^{2r_2 c_{b\Delta}}\right] \\
&\leq (1 + \tau_\xi)^{2r_2 c_{b\Delta}}\left[1 + \left(1 - \frac{2\xi}{\bar{l}}\right)^{r_1 c_{b\Delta}}\right]
\end{aligned}
$$

By combining the above inequalities with the identities (4.5) and (4.6), we have

$$
\begin{aligned}
P\left(\frac{b\Delta}{n}, \bar{l}, \xi\right) &= -\bar{l}\mathrm{H}_e\left(\xi/\bar{l}\right) - \xi\ln\tau_\xi \\
&\quad + \frac{t_{b\Delta}^{(1)}}{n}\ln\frac{g\left(\tau_\xi, r_1 c_{b\Delta}\right)}{2} + \frac{t_{b\Delta}^{(2)}}{n}\ln\frac{g\left(\tau_\xi, 2r_1 c_{b\Delta}\right)}{2} \\
&\quad + \frac{t_{b\Delta}^{(3)}}{n}\ln\frac{g\left(\tau_\xi, r_2 c_{b\Delta}\right)}{2} + \frac{t_{b\Delta}^{(4)}}{n}\ln\frac{g\left(\tau_\xi, 2r_2 c_{b\Delta}\right)}{2} \\
&\leq -\bar{l}\mathrm{H}_e\left(\xi/\bar{l}\right) - \xi\ln\tau_\xi + \bar{l}\ln\left(1 + \tau_\xi\right) - \frac{b\Delta}{n}\ln 2 + \frac{b\Delta}{n}\ln\left[1 + \left(1 - \frac{2\xi}{\bar{l}}\right)^{r_1 c_{b\Delta}}\right] \\
&\leq -\bar{l}\mathrm{H}_e\left(\xi/\bar{l}\right) - \xi\ln\frac{\xi}{\bar{l}-\xi} + \bar{l}\ln\left(1 + \tau_\xi\right) - \frac{b\Delta}{n}\ln 2 + \frac{b\Delta}{n}\ln\left[1 + \left(1 - \frac{2\xi}{\bar{l}}\right)^{r_1 c_{b\Delta}}\right] \\
&\leq -\bar{l}\mathrm{H}_e\left(\xi/\bar{l}\right) - \xi\ln\frac{\xi}{\bar{l}-\xi} + \bar{l}\ln\frac{\bar{l}}{\bar{l}-\xi} + \bar{l}\ln\left[1 + \frac{2\xi}{\bar{l}}\left(1 - \frac{2\xi}{\bar{l}}\right)^{c_{b\Delta} r_1 - 1}\right] \\
&\quad - \frac{b\Delta}{n}\ln 2 + \frac{b\Delta}{n}\ln\left[1 + \left(1 - \frac{2\xi}{\bar{l}}\right)^{r_1 c_{b\Delta}}\right] \\
&\leq -\frac{b\Delta}{n}\ln 2 + 2\xi\left(1 - \frac{2\xi}{\bar{l}}\right)^{c_{b\Delta} r_1 - 1} + \frac{b\Delta}{n}\left(1 - \frac{2\xi}{\bar{l}}\right)^{r_1 c_{b\Delta}} \\
&\leq -\frac{b\Delta}{n}\ln 2 + 2\xi\exp\left[-\frac{2\xi}{\bar{l}}\left(c_{b\Delta} r_1 - 1\right)\right] + \frac{b\Delta}{n}\exp\left(-\frac{2\xi}{\bar{l}}r_1 c_{b\Delta}\right)
\end{aligned}
$$

This completes the proof of Lemma 5.

**Lemma 6.** *Given $0 < \xi \leq \frac{\bar{l}}{2}$, the following properties hold:*

**S1** *for $1 \leq b \leq \frac{n}{\Delta} - 1$,*

$$
-\frac{\Delta}{n}\ln 2 \leq P\left(\frac{(b+1)\Delta}{n}, \bar{l}, \xi\right) - P\left(\frac{b\Delta}{n}, \bar{l}, \xi\right) \leq -\frac{\Delta}{n}Q_{\xi, \frac{(b+1)\Delta}{n}}
$$

*where*

$$
Q_{\xi, \frac{(b+1)\Delta}{n}} = \ln 2 - \ln\left(1 + \frac{2\left(1 - \frac{2\xi}{\bar{l}}\right)^{r_1 c_{(b+1)\Delta}}}{1 + \left(1 - \frac{2\xi}{\bar{l}}\right)^{2r_1 c_{(b+1)\Delta}}}\right)
$$

**S2** *for $0 < R_1 \leq R_2 \leq 1$,*

$$
-(R_2 - R_1)\ln 2 \leq P\left(R_2, \bar{l}, \xi\right) - P\left(R_1, \bar{l}, \xi\right) \leq -(R_2 - R_1)Q_{\xi, R_2}
$$

*where*

$$
Q_{\xi, R_2} = \ln 2 - \ln\left(1 + \frac{2\left(1 - \frac{2\xi}{\bar{l}}\right)^{r_1 c_{R_2}}}{1 + \left(1 - \frac{2\xi}{\bar{l}}\right)^{2r_1 c_{R_2}}}\right)
$$

*with $c_{R_2} = 2^{-\lceil \log_2 R_2 \rceil}$, which shows that $P(R, \bar{l}, \xi)$ strictly decreasing with respect to $R \in (0, 1]$, and Lipschitz-Continuous with constant $\ln 2$; and*

S3 $\lim_{R \to 0} P(R, \bar{l}, \xi) = 0$.

*Proof of **Lemma 6***: Let $P\left(\frac{b\Delta}{n}, \bar{l}, \xi, \tau\right)$ be the function defined in the proof of Lemma 4. Furthermore, let $\tau_b$ and $\tau_{b+1}$ be the solution to (4.2) for $b$ and $b+1$, respectively. From the proof of Lemma 4, it follows that given $\left(\frac{b\Delta}{n}, \bar{l}, \xi\right)$, the function $P\left(\frac{b\Delta}{n}, \bar{l}, \xi, \tau\right)$ achieves its minimum at $\tau = \tau_b$, and hence

$$
\begin{aligned}
P\left(\frac{(b+1)\Delta}{n}, \bar{l}, \xi\right) &= P\left(\frac{(b+1)\Delta}{n}, \bar{l}, \xi, \tau_{b+1}\right) \\
&\leq P\left(\frac{(b+1)\Delta}{n}, \bar{l}, \xi, \tau_b\right)
\end{aligned}
$$

Therefore,

$$
P\left(\frac{(b+1)\Delta}{n}, \bar{l}, \xi\right) - P\left(\frac{b\Delta}{n}, \bar{l}, \xi\right) \leq P\left(\frac{(b+1)\Delta}{n}, \bar{l}, \xi, \tau_b\right) - P\left(\frac{b\Delta}{n}, \bar{l}, \xi, \tau_b\right).
$$

Now we have

$$
c_{b\Delta} = 2^{-\lceil \log_2 \frac{b\Delta}{n} \rceil} \geq 2^{-\lceil \log_2 \frac{(b+1)\Delta}{n} \rceil} = c_{(b+1)\Delta} = 2^{-\lceil \log_2 \frac{b\Delta}{n} + \log_2 \frac{b+1}{b} \rceil} \geq 2^{-\lceil \log_2 \frac{b\Delta}{n} + 1 \rceil} = \frac{c_{b\Delta}}{2}.
$$

To continue, we distinguish between two cases: (1) $c_{b\Delta} = c_{(b+1)\Delta}$, and (2) $c_{b\Delta} = 2c_{(b+1)\Delta}$. In case (1), i.e., when $\lceil \log_2 \frac{b\Delta}{n} \rceil = \lceil \log_2 \frac{(b+1)\Delta}{n} \rceil$, we have

$$
\begin{aligned}
&P\left(\frac{(b+1)\Delta}{n}, \bar{l}, \xi, \tau_b\right) - P\left(\frac{b\Delta}{n}, \bar{l}, \xi, \tau_b\right) \\
=\ & \frac{t^{(1)}_{(b+1)\Delta} - t^{(1)}_{(b+1)\Delta}}{n} \ln \frac{g\left(\tau_b, r_1 c_{(b+1)\Delta}\right)}{2} + \frac{t^{(2)}_{(b+1)\Delta} - t^{(2)}_{b\Delta}}{n} \ln \frac{g\left(\tau_b, 2r_1 c_{(b+1)\Delta}\right)}{2} \\
&+ \frac{t^{(3)}_{(b+1)\Delta} - t^{(3)}_{b\Delta}}{n} \ln \frac{g\left(\tau_b, r_2 c_{(b+1)\Delta}\right)}{2} + \frac{t^{(4)}_{(b+1)\Delta} - t^{(4)}_{b\Delta}}{n} \ln \frac{g\left(\tau_b, 2r_2 c_{(b+1)\Delta}\right)}{2}.
\end{aligned}
$$

Meanwhile,

$$
\begin{aligned}
\frac{t^{(1)}_{(b+1)\Delta}}{n} &= \min\left\{ \frac{2(b+1)\Delta}{n} - 2^{\lceil \log_2 \frac{(b+1)\Delta}{n} \rceil}, R_1 2^{\lceil \log_2 \frac{(b+1)\Delta}{n} \rceil} \right\} \\
&= \min\left\{ \frac{2(b+1)\Delta}{n} - 2^{\lceil \log_2 \frac{b\Delta}{n} \rceil}, R_1 2^{\lceil \log_2 \frac{b\Delta}{n} \rceil} \right\} \\
&\geq \min\left\{ \frac{2b\Delta}{n} - 2^{\lceil \log_2 \frac{b\Delta}{n} \rceil}, R_1 2^{\lceil \log_2 \frac{b\Delta}{n} \rceil} \right\} \\
&= \frac{t^{(1)}_{b\Delta}}{n}.
\end{aligned}
$$

Furthermore, it can be verified that

$$
\frac{t^{(1)}_{(b+1)\Delta} + 2t^{(2)}_{(b+1)\Delta}}{n} = R_1 2^{\lceil \log_2 \frac{(b+1)\Delta}{n} \rceil} = R_1 2^{\lceil \log_2 \frac{b\Delta}{n} \rceil} = \frac{t^{(1)}_{b\Delta} + 2t^{(2)}_{b\Delta}}{n}.
$$

Therefore,

$$\frac{2\left(t_{b\Delta}^{(2)} - t_{(b+1)\Delta}^{(2)}\right)}{n} = \frac{t_{(b+1)\Delta}^{(1)} - t_{b\Delta}^{(1)}}{n} \geq 0.$$

Similarly, we have

$$\frac{2\left(t_{b\Delta}^{(4)} - t_{(b+1)\Delta}^{(4)}\right)}{n} = \frac{\left(t_{(b+1)\Delta}^{(3)} - t_{b\Delta}^{(3)}\right)}{n} \geq 0.$$

Consequently,

$$
\begin{aligned}
& P\left(\frac{(b+1)\Delta}{n}, \bar{l}, \xi, \tau_b\right) - P\left(\frac{b\Delta}{n}, \bar{l}, \xi, \tau_b\right) \\
= {} & \frac{\left(t_{b\Delta}^{(2)} - t_{(b+1)\Delta}^{(2)}\right)}{n}\left(-\ln 2 + \ln\frac{g^2\left(\tau_b, r_1 c_{(b+1)\Delta}\right)}{g\left(\tau_b, 2r_1 c_{(b+1)\Delta}\right)}\right) \\
& + \frac{\left(t_{b\Delta}^{(4)} - t_{(b+1)\Delta}^{(4)}\right)}{n}\left(-\ln 2 + \ln\frac{g^2\left(\tau_b, r_2 c_{(b+1)\Delta}\right)}{g\left(\tau_b, 2r_2 c_{(b+1)\Delta}\right)}\right).
\end{aligned}
$$

At the same time,

$$
\begin{aligned}
\frac{g^2\left(\tau_b, r_1 c_{(b+1)\Delta}\right)}{g\left(\tau_b, 2r_1 c_{(b+1)\Delta}\right)} &= \frac{\left[(1+\tau_b)^{r_1 c_{(b+1)\Delta}} + (1-\tau_b)^{r_1 c_{(b+1)\Delta}}\right]^2}{(1+\tau_b)^{2r_1 c_{(b+1)\Delta}} + (1-\tau_b)^{2r_1 c_{(b+1)\Delta}}} \\
&= 1 + \frac{2(1+\tau_b)^{r_1 c_{(b+1)\Delta}}(1-\tau_b)^{r_1 c_{(b+1)\Delta}}}{(1+\tau_b)^{2r_1 c_{(b+1)\Delta}} + (1-\tau_b)^{2r_1 c_{(b+1)\Delta}}} \\
&= 1 + \frac{2\left(1 - \frac{2\tau_b}{1+\tau_b}\right)^{r_1 c_{(b+1)\Delta}}}{1 + \left(1 - \frac{2\tau_b}{1+\tau_b}\right)^{2r_1 c_{(b+1)\Delta}}}.
\end{aligned}
$$

From the proof of Lemma 5,

$$0 \leq 1 - \frac{2\tau_b}{1+\tau_b} \leq 1 - \frac{2\xi}{\bar{l}} < 1.$$

On the other hand, it is easily verified that

$$f(x) = \frac{2x}{1+x^2}$$

is an increasing function for $x \in [0, 1)$. Therefore,

$$\frac{g^2\left(\tau_b, r_1 c_{(b+1)\Delta}\right)}{g\left(\tau_b, 2r_1 c_{(b+1)\Delta}\right)} \leq 1 + \frac{2\left(1 - \frac{2\xi}{\bar{l}}\right)^{r_1 c_{(b+1)\Delta}}}{1 + \left(1 - \frac{2\xi}{\bar{l}}\right)^{2r_1 c_{(b+1)\Delta}}}.$$

Similarly,

$$\frac{g^2\left(\tau_b, r_2 c_{(b+1)\Delta}\right)}{g\left(\tau_b, 2r_2 c_{(b+1)\Delta}\right)} \leq 1 + \frac{2\left(1-\frac{2\xi}{l}\right)^{r_2 c_{(b+1)\Delta}}}{1+\left(1-\frac{2\xi}{l}\right)^{2r_2 c_{(b+1)\Delta}}}$$

$$\leq 1 + \frac{2\left(1-\frac{2\xi}{l}\right)^{r_1 c_{(b+1)\Delta}}}{1+\left(1-\frac{2\xi}{l}\right)^{2r_1 c_{(b+1)\Delta}}}.$$

And finally,

$$P\left(\frac{(b+1)\Delta}{n}, \bar{l}, \xi, \tau_b\right) - P\left(\frac{b\Delta}{n}, \bar{l}, \xi, \tau_b\right)$$

$$\leq -\frac{\left(t_{b\Delta}^{(2)} - t_{(b+1)\Delta}^{(2)}\right) + \left(t_{b\Delta}^{(4)} - t_{(b+1)\Delta}^{(4)}\right)}{n} Q_{\xi, \frac{(b+1)\Delta}{n}}$$

$$= -\frac{\left(t_{(b+1)\Delta}^{(1)} - t_{b\Delta}^{(1)}\right) - \left(t_{b\Delta}^{(2)} - t_{(b+1)\Delta}^{(2)}\right) + \left(t_{(b+1)\Delta}^{(3)} - t_{b\Delta}^{(3)}\right) - \left(t_{b\Delta}^{(4)} - t_{(b+1)\Delta}^{(4)}\right)}{n} Q_{\xi, \frac{(b+1)\Delta}{n}}$$

$$= -\frac{\left(t_{(b+1)\Delta}^{(1)} + t_{(b+1)\Delta}^{(2)} + t_{(b+1)\Delta}^{(3)} + t_{(b+1)\Delta}^{(4)}\right) - \left(t_{b\Delta}^{(1)} + t_{b\Delta}^{(2)} + t_{b\Delta}^{(3)} + t_{b\Delta}^{(4)}\right)}{n} Q_{\xi, \frac{(b+1)\Delta}{n}}$$

$$= -\frac{\Delta}{n} Q_{\xi, \frac{(b+1)\Delta}{n}}.$$

Using a similar argument, we can show that

$$P\left(\frac{(b+1)\Delta}{n}, \bar{l}, \xi\right) - P\left(\frac{b\Delta}{n}, \bar{l}, \xi\right) \geq P\left(\frac{(b+1)\Delta}{n}, \bar{l}, \xi, \tau_{b+1}\right) - P\left(\frac{b\Delta}{n}, \bar{l}, \xi, \tau_{b+1}\right)$$

$$= \frac{\left(t_{b\Delta}^{(2)} - t_{(b+1)\Delta}^{(2)}\right)}{n}\left(-\ln 2 + \ln \frac{g^2\left(\tau_{b+1}, r_1 c_{(b+1)\Delta}\right)}{g\left(\tau_{b+1}, 2r_1 c_{(b+1)\Delta}\right)}\right)$$

$$+ \frac{\left(t_{b\Delta}^{(4)} - t_{(b+1)\Delta}^{(4)}\right)}{n}\left(-\ln 2 + \ln \frac{g^2\left(\tau_{b+1}, r_2 c_{(b+1)\Delta}\right)}{g\left(\tau_{b+1}, 2r_2 c_{(b+1)\Delta}\right)}\right)$$

$$\geq -\frac{\left(t_{b\Delta}^{(2)} - t_{(b+1)\Delta}^{(2)}\right) + \left(t_{b\Delta}^{(4)} - t_{(b+1)\Delta}^{(4)}\right)}{n}\ln 2$$

$$= -\frac{\Delta}{n}\ln 2.$$

This completes the proof of Property S1 in case (1).

In case (2), i.e. when $\lceil \log_2 \frac{b\Delta}{n}\rceil = \lceil \log_2 \frac{(b+1)\Delta}{n}\rceil - 1$, we have

$$P\left(\frac{(b+1)\Delta}{n}, \bar{l}, \xi, \tau_b\right) - P\left(\frac{b\Delta}{n}, \bar{l}, \xi, \tau_b\right)$$

$$= \frac{t_{(b+1)\Delta}^{(1)}}{n}\ln\frac{g\left(\tau_b, r_1 c_{(b+1)\Delta}\right)}{2} + \frac{t_{(b+1)\Delta}^{(2)} - t_{b\Delta}^{(1)}}{n}\ln\frac{g\left(\tau_b, 2r_1 c_{(b+1)\Delta}\right)}{2} - \frac{t_{b\Delta}^{(2)}}{n}\ln\frac{g\left(\tau_b, 4r_1 c_{(b+1)\Delta}\right)}{2}$$

$$+ \frac{t^{(3)}_{(b+1)\Delta}}{n} \ln \frac{g\left(\tau_b, r_2 c_{(b+1)\Delta}\right)}{2} + \frac{t^{(4)}_{(b+1)\Delta} - t^{(3)}_{b\Delta}}{n} \ln \frac{g\left(\tau_b, 2r_2 c_{(b+1)\Delta}\right)}{2} - \frac{t^{(4)}_{b\Delta}}{n} \ln \frac{g\left(\tau_b, 4r_2 c_{(b+1)\Delta}\right)}{2}.$$

On the other hand,

$$\frac{t^{(1)}_{(b+1)\Delta} + 2t^{(2)}_{(b+1)\Delta}}{n} = R_1 2^{\left\lceil \log_2 \frac{(b+1)\Delta}{n}\right\rceil} = R_1 2^{\left\lceil \log_2 \frac{b\Delta}{n}\right\rceil + 1} = \frac{2t^{(1)}_{b\Delta} + 4t^{(2)}_{b\Delta}}{n}$$

which implies that

$$\frac{t^{(2)}_{(b+1)\Delta} - t^{(1)}_{b\Delta}}{n} = \frac{2t^{(2)}_{b\Delta} - t^{(1)}_{(b+1)\Delta}/2}{n}.$$

Similarly,

$$\frac{t^{(4)}_{(b+1)\Delta} - t^{(3)}_{b\Delta}}{n} = \frac{2t^{(4)}_{b\Delta} - t^{(3)}_{(b+1)\Delta}/2}{n}$$

and therefore,

$$P\left(\frac{(b+1)\Delta}{n}, \bar{l}, \xi, \tau_b\right) - P\left(\frac{b\Delta}{n}, \bar{l}, \xi, \tau_b\right)$$

$$= \frac{t^{(1)}_{(b+1)\Delta}/2}{n}\left[-\ln 2 + \ln \frac{g^2\left(\tau_b, r_1 c_{(b+1)\Delta}\right)}{g\left(\tau_b, 2r_1 c_{(b+1)\Delta}\right)}\right] + \frac{t^{(2)}_{b\Delta}}{n}\left[-\ln 2 + \ln \frac{g^2\left(\tau_b, 2r_1 c_{(b+1)\Delta}\right)}{g\left(\tau_b, 4r_1 c_{(b+1)\Delta}\right)}\right]$$

$$+ \frac{t^{(3)}_{(b+1)\Delta}/2}{n}\left[-\ln 2 + \ln \frac{g^2\left(\tau_b, r_2 c_{(b+1)\Delta}\right)}{g\left(\tau_b, 2r_2 c_{(b+1)\Delta}\right)}\right] + \frac{t^{(4)}_{b\Delta}}{n}\left[-\ln 2 + \ln \frac{g^2\left(\tau_b, 2r_2 c_{(b+1)\Delta}\right)}{g\left(\tau_b, 4r_2 c_{(b+1)\Delta}\right)}\right]$$

$$\leq -\frac{t^{(1)}_{(b+1)\Delta}/2 + t^{(2)}_{b\Delta} + t^{(3)}_{(b+1)\Delta}/2 + t^{(4)}_{b\Delta}}{n} Q_{\xi, \frac{(b+1)\Delta}{n}}$$

$$= -\frac{\Delta}{n} Q_{\xi, \frac{(b+1)\Delta}{n}}$$

where the last step is due to the fact that

$$\frac{t^{(1)}_{(b+1)\Delta}/2 + t^{(2)}_{b\Delta} + t^{(3)}_{(b+1)\Delta}/2 + t^{(4)}_{b\Delta}}{n}$$

$$= \frac{t^{(1)}_{(b+1)\Delta} - \left(t^{(1)}_{b\Delta} - t^{(2)}_{(b+1)\Delta}\right) - t^{(2)}_{b\Delta} + t^{(3)}_{(b+1)\Delta} - \left(t^{(3)}_{b\Delta} - t^{(4)}_{(b+1)\Delta}\right) - t^{(4)}_{b\Delta}}{n}$$

$$= \frac{\left(t^{(1)}_{(b+1)\Delta} + t^{(2)}_{(b+1)\Delta} + t^{(3)}_{(b+1)\Delta} + t^{(4)}_{(b+1)\Delta}\right) - \left(t^{(1)}_{b\Delta} + t^{(2)}_{b\Delta} + t^{(3)}_{b\Delta} + t^{(4)}_{b\Delta}\right)}{n}$$

$$= \frac{\Delta}{n}.$$

In a similar manner, we have

$$
\begin{aligned}
P&\left(\frac{(b+1)\Delta}{n},\bar{l},\xi\right) - P\left(\frac{b\Delta}{n},\bar{l},\xi\right)\\
\geq\ & P\left(\frac{(b+1)\Delta}{n},\bar{l},\xi,\tau_{b+1}\right) - P\left(\frac{b\Delta}{n},\bar{l},\xi,\tau_{b+1}\right)\\
=\ & \frac{t^{(1)}_{(b+1)\Delta/2}}{n}\left[-\ln 2 + \ln\frac{g^2\left(\tau_{b+1},r_1 c_{(b+1)\Delta}\right)}{g\left(\tau_{b+1},2r_1 c_{(b+1)\Delta}\right)}\right] + \frac{t^{(2)}_{b\Delta}}{n}\left[-\ln 2 + \ln\frac{g^2\left(\tau_{b+1},2r_1 c_{(b+1)\Delta}\right)}{g\left(\tau_{b+1},4r_1 c_{(b+1)\Delta}\right)}\right]\\
&+ \frac{t^{(3)}_{(b+1)\Delta/2}}{n}\left[-\ln 2 + \ln\frac{g^2\left(\tau_{b+1},r_2 c_{(b+1)\Delta}\right)}{g\left(\tau_{b+1},2r_2 c_{(b+1)\Delta}\right)}\right] + \frac{t^{(4)}_{b\Delta}}{n}\left[-\ln 2 + \ln\frac{g^2\left(\tau_{b+1},2r_2 c_{(b+1)\Delta}\right)}{g\left(\tau_{b+1},4r_2 c_{(b+1)\Delta}\right)}\right]\\
\geq\ & -\frac{t^{(1)}_{(b+1)\Delta/2} + t^{(2)}_{b\Delta} + t^{(3)}_{(b+1)\Delta/2} + t^{(4)}_{b\Delta}}{n}\ln 2\\
=\ & -\frac{\Delta}{n}\ln 2 \ .
\end{aligned}
$$

The completes the proof of Property S1 in case (2).

Property S2 can be proved in a similar manner.

Now let us move to the proof of Property S3. By Lemma 4, for $\xi\in(0,\bar{l}/2]$,

$$
P(R,\bar{l},\xi) \geq P(R,\bar{l},\bar{l}/2) = -R\ln 2
$$

which implies that

$$
\lim_{R\to 0} P(R,\bar{l},\xi) \geq 0 \ .
$$

At the same time, let $\tau_R$ be the solution to the equation (4.2) with $\frac{b\Delta}{n}=R$, we have

$$
\begin{aligned}
P(R,\bar{l},\xi) &= P(R,\bar{l},\xi,\tau_R)\\
&\leq P\left(R,\bar{l},\xi,\frac{\xi}{\bar{l}-\xi}\right)\\
&\leq -\bar{l}\mathrm{H}_e(\xi/\bar{l}) - \xi\ln\left(\frac{\xi}{\bar{l}-\xi}\right) + \bar{l}\ln\left(1+\frac{\xi}{\bar{l}-\xi}\right)\\
&= 0
\end{aligned}
$$

where the third step follows the fact that $\frac{\xi}{\bar{l}-\xi}\leq 1$ and $\left(1-\frac{\xi}{\bar{l}-\xi}\right)^k \leq \left(1+\frac{\xi}{\bar{l}-\xi}\right)^k$ for any positive integer $k$. And therefore,

$$
\lim_{R\to 0} P(R,\bar{l},\xi) \leq 0
$$

which further yields

$$
\lim_{R\to 0} P(R,\bar{l},\xi) = 0 \ .
$$

This completes the proof of Lemma 6.

**Lemma 7.** *Suppose that $\bar{l}$ is an odd integer. Then for any given $\frac{b\Delta}{n} \geq 0.75$, $P\left(\frac{b\Delta}{n}, \bar{l}, \xi\right)$ is a strictly decreasing function of $\xi$ in the range $\left(\frac{\bar{l}}{2}, \bar{l} - \frac{t_{b\Delta}^{(1)}}{n}\right]$.*

*Proof of Lemma 7*: Since $\bar{l}$ is an odd integer, we have $R_1 = 1$, $R_2 = 0$, and hence $t_{b\Delta}^{(3)} = t_{b\Delta}^{(4)} = 0$. Furthermore, whenever $\frac{b\Delta}{n} \geq 0.75 > 0.5$, one has $c_{b\Delta} = 1$, which, coupled with $R_1 = 1$, implies

$$\frac{t_{b\Delta}^{(1)}}{n} = \min\left\{\frac{2b\Delta}{n} - 1, 1\right\} = \frac{2b\Delta}{n} - 1$$

and

$$\frac{t_{b\Delta}^{(2)}}{n} = \frac{b\Delta}{n} - \frac{t_{b\Delta}^{(1)}}{n} = 1 - \frac{b\Delta}{n}.$$

In view of (B.6), it suffices to show that

$$\tau > \frac{\xi/\bar{l}}{1 - \xi/\bar{l}}$$

or equivalently,

$$\frac{1}{1+\tau} < 1 - \xi/\bar{l}$$

for $\xi \in \left(\frac{\bar{l}}{2}, \bar{l} - \frac{t_{b\Delta}^{(1)}}{n}\right]$, where $\tau$ is the solution to the equation (4.2). By Lemma 2 and the fact that $\tau = 1$ when $\xi = \frac{\bar{l}}{2}$, we have $\tau > 1$ for $\xi \in \left(\frac{\bar{l}}{2}, \bar{l} - \frac{t_{b\Delta}^{(1)}}{n}\right]$. Moreover, according to the discussion above, equation (4.2) can be further simplified as

$$\left(\frac{2b\Delta}{n} - 1\right)\frac{g(\tau, \bar{l} - 1)}{g(\tau, \bar{l})} + \left(2 - \frac{2b\Delta}{n}\right)\frac{g(\tau, 2\bar{l} - 1)}{g(\tau, 2\bar{l})} = 1 - \xi/\bar{l}$$

or

$$\frac{1}{1+\tau}\left[\left(\frac{2b\Delta}{n} - 1\right)\frac{1 + \left(\frac{\tau-1}{\tau+1}\right)^{\bar{l}-1}}{1 - \left(\frac{\tau-1}{\tau+1}\right)^{\bar{l}}} + \left(2 - \frac{2b\Delta}{n}\right)\frac{1 - \left(\frac{\tau-1}{\tau+1}\right)^{2\bar{l}-1}}{1 + \left(\frac{\tau-1}{\tau+1}\right)^{2\bar{l}}}\right] = 1 - \xi/\bar{l}\,.$$

Let $z = \frac{\tau-1}{\tau+1}$, and the lemma is proved by showing that

$$\left(\frac{2b\Delta}{n} - 1\right)\frac{1 + z^{\bar{l}-1}}{1 - z^{\bar{l}}} + \left(2 - \frac{2b\Delta}{n}\right)\frac{1 - z^{2\bar{l}-1}}{1 + z^{2\bar{l}}} > 1$$

for $z \in (0, 1)$. Towards this, note that

$$\frac{1 + z^{\bar{l}-1}}{1 - z^{\bar{l}}} > 1 > \frac{1 - z^{2\bar{l}-1}}{1 + z^{2\bar{l}}}$$

and

$$\left(\frac{2b\Delta}{n} - 1\right)\frac{1 + z^{\bar{l}-1}}{1 - z^{\bar{l}}} + \left(2 - \frac{2b\Delta}{n}\right)\frac{1 - z^{2\bar{l}-1}}{1 + z^{2\bar{l}}}$$

$$= \quad \frac{1}{2}\left(\frac{1 + z^{\bar{l}-1}}{1 - z^{\bar{l}}} + \frac{1 - z^{2\bar{l}-1}}{1 + z^{2\bar{l}}}\right) + \left(\frac{2b\Delta}{n} - \frac{3}{2}\right)\left(\frac{1 + z^{\bar{l}-1}}{1 - z^{\bar{l}}} - \frac{1 - z^{2\bar{l}-1}}{1 + z^{2\bar{l}}}\right)$$

$$\geq \quad \frac{1}{2}\left(\frac{1 + z^{\bar{l}-1}}{1 - z^{\bar{l}}} + \frac{1 - z^{2\bar{l}-1}}{1 + z^{2\bar{l}}}\right)$$

when $\frac{b\Delta}{n} \geq 0.75$. Furthermore,

$$\frac{1 + z^{\bar{l}-1}}{1 - z^{\bar{l}}} + \frac{1 - z^{2\bar{l}-1}}{1 + z^{2\bar{l}}} \quad \geq \quad 2\sqrt{\frac{1 + z^{\bar{l}-1}}{1 - z^{\bar{l}}}\frac{1 - z^{2\bar{l}-1}}{1 + z^{2\bar{l}}}}$$

$$= \quad 2\sqrt{\frac{1 + z^{\bar{l}-1}}{1 + z^{2\bar{l}}}}\sqrt{\frac{1 - z^{2\bar{l}-1}}{1 - z^{\bar{l}}}} > 2$$

since $0 < z < 1$. This completes the proof of Lemma 7.

## Appendix C

### Proof of Theorem 1

Given $x^n$ and $y^n$, let $j = j(x^n, y^n)$ be the number of interactions at the time the decoder sends bit 1 to the encoder. From (3.1) and (3.2), it follows that

$$r_f(x^n, y^n | \mathcal{I}_n) = \begin{cases} \frac{j\Delta}{n} + \mathrm{H}(\epsilon) + \frac{\Delta}{n} & \text{if } j \leq \Delta/n \\ 1 + \eta_n + \mathrm{H}(\epsilon) + \frac{\Delta}{n} & \text{otherwise} \end{cases} \tag{C.1}$$

and

$$r_b(x^n, y^n | \mathcal{I}_n) = \frac{j}{n}. \tag{C.2}$$

Since $\Delta \sim \sqrt{n}$ and $j \leq \frac{n}{\Delta} + 1$ according to Algorithm 1, (4.8) follows immediately.

In view of the description of Algorithm 1, it is not hard to see that at the $(j-1)$th interaction, one always has

$$\Gamma_{j-1} < h_n(x^n | y^n) . \tag{C.3}$$

We now distinguish between two cases: (1) $h_n(x^n | y^n) \leq \Gamma_{\frac{n}{\Delta}}$, and (2) $h_n(x^n | y^n) > \Gamma_{\frac{n}{\Delta}}$. In case (1), it follows from (C.3) that

$$j \leq \frac{n}{\Delta} \tag{C.4}$$

and

$$\frac{1}{\ln 2}\left[-P\left(\frac{(j-1)\Delta}{n}, \bar{l}, l_1\epsilon\right) - \frac{3\lceil\bar{l}\rceil}{\Delta}\ln\frac{n\bar{l}}{2} - \frac{1}{2n}\ln\frac{n\bar{l}}{4}\right] - \frac{\Delta}{n} < h_n(x^n | y^n)$$

or equivalently

$$-P\left(\frac{(j-1)\Delta}{n}, \bar{l}, l_1\epsilon\right) < \left[h_n(x^n|y^n) + \frac{\Delta}{n}\right]\ln 2 + \frac{3\lceil\bar{l}\rceil}{\Delta}\ln\frac{n\bar{l}}{2} + \frac{1}{2n}\ln\frac{n\bar{l}}{4}$$

$$= -P\left(R_{L(z)}^{(\Delta)}\left(\epsilon, h_n(x^n|y^n)\right), \bar{l}, l_1\epsilon\right) .$$

By Lemma 6, $P\left(R, \bar{l}, l_1\epsilon\right)$ is strictly decreasing with respect to $R$. Therefore,

$$\frac{(j-1)\Delta}{n} < R_{L(z)}^{(\Delta)}\left(\epsilon, h_n(x^n|y^n)\right) . \tag{C.5}$$

Combining (C.1), (C.4), and (C.5) together yields

$$r_f(x^n, y^n|\mathcal{I}_n) \le R_{L(z)}^{(\Delta)}\left(\epsilon, h_n(x^n|y^n)\right) + \mathrm{H}(\epsilon) + \frac{2\Delta}{n} .$$

This completes the proof of (4.7) in case (1).

In case (2), $j$ could be strictly greater than $\frac{n}{\Delta}$. Regardless of the value of $j$, in case (2), one always has

$$r_f(x^n, y^n|\mathcal{I}_n) \le 1 + \eta_n + \mathrm{H}(\epsilon) + \frac{\Delta}{n}$$

$$= R_{L(z)}^{(\Delta)}\left(\epsilon, h_n(x^n|y^n)\right) + \mathrm{H}(\epsilon) + \frac{2\Delta}{n} .$$

This completes the proof of (4.7) in case (2).

Towards bounding the error probability, for any $x^n \in \mathcal{B}^n$ and $0 < \epsilon < 0.5$, define

$$B(\epsilon, x^n) = \left\{z^n \in \mathcal{B}^n : \frac{1}{n}wt(z^n - x^n) < \epsilon \text{ or } \frac{1}{n}wt(z^n - x^n) > 1 - \epsilon\right\} .$$

To proceed,

$$P_e\{\mathcal{I}_n|x^n, y^n\} = \Pr\{\tilde{x}^n \ne x^n\}$$

$$= \Pr\{\hat{x}^n \in B(\epsilon, x^n)\}\Pr\{\tilde{x}^n \ne x^n | \hat{x}^n \in B(\epsilon, x^n)\}$$

$$+ \Pr\{\hat{x}^n \notin B(\epsilon, x^n)\}\Pr\{\tilde{x}^n \ne x^n | \hat{x}^n \notin B(\epsilon, x^n)\}$$

$$\le \Pr\{\tilde{x}^n \ne x^n | \hat{x}^n \in B(\epsilon, x^n)\} + \Pr\{\hat{x}^n \notin B(\epsilon, x^n)\} .$$

We first consider $\Pr\{\hat{x}^n \notin B(\epsilon, x^n)\}$. By the union bound,

$$\Pr\{\hat{x}^n \notin B(\epsilon, x^n)\}$$

$$\le \Pr\left\{\exists z^n \notin B(\epsilon, x^n) : \mathbf{H}_{b\Delta\times n}^{(b\Delta)}z^n = \mathbf{H}_{b\Delta\times n}^{(b\Delta)}x^n, h_n(z^n|y^n) \le \Gamma_b \text{ for some } b, 1 \le b \le \frac{n}{\Delta}\right\}$$

$$+ \Pr\left\{\exists z^n \notin B(\epsilon, x^n) : \mathbf{H}_{n\times n}z^n = \mathbf{H}_{n\times n}x^n, \mathbf{H}'_{\eta_n n\times n}z^n = \mathbf{H}'_{\eta_n n\times n}x^n\right\}$$

$$\le \sum_{b=1}^{\frac{n}{\Delta}}\Pr\left\{\exists z^n \notin B(\epsilon, x^n) : \mathbf{H}_{b\Delta\times n}^{(b\Delta)}z^n = \mathbf{H}_{b\Delta\times n}^{(b\Delta)}x^n, h_n(z^n|y^n) \le \Gamma_b\right\}$$

$$+ \Pr\left\{\exists z^n \notin B(\epsilon, x^n) : \mathbf{H}_{n\times n}z^n = \mathbf{H}_{n\times n}x^n, \mathbf{H}'_{\eta_n n\times n}z^n = \mathbf{H}'_{\eta_n n\times n}x^n\right\} .$$

Now by Lemma 1, for $1 \leq b \leq \frac{n}{\Delta}$,

$$\Pr\left\{\mathbf{H}_{b\Delta \times n}^{(b\Delta)} z^n = \mathbf{H}_{b\Delta \times n}^{(b\Delta)} x^n\right\}$$

$$= \Pr\left\{\mathbf{H}_{b\Delta \times n}^{(b\Delta)}(z^n - x^n) = 0^{b\Delta}\right\}$$

$$\leq \exp\left\{nP\left(\frac{b\Delta}{n}, \bar{l}, \xi\right) + \frac{3n\lceil \bar{l} \rceil}{b\Delta}\ln(n\hat{\xi}) + \frac{1}{2}\ln n\xi\left(1 - \frac{\xi}{\bar{l}}\right) + O(1)\right\}$$

$$\leq \exp\left\{n\left[P\left(\frac{b\Delta}{n}, \bar{l}, \xi\right) + \frac{3\lceil \bar{l} \rceil}{\Delta}\ln\frac{n\bar{l}}{2} + \frac{1}{2n}\ln\frac{n\bar{l}}{4}\right] + O(1)\right\}$$

while

$$\Pr\left\{\mathbf{H}_{n \times n} z^n = \mathbf{H}_{n \times n} x^n\right\}$$

$$= \Pr\left\{\mathbf{H}_{n \times n}(z^n - x^n) = 0^n\right\}$$

$$\leq \exp\left\{nP\left(1, \bar{l}, \xi\right) + 3\lceil \bar{l} \rceil \ln(n\hat{\xi}) + \frac{1}{2}\ln n\xi\left(1 - \frac{\xi}{\bar{l}}\right) + O(1)\right\}$$

$$\leq \exp\left\{n\left[P\left(1, \bar{l}, \xi\right) + \frac{3\lceil \bar{l} \rceil}{n}\ln\frac{n\bar{l}}{2} + \frac{1}{2n}\ln\frac{n\bar{l}}{4}\right] + O(1)\right\},$$

where $\xi = \bar{l}^{\succ (z^n - x^n)}$ and $\hat{\xi} = \max\left\{\frac{1}{n}, \min\left\{\xi, \bar{l} - \xi\right\}\right\}$. Simple calculation reveals that $l_1\epsilon \leq \xi \leq \bar{l} - l_1\epsilon$ for $z^n \notin B(\epsilon, x^n)$, which, together with Lemmas 3 and 4, further implies that

$$\Pr\left\{\mathbf{H}_{b\Delta \times n}^{(b\Delta)} z^n = \mathbf{H}_{b\Delta \times n}^{(b\Delta)} x^n\right\}$$

$$\leq \exp\left\{n\left[P\left(\frac{b\Delta}{n}, \bar{l}, l_1\epsilon\right) + \frac{3\lceil \bar{l} \rceil}{\Delta}\ln\frac{n\bar{l}}{2} + \frac{1}{2n}\ln\frac{n\bar{l}}{4}\right] + O(1)\right\} = 2^{-n\Gamma_b - \Delta + O(1)}$$

and

$$\Pr\left\{\mathbf{H}_{n \times n} z^n = \mathbf{H}_{n \times n} x^n\right\}$$

$$\leq \exp\left\{n\left[P\left(1, \bar{l}, l_1\epsilon\right) + \frac{3\lceil \bar{l} \rceil}{n}\ln\frac{n\bar{l}}{2} + \frac{1}{2n}\ln\frac{n\bar{l}}{4}\right] + O(1)\right\} = 2^{-n(1 - \eta_n) - \Delta + O(1)}.$$

Now by the union bound again, for $1 \leq b \leq \frac{n}{\Delta}$,

$$\Pr\left\{\exists z^n \notin B(\epsilon, x^n) : \mathbf{H}_{b\Delta \times n}^{(b\Delta)} z^n = \mathbf{H}_{b\Delta \times n}^{(b\Delta)} x^n, h_n(z^n | y^n) \leq \Gamma_b\right\}$$

$$\leq \left|z^n \notin B(\epsilon, x^n) : h_n(z^n | y^n) \leq \Gamma_b\right| 2^{-n\Gamma_b - \Delta + O(1)}$$

$$\leq \left|z^n : h_n(z^n | y^n) \leq \Gamma_b\right| 2^{-n\Gamma_b - \Delta + O(1)}.$$

At this point, we invoke the following lemma, which is from [2]:

**Lemma 8.** *For any $y^n \in \mathcal{Y}^n$ and any $0 \leq \alpha \leq 1$,*

$$\left|z^n : h_n(z^n | y^n) \leq \alpha\right| \leq 2^{n\alpha}$$

*where $h_n(\cdot|\cdot)$ is the code length function of any decodable code.*

Therefore, we have

$$\Pr\left\{\exists z^n \notin B(\epsilon, x^n) : \mathbf{H}_{b\Delta \times n}^{(b\Delta)} z^n = \mathbf{H}_{b\Delta \times n}^{(b\Delta)} x^n, h_n(z^n|y^n) \leq \Gamma_b\right\} \leq 2^{-\Delta + O(1)}.$$

At the same time,

$$\Pr\left\{\exists z^n \notin B(\epsilon, x^n) : \mathbf{H}_{n \times n} z^n = \mathbf{H}_{n \times n} x^n, \mathbf{H}'_{\eta_n n \times n} z^n = \mathbf{H}'_{\eta_n n \times n} x^n\right\}$$

$$\leq \sum_{z^n \notin B(\epsilon, x^n)} \Pr\left\{\mathbf{H}_{n \times n}(z^n - x^n) = 0^n\right\} \Pr\left\{\mathbf{H}'_{\eta_n n \times n}(z^n - x^n) = 0^{\eta_n n}\right\}$$

$$\leq \sum_{z^n \notin B(\epsilon, x^n)} 2^{-n(1-\eta_n) - \Delta + O(1)} 2^{-\eta_n n}$$

$$\leq 2^{-\Delta + O(1)}.$$

To sum up, we have shown that

$$\Pr\left\{\hat{x}^n \notin B(\epsilon, x^n)\right\} \leq 2^{-\Delta + \log_2\left(\frac{n}{\Delta} + 1\right) + O(1)}.$$

Before moving to the next target $\Pr\left\{\tilde{x}^n \neq x^n \,|\, \hat{x}^n \in B(\epsilon, x^n)\right\}$, it is not hard to verify the following bound on $|B(\epsilon, x^n)|$:

$$|B(\epsilon, x^n)| = 2 \sum_{d=0}^{\lfloor n\epsilon \rfloor} \binom{n}{d}$$

$$\leq 2 \cdot 2^{n\mathrm{H}\left(\frac{\lfloor n\epsilon \rfloor}{n}\right)} \leq 2^{n\mathrm{H}(\epsilon) + 1}.$$

Now suppose $\hat{x}^n \in B(\epsilon, x^n)$, then $x^n \in B(\epsilon, \hat{x}^n)$, which, according to Algorithm 1, implies that

$$\Pr\left\{\tilde{x}^n \neq x^n \,|\, \hat{x}^n \in B(\epsilon, x^n)\right\} = \Pr\left\{\exists z^n \in B(\epsilon, \hat{x}^n)/\{x^n\} : \mathbf{H}''_{(n\mathrm{H}(\epsilon) + \Delta) \times n} z^n = \mathbf{H}''_{(n\mathrm{H}(\epsilon) + \Delta) \times n} x^n\right\}$$

$$\leq |B(\epsilon, \hat{x}^n)| 2^{-n\mathrm{H}(\epsilon) + \Delta}$$

$$\leq 2^{-\Delta + O(1)}.$$

In summary,

$$P_e\left\{\mathcal{I}_n | x^n, y^n\right\} \leq \Pr\left\{\tilde{x}^n \neq x^n \,|\, \hat{x}^n \in B(\epsilon, x^n)\right\} + \Pr\left\{\hat{x}^n \notin B(\epsilon, x^n)\right\}$$

$$\leq 2^{-\Delta + O(1)} + 2^{-\Delta + \log_2\left(\frac{n}{\Delta} + 1\right) + O(1)}$$

$$\leq 2^{-\Delta + \log_2\left(\frac{n}{\Delta} + 1\right) + O(1)}.$$

The theorem is proved.

## Appendix D

## Doubly Asymptotical Performance

In the appendix, we prove Propositions 1 and 2 and Theorem 2.

Proof of **Proposition 1**: In view of Lemma 6, it follows from the definition of $R_{L(z)}(\epsilon, h)$ that $R_{L(z)}(\epsilon, h)$ is the solution to

$$-P\left(R, \bar{l}, l_1\epsilon\right) = h \ln 2$$

if $h \ln 2 < -P\left(1, \bar{l}, l_1\epsilon\right)$, and

$$R_{L(z)}(\epsilon, h) = 2 + \frac{1}{\ln 2} P\left(1, \bar{l}, l_1\epsilon\right)$$

otherwise. On the other hand, in view of the fact that $l_1\epsilon \geq \frac{\bar{l}}{\lfloor \bar{l} \rfloor}$ and of Lemma 5, for $R \in (0, 1]$,

$$
\begin{aligned}
P\left(R, \bar{l}, l_1\epsilon\right) &\leq -R\ln 2 + 2l_1\epsilon \exp\left[-\frac{2l_1\epsilon}{\bar{l}}\left(c_R r_1 - 1\right)\right] + R\exp\left(-\frac{2l_1\epsilon}{\bar{l}} r_1 c_R\right) \\
&\leq -R\ln 2 + 2l_1\epsilon \exp\left[-\frac{2l_1\epsilon}{\bar{l}}\left(\lfloor \bar{l} \rfloor - 1\right)\right] + \exp\left(-\frac{2l_1\epsilon}{\bar{l}}\lfloor \bar{l} \rfloor\right)
\end{aligned}
$$

where $c_R \overset{\Delta}{=} 2^{-\lceil \log_2 R \rceil} \geq 1$. Now if $h \ln 2 \geq -P\left(1, \bar{l}, l_1\epsilon\right)$, then

$$
\begin{aligned}
r_{L(z)}(\epsilon, h) &= R_{L(z)}(\epsilon, h) + \mathrm{H}(\epsilon) - h \\
&\leq 2 + \frac{2}{\ln 2} P\left(1, \bar{l}, l_1\epsilon\right) + \mathrm{H}(\epsilon) \\
&\leq \frac{4l_1\epsilon}{\ln 2}\exp\left[-\frac{2l_1\epsilon}{\bar{l}}\left(\lfloor \bar{l} \rfloor - 1\right)\right] + \frac{2}{\ln 2}\exp\left(-\frac{2l_1\epsilon}{\bar{l}}\lfloor \bar{l} \rfloor\right) + \mathrm{H}(\epsilon).
\end{aligned}
\tag{D.1}
$$

If $h \ln 2 < -P\left(1, \bar{l}, l_1\epsilon\right)$, then

$$
\begin{aligned}
h \ln 2 &= -P\left(R_{L(z)}(\epsilon, h), \bar{l}, l_1\epsilon\right) \\
&\geq R_{L(z)}(\epsilon, h)\ln 2 - 2l_1\epsilon \exp\left[-\frac{2l_1\epsilon}{\bar{l}}\left(\lfloor \bar{l} \rfloor - 1\right)\right] - \exp\left(-\frac{2l_1\epsilon}{\bar{l}}\lfloor \bar{l} \rfloor\right)
\end{aligned}
$$

which implies that

$$R_{L(z)}(\epsilon, h) \leq h + \frac{2l_1\epsilon}{\ln 2}\exp\left[-\frac{2l_1\epsilon}{\bar{l}}\left(\lfloor \bar{l} \rfloor - 1\right)\right] + \frac{1}{\ln 2}\exp\left(-\frac{2l_1\epsilon}{\bar{l}}\lfloor \bar{l} \rfloor\right).$$

Therefore,

$$
\begin{aligned}
r_{L(z)}(\epsilon, h) &= R_{L(z)}(\epsilon, h) + \mathrm{H}(\epsilon) - h \\
&\leq \frac{2l_1\epsilon}{\ln 2}\exp\left[-\frac{2l_1\epsilon}{\bar{l}}\left(\lfloor \bar{l} \rfloor - 1\right)\right] + \frac{1}{\ln 2}\exp\left(-\frac{2l_1\epsilon}{\bar{l}}\lfloor \bar{l} \rfloor\right) + \mathrm{H}(\epsilon).
\end{aligned}
\tag{D.2}
$$

Combining (D.1) with (D.2) completes the proof of Proposition 1.

Proof of **Proposition 2**: Note that $k \geq e^{\frac{2}{l_1}}$, which implies that

$$kl_1 \frac{\ln k}{k} = l_1 \ln k \geq 2 \geq \frac{k\bar{l}}{\lfloor k\bar{l} \rfloor},$$

and therefore, we can apply Proposition 1 on $r_{L(z^k)}\left(\frac{\ln k}{k}, h\right)$, resulting in

$$r_{L(z^k)}\left(\frac{\ln k}{k}, h\right) \leq \frac{4l_1 \ln k}{\ln 2} \exp\left[-\frac{2l_1\left(\lfloor k\bar{l} \rfloor - 1\right)}{k\bar{l}} \ln k\right] + \frac{2}{\ln 2} \exp\left(-\frac{2l_1 \lfloor k\bar{l} \rfloor}{k\bar{l}} \ln k\right) + \mathrm{H}\left(\frac{\ln k}{k}\right)$$

It is easily verified that

$$\mathrm{H}\left(\frac{\ln k}{k}\right) = O\left(\frac{\ln^2 k}{k}\right).$$

On the other hand,

$$\frac{2l_1 \lfloor k\bar{l} \rfloor}{k\bar{l}} \geq \frac{2l_1\left(\lfloor k\bar{l} \rfloor - 1\right)}{k\bar{l}} \geq \frac{4\left(\lfloor k\bar{l} \rfloor - 1\right)}{k\bar{l}} \geq 1.$$

Therefore,

$$r_{L(z^k)}\left(\frac{\ln k}{k}, h\right) = O\left(\frac{\ln k}{k}\right) + O\left(\frac{1}{k}\right) + O\left(\frac{\ln^2 k}{k}\right) = O\left(\frac{\ln^2 k}{k}\right).$$

Proof of **Theorem 2**: In view of Theorem 1, (4.14) and (4.15) follow immediately. Thus it suffices to prove (4.13). From Theorem 1 again, we have

$$r_f\left(X^n, Y^n | \mathcal{I}_n\left(L(z^k), \frac{\ln k}{2k}\right)\right) \leq R_{L(z^k)}^{(\Delta)}\left(\frac{\ln k}{2k}, h_n(x^n | y^n)\right) + \mathrm{H}\left(\frac{\ln k}{2k}\right) + \frac{2\Delta}{n}. \tag{D.3}$$

Let $\delta > 0$ be a small number to be specified later. In view of the definition of $R_{L(z)}^{(\Delta)}\left(\epsilon, h_n(x^n | y^n)\right)$ and Lemma 6, it is not hard to verify that $R_{L(z)}^{(\Delta)}\left(\epsilon, h_n(x^n | y^n)\right)$ is non-decreasing as $h_n(x^n | y^n)$ increases. This, coupled with (D.3) and (4.12), implies that with probability one

$$r_f\left(X^n, Y^n | \mathcal{I}_n\left(L(z^k), \frac{\ln k}{2k}\right)\right) \leq R_{L(z^k)}^{(\Delta)}\left(\frac{\ln k}{2k}, \mathrm{H}(X|Y) + \delta\right) + \mathrm{H}\left(\frac{\ln k}{2k}\right) + \frac{2\Delta}{n} \tag{D.4}$$

for sufficiently large $n$. Applying Propositions 1 and 2 to (D.4), we have

$$\begin{aligned}
\limsup_{n \to \infty} r_f\left(X^n, Y^n | \mathcal{I}_n\left(L(z^k), \frac{\ln k}{2k}\right)\right) &\leq \mathrm{H}(X|Y) + \delta + r_{L(z^k)}\left(\frac{\ln k}{2k}, \mathrm{H}(X|Y) + \delta\right) \\
&= \mathrm{H}(X|Y) + \delta + O\left(\frac{\ln^2 k}{k}\right)
\end{aligned} \tag{D.5}$$

with probability one. Letting $\delta \to 0$ and then $k \to \infty$ in (D.5) yields

$$\limsup_{k \to \infty} \limsup_{n \to \infty} r_f\left(X^n, Y^n \left| \mathcal{I}_n\left(L(z^k), \frac{\ln k}{2k}\right)\right.\right) \leq \mathrm{H}(X|Y)$$

with probability one. This, coupled with the converse [2, Theorem 3], implies (4.13). This competes the proof of Theorem 2.

## Appendix E

## Proof of Theorem 3

In view of Theorem 1, it suffices to prove (5.2) and (5.4). Note that from the proof of Theorem 1 and the description of Algorithm 3, it can be seen that for any sequence of source-side information pairs $(X^n, Y^n)$,

$$r_f\left(X^n, Y^n | \tilde{\mathcal{I}}_n\right) \leq \frac{\Delta}{n} + \begin{cases} R_{L(z)}^{(\Delta)}\left(\epsilon, \mathrm{H}\left(\frac{1}{n}wt(X^n - Y^n) + \frac{\ln n + 1}{n}\right)\right) & \text{if } wt(X^n - Y^n) \leq 0.5n \\ R_{L(z)}^{(\Delta)}\left(\epsilon, 1\right) & \text{otherwise.} \end{cases}$$

Therefore,

$$\begin{aligned} r_f(\tilde{\mathcal{I}}_n) &\leq \frac{\Delta}{n} + \Pr\left\{\frac{1}{n}wt\left(X^n - Y^n\right) \leq p_0 + \sqrt{\frac{\ln n}{n}}\right\} \\ &\quad \times \mathbb{E}\left[R_{L(z)}^{(\Delta)}\left(\epsilon, \mathrm{H}\left(\frac{1}{n}wt(X^n - Y^n)\right) + \frac{\ln n + 1}{n}\right) \bigg| \frac{1}{n}wt\left(X^n - Y^n\right) \leq p_0 + \sqrt{\frac{\ln n}{n}}\right] \\ &\quad + \Pr\left\{p_0 + \sqrt{\frac{\ln n}{n}} < \frac{1}{n}wt(\hat{X}^n - Y^n) \leq 0.5\right\} \\ &\quad \times \mathbb{E}\left[R_{L(z)}^{(\Delta)}\left(\epsilon, \mathrm{H}\left(\frac{1}{n}wt(X^n - Y^n)\right) + \frac{\ln n + 1}{n}\right) \bigg| p_0 + \sqrt{\frac{\ln n}{n}} < \frac{1}{n}wt\left(X^n - Y^n\right) \leq 0.5\right] \\ &\quad + \Pr\left\{\frac{1}{n}wt\left(X^n - Y^n\right) > 0.5\right\} R_{L(z)}^{(\Delta)}(\epsilon, 1) \\ &\leq \frac{\Delta}{n} + \mathbb{E}\left[R_{L(z)}^{(\Delta)}\left(\epsilon, \mathrm{H}\left(\frac{1}{n}wt(X^n - Y^n)\right) + \frac{\ln n + 1}{n}\right) \bigg| \frac{1}{n}wt\left(X^n - Y^n\right) \leq p_0 + \sqrt{\frac{\ln n}{n}}\right] \\ &\quad + \Pr\left\{\frac{1}{n}wt\left(X^n - Y^n\right) > p_0 + \sqrt{\frac{\ln n}{n}}\right\} R_{L(z)}^{(\Delta)}(\epsilon, 1) \end{aligned}$$

where we assume that

$$p_0 < 0.5 - \sqrt{\frac{\ln n}{n}}$$

which always holds for sufficiently large $n$ as $p_0 < 0.5$. On one hand, given

$$\frac{1}{n}wt\left(X^n - Y^n\right) \leq p_0 + \sqrt{\frac{\ln n}{n}} < 0.5$$

we have

$$\begin{aligned} \mathrm{H}\left(\frac{1}{n}wt(X^n - Y^n)\right) &\leq \mathrm{H}\left(p_0 + \sqrt{\frac{\ln n}{n}}\right) \\ &\leq \mathrm{H}(p_0) + \log_2\left(\frac{1 - p_0}{p_0}\right)\sqrt{\frac{\ln n}{n}} \end{aligned}$$

which further implies that

$$\mathbb{E}\left[R_{L(z)}^{(\Delta)}\left(\epsilon, \mathrm{H}\left(\frac{1}{n}wt(X^n - Y^n)\right) + \frac{\ln n + 1}{n}\right)\bigg| \frac{1}{n}wt(X^n - Y^n) \le p_0 + \sqrt{\frac{\ln n}{n}}\right]$$

$$\le R_{L(z)}^{(\Delta)}\left(\epsilon, \mathrm{H}(p_0) + \frac{\ln n + 1}{n} + \log_2\left(\frac{1 - p_0}{p_0}\right)\sqrt{\frac{\ln n}{n}}\right).$$

On the other hand, by Hoeffding's inequality,

$$\Pr\left\{\frac{1}{n}wt(X^n - Y^n) > p_0 + \sqrt{\frac{\ln n}{n}}\right\} \le n^{-2}$$

from which (5.2) is proved.

Towards showing (5.4), we have

$$
\begin{aligned}
P_b(\tilde{\mathcal{I}}_n) &= \mathbb{E}\left[\frac{1}{n}wt(\hat{X}^n - X^n)\right] \\
&= \mathbb{E}\left[\mathbb{E}\left[\frac{1}{n}wt(\hat{X}^n - X^n)\bigg| X^n, Y^n\right]\right] \\
&= \sum_{(x^n, y^n):\frac{1}{n}wt(x^n - y^n) \le 0.5} \Pr\{X^n = x^n, Y^n = y^n\}\mathbb{E}\left[\frac{1}{n}wt(\hat{X}^n - x^n)\bigg| x^n, y^n\right] \\
&\quad + \sum_{(x^n, y^n):\frac{1}{n}wt(x^n - y^n) > 0.5} \Pr\{X^n = x^n, Y^n = y^n\}\mathbb{E}\left[\frac{1}{n}wt(\hat{X}^n - x^n)\bigg| x^n, y^n\right] \\
&\le \sum_{(x^n, y^n):\frac{1}{n}wt(x^n - y^n) \le 0.5} \Pr\{X^n = x^n, Y^n = y^n\}\mathbb{E}\left[\frac{1}{n}wt(\hat{X}^n - x^n)\bigg| x^n, y^n\right] \\
&\quad + \Pr\left\{\frac{1}{n}wt(X^n - Y^n) > 0.5\right\}.
\end{aligned}
$$
(E.1)

By Hoeffding's inequality,

$$\Pr\left\{\frac{1}{n}wt(X^n - Y^n) > 0.5\right\} \le e^{-2n(0.5 - p_0)^2}.$$
(E.2)

On the other hand,

$$
\begin{aligned}
&\mathbb{E}\left[\frac{1}{n}wt(\hat{X}^n - x^n)\bigg| x^n, y^n\right] \\
&= \Pr\left\{\frac{1}{n}wt(\hat{X}^n - x^n) \le \epsilon\bigg| x^n, y^n\right\}\mathbb{E}\left[\frac{1}{n}wt(\hat{X}^n - x^n)\bigg| \frac{1}{n}wt(\hat{X}^n - x^n) \le \epsilon, x^n, y^n\right] \\
&\quad + \Pr\left\{\frac{1}{n}wt(\hat{X}^n - x^n) > \epsilon\bigg| x^n, y^n\right\}\mathbb{E}\left[\frac{1}{n}wt(\hat{X}^n - x^n)\bigg| \frac{1}{n}wt(\hat{X}^n - x^n) > \epsilon, x^n, y^n\right] \\
&\le \epsilon + \Pr\left\{\frac{1}{n}wt(\hat{X}^n - x^n) > \epsilon\bigg| x^n, y^n\right\}
\end{aligned}
$$
(E.3)

Now we would like to bound

$$\Pr\left\{\frac{1}{n}wt(\hat{X}^n - x^n) > \epsilon \,\middle|\, x^n, y^n\right\}$$

when $\frac{1}{n}wt(x^n - y^n) \leq 0.5$. By the argument in the proof of Theorem 1,

$$\Pr\left\{\frac{1}{n}wt(\hat{X}^n - x^n) > \epsilon \,\middle|\, x^n, y^n\right\}$$

$$\leq \Pr\left\{\exists \hat{x}^n, \frac{1}{n}wt(\hat{x}^n - x^n) > \epsilon, \mathbf{H}_{b\Delta \times n}^{(b\Delta)}(\hat{x}^n - x^n) = 0^{b\Delta}, \gamma(\hat{x}^n, y^n) \leq \Gamma_b \text{ for some } b, 1 \leq b \leq \frac{n}{\Delta}\right\}$$

$$+ \Pr\left\{\exists \hat{x}^n, \frac{1}{n}wt(\hat{x}^n - x^n) > \epsilon, \mathbf{H}_{n \times n}(\hat{x}^n - x^n) = 0^n, \mathbf{H}'_{\eta_n n \times n}(\hat{x}^n - x^n) = 0^{\eta_n n}\right\}$$

$$\leq \sum_{b=1}^{\lfloor \frac{0.75n}{\Delta} \rfloor} \Pr\left\{\exists \hat{x}^n, \frac{1}{n}wt(\hat{x}^n - x^n) > \epsilon, \mathbf{H}_{b\Delta \times n}^{(b\Delta)}(\hat{x}^n - x^n) = 0^{b\Delta}, \gamma(\hat{x}^n, y^n) \leq \Gamma_b\right\}$$

$$+ \sum_{b=\lfloor \frac{0.75n}{\Delta} \rfloor + 1}^{\frac{n}{\Delta}} \Pr\left\{\exists \hat{x}^n, \frac{1}{n}wt(\hat{x}^n - x^n) > \epsilon, \mathbf{H}_{b\Delta \times n}^{(b\Delta)}(\hat{x}^n - x^n) = 0^{b\Delta}, \gamma(\hat{x}^n, y^n) \leq \Gamma_b\right\}$$

$$+ \Pr\left\{\exists \hat{x}^n, \frac{1}{n}wt(\hat{x}^n - x^n) > \epsilon, \mathbf{H}_{n \times n}(\hat{x}^n - x^n) = 0^n, \mathbf{H}'_{\eta_n n \times n}(\hat{x}^n - x^n) = 0^{\eta_n n}\right\}. \quad (E.4)$$

For $1 \leq b \leq \lfloor \frac{0.75n}{\Delta} \rfloor$, $\frac{b\Delta}{n} \leq 0.75$ and therefore,

$$\gamma(\hat{x}^n, y^n) \leq \Gamma_b \leq \frac{b\Delta}{n} \leq 0.75$$

which, together with (5.1), further implies that

$$\frac{1}{n}wt(\hat{x}^n - y^n) < \mathrm{H}^{-1}(0.75)$$

and

$$\begin{aligned}
\frac{1}{n}wt(\hat{x}^n - x^n) &\leq \frac{1}{n}wt(x^n - y^n) + \frac{1}{n}wt(\hat{x}^n - y^n) \\
&< 0.5 + \mathrm{H}^{-1}(0.75) \\
&\leq 1 - \epsilon
\end{aligned}$$

since $\epsilon \leq 0.5 - \mathrm{H}^{-1}(0.75)$. Consequently, we have for any $1 \leq b \leq \lfloor \frac{0.75n}{\Delta} \rfloor$

$$\Pr\left\{\exists \hat{x}^n, \frac{1}{n}wt(\hat{x}^n - x^n) > \epsilon, \mathbf{H}_{b\Delta \times n}^{(b\Delta)}(\hat{x}^n - x^n) = 0^{b\Delta}, \gamma(\hat{x}^n, y^n) \leq \Gamma_b\right\}$$

$$= \Pr\left\{\exists \hat{x}^n, \epsilon < \frac{1}{n}wt(\hat{x}^n - x^n) < 1 - \epsilon, \mathbf{H}_{b\Delta \times n}^{(b\Delta)}(\hat{x}^n - x^n) = 0^{b\Delta}, \gamma(\hat{x}^n, y^n) \leq \Gamma_b\right\}$$

$$\leq 2^{-\Delta + O(1)} \quad (E.5)$$

where the inequality above has been proved in Appendix C. For $b \geq \lfloor \frac{0.75n}{\Delta} \rfloor + 1$, by Lemmas 3 and 7, $P\left(\frac{b\Delta}{n}, \bar{l}, \xi\right)$ is a strictly decreasing function of $\xi$ in the range $\left(0, \bar{l} - \frac{t_{b\Delta}^{(1)}}{n}\right]$. In view of this, it can be shown by the same technique as in Appendix C that for any $b \geq \lfloor \frac{0.75n}{\Delta} \rfloor + 1$

$$\Pr\left\{\exists \hat{x}^n, \frac{1}{n} wt(\hat{x}^n - x^n) > \epsilon, \mathbf{H}_{b\Delta \times n}^{(b\Delta)}(\hat{x}^n - x^n) = 0^{b\Delta}, \gamma(\hat{x}^n, y^n) \leq \Gamma_b\right\} \leq 2^{-\Delta + O(1)} \tag{E.6}$$

and

$$\Pr\left\{\exists \hat{x}^n, \frac{1}{n} wt(\hat{x}^n - x^n) > \epsilon, \mathbf{H}_{n \times n}(\hat{x}^n - x^n) = 0^n, \mathbf{H}'_{\eta_n n \times n}(\hat{x}^n - x^n) = 0^{\eta_n n}\right\} \leq 2^{-\Delta + O(1)}. \tag{E.7}$$

Plugging (E.5), (E.6), and (E.7) into (E.4) yields

$$\Pr\left\{\frac{1}{n} wt(\hat{X}^n - x^n) > \epsilon \,\middle|\, x^n, y^n\right\} \leq 2^{-\Delta + \log_2\left(\frac{n}{\Delta} + 1\right) + O(1)} \tag{E.8}$$

for any $(x^n, y^n)$ with $\frac{1}{n} wt(x^n - y^n) \leq 0.5$. This, combined with (E.3), (E.2), and (E.1), implies

$$P_b(\tilde{\mathcal{I}}_n) \leq \epsilon + 2^{-\Delta + \log_2\left(\frac{n}{\Delta} + 1\right) + O(1)} + e^{-2n(0.5 - p_0)^2}$$

which completes the proof of (5.4) and hence of Theorem 3.

## Appendix F

### Proof of Theorem 4

Note that (5.2) applies to any value of $\bar{l}$, since its proof in Appendix E does not rely on the condition that $\bar{l}$ be an odd integer. Then by using Proposition 3 and following the same approach as that in the proof of Theorem 2, (5.5) is proved, while (5.6) is obvious.

What remains is to prove (5.7). To this end, let $\epsilon = \frac{1}{2\sqrt{k}}$. Then $p_0 < \frac{1-\epsilon}{2}$ as $k > \left(\frac{1}{2(1-2p_0)}\right)^2$. By the same argument as in Appendix E,

$$
\begin{aligned}
P_b(\tilde{\mathcal{I}}_n) &= \mathbb{E}\left[\frac{1}{n} wt(\hat{X}^n - X^n)\right] \\
&\leq \sum_{(x^n, y^n): \frac{1}{n} wt(x^n - y^n) \leq \frac{1-\epsilon}{2}} \Pr\{X^n = x^n, Y^n = y^n\} \mathbb{E}\left[\frac{1}{n} wt(\hat{X}^n - x^n) \,\middle|\, x^n, y^n\right] \\
&\quad + \Pr\left\{\frac{1}{n} wt(X^n - Y^n) > \frac{1-\epsilon}{2}\right\}
\end{aligned}
$$

and

$$\mathbb{E}\left[\frac{1}{n} wt(\hat{X}^n - x^n) \,\middle|\, x^n, y^n\right] \leq \epsilon + \Pr\left\{\frac{1}{n} wt(\hat{X}^n - x^n) > \epsilon \,\middle|\, x^n, y^n\right\}$$

given $\frac{1}{n} wt(x^n - y^n) \leq \frac{1-\epsilon}{2}$. At the same time, by the decoding procedure of Algorithm 3,

$$\gamma(\hat{X}^n, y^n) \leq \gamma(X^n, y^n)$$

and therefore

$$\frac{1}{n}wt(\hat{X}^n - y^n) \leq \frac{1}{n}wt(x^n - y^n)$$

which further implies that

$$\frac{1}{n}wt(\hat{X}^n - x^n) \leq \frac{1}{n}wt(\hat{X}^n - y^n) + \frac{1}{n}wt(x^n - y^n) \leq 1 - \epsilon.$$

Consequently, for any $(x^n, y^n)$ with $\frac{1}{n}wt(x^n - y^n) \leq \frac{1-\epsilon}{2}$,

$$\Pr\left\{\frac{1}{n}wt(\hat{X}^n - x^n) > \epsilon \,\middle|\, x^n, y^n\right\} = \Pr\left\{\epsilon < \frac{1}{n}wt(\hat{X}^n - x^n) \leq 1 - \epsilon \,\middle|\, x^n, y^n\right\}$$

$$\leq 2^{-\Delta + \log_2\left(\frac{n}{\Delta} + 1\right) + O(1)}$$

where the last inequality has been proved in Appendix C. The inequality (5.7) now follows from the fact that

$$\Pr\left\{\frac{1}{n}wt(X^n - Y^n) > \frac{1-\epsilon}{2}\right\} \leq e^{-2n\left(\frac{1-\epsilon}{2} - p_0\right)^2} = e^{-2n\left(0.5 - \frac{1}{4\sqrt{k}} - p_0\right)^2}.$$

This completes the proof of Theorem 4.

## REFERENCES

[1] E.-H. Yang and D.-K. He, "On interactive encoding and decoding for lossless source coding with decoder only side information," in *Proc. of ISIT'08*, July 2008, pp. 419–423.

[2] ——, "Interactive encoding and decoding for one way learning: Near lossless recovery with side information at the decoder," *IEEE Trans. Inf. Theory*, vol. 56, no. 4, pp. 1808–1824, 2010.

[3] E.-H. Yang, A. Kaltchenko, and J. C. Kieffer, "Universal lossless data compression with side information by using a conditional mpm grammar transform," *IEEE Trans. Inform. Theory*, vol. 47, pp. 2130–2150, 2001.

[4] J. Meng, E.-H. Yang, and D.-K. He, "Linear interactive encoding and decoding for lossless source coding with decoder only side information," *IEEE Trans. Inf. Theory*, vol. 57, no. 8, pp. 5281–5297, Aug. 2011.

[5] M. Sartipi and F. Fekri, "Distributed source coding in wireless sensor networks using ldpc coding: The entire slepian-wolf rate region," in *Proc. Wireless Communications and Networking Conference*, 2005.

[6] D. Schonberg, K. Ramchandran, and S. S. Pradhan, "Distributed code constructions for the entire slepian-wolf rate region for arbitarily correlated sources," in *Proc. IEEE Data Compression Conference*, 2004.

[7] ——, "Ldpc codes can approach the slepian-wolf bound for general binary sources," in *Proc. of fortieth Annual Allerton Conference*, Urbana-Champaign, IL, Oct. 2002.

[8] A. D. Liveris, Z. Xiong, and C. N. Georghiades, "Compression of binary sources with side information at the decoder using ldpc codes," *IEEE Comm. Letters*, vol. 6, pp. 440–442, Oct. 2002.

[9] J. Jiang, D. He, and A. Jagmohan, "Rateless slepian-wolf coding based on rate adaptive low-density-parity-check codes," in *Proc. of ISIT'07*, 2007, pp. 1316 –1320.

[10] A. W. Eckford and W. Yu, "Rateless slepian-wolf codes," in *Proc. of Asilomar Conf. on Signals, Syst., Comput'05*, 2005.

[11] D. Varodayan, A. Aaron, and B. Girod, "Rate-adaptive distributed source coding using low-denstiy-parity-check codes," in *Thirty-Ninth Asilomar Conference on Signals, Systems and Computers*, Oct. 2005, pp. 1203–1207.

[12] R. M. Tanner, "A recursive approach to low complexity codes," *IEEE Trans. Inform. Theory*, vol. 27, pp. 533–547, 1981.

[13] T. Richardson and R. Urbanke, *Modern Coding Theory*. Cambridge University Press, 2008.

[14] J. Ziv and A. Lempel, "A universal algorithm for sequential data compression," *IEEE Trans. Inf. Theory*, vol. IT-23, no. No. 3, pp. 337–343, May 1977.

[15] ——, "Compression of lndiwdual sequences via variable-rate coding," *IEEE Trans. Inf. Theory*, vol. IT-24, no. 5, pp. 530–536, Sep. 1978.

[16] J.-C. Kieffer and E.-H. Yang, "Grammar based codes: A new class of universal lossless source codes," *IEEE Trans. Inf. Theory*, vol. IT-46, no. 3, pp. 737–754, May 2000.

[17] E.-H. Yang and J.-C. Kieffer, "Effcient universal lossless compression algorithms based on a greedy sequential grammar transform-part one: Without context models," *IEEE Trans. Inf. Theory*, vol. IT-46, no. 3, pp. 755–777, May 2000.

[18] F. R. Kschischang, B. J. Frey, and H. A. Leoliger, "Factor graphs and the sum-product algorithm," *IEEE Trans. Inf. Theory*, vol. IT-47, pp. 498–519, Feb. 2001.

[19] A. Amraouli, "Lthc: Ldpcopt," online available at the website: http://lthcwww.epfl.ch/research/ldpcopt.

[20] S. Litsyn and V. Shevelev, "On ensembles of low-density parity-check codes: Asymptotic distance distributions," *IEEE Trans. Inf. Theory*, vol. 48, no. 4, pp. 887–908, April 2002.

[21] ——, "Distance distributions in ensembles of irregular low-density parity-check codes," *IEEE Trans. Inf. Theory*, vol. 49, no. 12, pp. 3140–3159, Dec. 2003.

[22] C. Di, T. J. Richardson, and R. L. Urbanke, "Weight distribution of low-density parity-check codes," *IEEE Trans. Inf. Theory*, vol. 52, no. 11, pp. 4839–4855, Nov. 2006.

[23] G. Miller and D. Burshtein, "Asymptotical enumeration method for analyzing ldpc codes," *IEEE Trans. Inf. Theory*, vol. 50, no. 6, pp. 1115–1131, June 2004.

[24] M. P. Mineev and A. I. Pavlov, "On the number of (0,1)-matrices with prescribed sums of rows and columns," *Doc. Akad. Nauk SSSR*, vol. 230, pp. 1276–1282, 1976.

[25] B. McKay, "Asymptotics for 0-1 matrices with prescribed line sums," *Enumeration and Design*, pp. 225–238, 1984.

[26] I. Csiszar and J. Korner, *Information Theory: Coding Theorems for Discrete Memoryless Systems*. Academic Press, INC, 1981.